\documentclass[twocolumn,trackchanges]{aastex631}
\usepackage{url}
\newcommand{\addedc}[1]{#1}

\begin{document}

\title{Detectability of Late-time Supernova Neutrinos with Fallback Accretion onto Protoneutron star}

\author[0000-0002-9234-813X]{Ryuichiro Akaho}
\affiliation{Graduate School of Advanced Science and Engineering,Waseda University,\\
3-4-1 Okubo, Shinjuku, Tokyo 169-8555, Japan}

\author[0000-0002-7205-6367]{Hiroki Nagakura}
\affiliation{Division of Science, National Astronomical Observatory of Japan, \\ 2-21-1 Osawa, Mitaka, Tokyo 181-8588, Japan}

\author[0000-0001-5869-8518]{Thierry Foglizzo}
\affiliation{Université Paris-Saclay, Université Paris Cité, CEA, CNRS, AIM, 91191, Gif-sur-Yvette, France}



\begin{abstract}
We investigate the late-time neutrino emission powered by fallback mass accretion onto a proto-neutron star (PNS), using neutrino radiation-hydrodynamic simulations with full Boltzmann neutrino transport. We follow the time evolution of the accretion flow onto the PNS until the system reaches a quasi-steady state. A standing shock wave is commonly formed in the accretion flow, whereas the shock radius varies depending on the mass accretion rate and the PNS mass. A sharp increase in temperature emerges in the vicinity of the PNS ($\sim 10$ km), which characterizes neutrino emission. Both the neutrino luminosity and the average energy become higher with increasing mass accretion rate and PNS mass. The mean energy of the emitted neutrinos is in the range of $10\lesssim\epsilon\lesssim20\,\mathrm{MeV}$,
which is higher than that estimated from PNS cooling models ($\lesssim10\,\mathrm{MeV}$).
Assuming a distance to core-collapse supernova of $10\,\mathrm{kpc}$, we quantify neutrino event rates for Super-Kamiokande (Super-K) and DUNE.
The estimated detection rates are well above the background, and their energy-dependent features are qualitatively different from those expected from PNS cooling models.
Another notable feature is that the neutrino emission is strongly flavor dependent, exhibiting that the neutrino event rate hinges on the neutrino oscillation model. We estimate them in the case with adiabatic Mikheev-Smirnov-Wolfenstein model, and show that the normal- and inverted mass hierarchy offer a large number of neutrino detections in Super-K and DUNE, respectively. Hence the simultaneous observation with Super-K and DUNE of fallback neutrinos will provide a strong constraint on the neutrino mass hierarchy. 
\end{abstract}



\section{Introduction} \label{sec:intro}
Most massive stars with zero-age main sequence mass $\gtrsim 8 M_{\sun}$ end their lives as core-collapse supernovae (CCSNe). In the central region, a protoneutron star (PNS) is formed as a consequence of gravitational collapse of its iron core. A huge gravitational energy of the PNS is released via neutrinos. The neutrinos absorbed behind the stalled shock wave can foster shock expansion. 

After shock revival, the
evolution of the PNS proceeds on a Kelvin-Helmholtz timescale of neutrino cooling (of the order of ten seconds) \citep{Burrows1986}, that is much longer than the timescale on which shock revival is initiated by the neutrino-heating mechanism ($\lesssim 1$s) \citep[see for a recent review, e.g.,][]{Burrows2021}. The neutrino signal thus carries imprints of not only the explosion mechanism but also the subsequent evolution of the remnant system. This argument is in line with neutrino data of SN 1987A \citep{Hirata1987,Bionta1987}.
On the other hand, the low-statistics neutrino data from SN 1987A provided little constraint on the detailed features of neutrinos such as flavor-dependent features and their time structure. With current neutrino detectors the next nearby CCSN will provide high-statistics signals placing constraints on the CCSN dynamics.

Significant progress has been made in the last decades in neutrino detection techniques.
Various types of neutrino detectors such as water Cherenkov \citep{Ikeda2007,Abe2018,Abbasi2011}, liquid argon \citep{Abi2021},
liquid scintillators \citep{Asakura2016, An2016}, and dark-matter detectors \citep{Lang2016} 
are currently operating or planned, offering the means to distinguish neutrino flavors \citep{Horiuchi2018}. The sensitivity of detectors has steadily improved, and the detector size has also increased by more than an order of magnitude compared to those used in the 1980s, suggesting that the long-term neutrino signal will be detectable for nearby CCSNe \citep[see, e.g.,][]{Suwa2019,Li2021}. 

This has motivated the CCSN community to develop theoretical models for long-term evolution of PNS cooling. Numerical simulations have also been performed
\citep[see, e.g.,][]{Fischer2010,Roberts2012a,Roberts2012b}, and various physical quantities relevant to PNS cooling have also been investigated: the nuclear matter equation of state \citep{Nakazato2019,Nakazato2020,Nakazato2022,Sumiyoshi2023}, progenitor dependence \citep{Nakazato2013}, and neutrino matter interactions
\citep{Fischer2012,Martinez2012,Fischer2020,Pascal2022,Sugiura2022}.
These works will play pivotal roles to place a constraint on microphysical parameters in supranuclear density of PNS in real observations.

In this paper, we discuss the late time neutrino emission of CCSN from a different perspective: fallback accretion (FBA) onto the PNS. Most previous studies have a priori assumed that FBA has no influence on the neutrino signal. One thing we do notice here is, however, that large amounts of FBA have been observed rather commonly in recent multi-dimensional (multi-D) CCSN simulations \citep[see, e.g.,][]{Burrows2021, Bollig2021, Nagakura2021a}. More interestingly, they may last a very long time ($\gg 10$s) \citep[see, e.g., Fig.2 in][]{Janka2022} due to the shock deceleration or reverse shock that occurs after the shock wave passes the CO/He-core interface \citep{Fryxell1991} and He/H interface \citep{Chevalier1989}. This suggests that the neutrino emission from FBA potentially overwhelms those radiated from PNS.

The impact of FBA in the late time neutrino emission was investigated by the pioneering work of \citet{Fryer2009}. This study showed that FBA has a large influence on the neutrino luminosity and their average energy. It should be noted, however, that there are potential systematic uncertainties in their models; for instances, the inner boundary of the computational domain in the simulations is located much outside the neutrino sphere (which will be shown in Sec.~\ref{subsec:neutrinodistri} and see also Table~2 in \citet{Fryer2009}), and the neutrino transport was handled with a gray flux-limited diffusion approximation \citep{Herant1994}. These simplifications prevented them from studying detailed features of neutrinos from FBA, and they may discard some important properties inherent in FBA.

In this paper, we investigate the neutrino emission driven by FBA onto the PNS by performing neutrino radiation-hydrodynamics simulations covering the optically thick region in the computational domain. We analyze the neutrino emission by systematically changing the accretion rate (at the outer boundary) and the PNS mass. Based on the numerical simulations, we provide some key features of the neutrino signal from FBA, and then the neutrino event rates in some representative CCSN neutrino detectors are estimated. 

This paper is organized as follows.
In Sec.~\ref{sec:fallback}, we start with providing an overview of FBA onto PNS in the post explosion phase, and then we describe our approach to study the neutrino signal powered by FBA. The numerical methods and the PNS models are described in Sec.~\ref{sec:method}. All results of our numerical simulations are encapsulated in Sec.~\ref{sec:results}.
The detectability of neutrinos is discussed in Sec.~\ref{sec:detectability}. We summarize our findings in section \ref{sec:summary}. 

\section{Fallback Accretion in CCSNe}
\label{sec:fallback}
Even after the shock wave begins its runaway expansion, a certain amount of post-shock matter is bound by the gravity of the PNS, and it eventually returns back to the PNS.
Such FBA in CCSNe has been studied in the literature from the early 1970s. The importance of FBA was first pointed out by \citet{Colgate1971}. They suggested that FBA is necessary to explain the consistent amount of nucleosynthetic yields. From the observational point of view, some previous studies suggested that FBA has an influence on both electromagnetic- \citep{Dexter2013} and neutrino emission \citep{Fryer2009} in the late phase. We also note that FBA potentially accounts for some peculiar energetic \citep{Moriya2018,Moriya2019} and weak CCSN explosions \citep{Moriya2010}.
If FBA leads to an oversupply of mass onto the PNS, it may trigger a black hole formation \citep{Zhang2008,Chan2018}.
If the core is rapidly rotating, gamma-ray burst would occur following the collapsar scenario \citep{MacFadyen1999,MacFadyen2001,Perna2014}. 
FBA can also affect the PNS spin \citep{Barrere2022,Ronchi2022,Coleman2022} and its spin-kick alignment indicated by some pulsar observations \citep{Johnson2005,Johnson2007,Ng2007,Janka2022}.

FBA in CCSNe can be categorized into several phases \citep{Chevalier1989}. 
In the early post shock revival phase, it would be chaotic due to the turbulent accretion flows originated from multi-D fluid instabilities in the post-shock region. It should also be mentioned that asymmetric shock revival can lead to large FBA from the angular region where the shock expansion is weaker \citep{Nagakura2021a,Bollig2021}.
In the very later phase, which is referred to as the uniform expansion phase $\gtrsim10^3\,\mathrm{s}$, the accretion rate simply scales as $\dot M\propto t^{-5/3}$ \citep{Chevalier1989}. This scaling is verified by various numerical studies \citep{Zhang2008,Dexter2013,Janka2022}.
Note that the accretion rate on this free-expansion phase may be significantly enhanced by the arrival of the reverse shock onto PNS. 
It has also been suggested that strong FBA can be formed by the deceleration of the shock at the CO/He-core or He/H interfaces \citep{Janka2022} (see also Fig.~2 of \citet{Zhang2008}, in which the enhancement of FBA is clearly visible).

In this paper, we pay attention to the phase of $\gtrsim 10\,\mathrm{s}$ after core bounce. In this phase, the PNS temperature at the surface becomes less than $\sim 3$ MeV \citep{Roberts2012b,Nakazato2013}, and the neutrino emission gradually subsides in the Kelvin-Helmholtz timescale. This suggests that the neutrino emission can be dominated by FBA, inferred from the previous works \citep{Fryer2009,Nagakura2020,Nagakura2021a,Bollig2021}.
\addedc{
It should also be noted that we develop a general discussion of FBA neutrino emission without specifying any late phases in this study, since our approach can be applied to different situations. Nevertheless, the increase of FBA by a reverse shock created at the CO/He-core or He/H interfaces is an intriguing phase, since a large FBA may happen at a very late phase of CCSNe ($\gtrsim10^3\,\mathrm{s}$) \citep{Zhang2008}.
}

\section{Methods and Models}
\label{sec:method}
\subsection{Numerical Method}
\label{sec:numerical}
We perform neutrino radiation-hydrodynamics simulation of FBA onto the PNS. We employ a general relativistic Boltzmann radiation-hydrodynamics code, in which we solve the hydrodynamic equation and the Boltzmann equation for neutrino transport, both in general relativity. The numerical details can be found in a series of papers \citep{Nagakura2014,Nagakura2017,Nagakura2019,Akaho2021,Akaho2023}.
\addedc{Although our code can treat multi-dimension in space, we assume spherically symmetry in this study.}

\addedc{The Boltzmann neutrino transport directly solves the neutrino distribution function 
under multi-species, multi-energy, and multi-angle treatments. This allows us to develop accurate models of the neutrino radiation field even in semi-transparent regions, whereas the accuracy is not guaranteed in other approximate methods (e.g., MGFLD and two-moment methods). The interesting issue of quantifying the error for each approximate transport method is beyond the scope of this paper.
}

\addedc{
Assuming spherical symmetry, the distribution function can be expressed with four variables; time ($t$), radius ($r$), neutrino energy ($\epsilon$), and the zenith angle in the momentum space ($\theta_\nu$).
We note that the spacetime metric is determined by solving
the Tolman-Oppenheimer-Volkov (TOV) equations used to construct the NS, as we explain later in section \ref{sec:PNSmodel}. 
The explicit form of the Boltzmann equation in the conserved form can be written
as \citep{Shibata2014}
\begin{eqnarray}
\label{eq_conservBoltz}
 \frac{1}{\alpha}\frac{\partial f}{\partial t} 
&+& \frac{\mathrm{cos}\theta_\nu}{\alpha r^2 \sqrt{\gamma_{rr}}}\frac{\partial}{\partial r}\left(\alpha r^2 f\right)
-\frac{1}{\epsilon^2}\frac{\partial}{\partial\epsilon}\left(\epsilon^3f\omega_{(t)}\right) \nonumber \\
&+& \frac{1}{\mathrm{\mathrm{sin}\theta_\nu}}\frac{\partial}{\partial\theta_\nu}\left(\mathrm{sin}\theta_\nu f\omega_{(\theta_\nu)}\right) = S_{\rm rad},
\end{eqnarray}
where $\alpha$, and $\gamma_{ij}$ denote the lapse function and the spatial metric, respectively. 
The factors $\omega_{(t)}$ and $\omega_{(\theta_\nu)}$ are defined as
\begin{eqnarray}
\omega_{(t)} &\equiv& \epsilon^{-2} p^\mu p_\nu \nabla_\mu e_{(t)}^\nu, \nonumber \\
\omega_{(\theta_\nu)} &\equiv& - \epsilon^{-2} p^\mu p_\nu \nabla_\mu e_{(r)}^\nu\mathrm{sin}\theta_\nu, 
\end{eqnarray}
where $p^\mu$ denotes the neutrino four-momentum.
The tetrad bases are given as
\begin{eqnarray}
e^\mu_{(t)} & \equiv& \left(\alpha^{-1}, 0,0,0\right),
\nonumber \\
e^\mu_{(r)} & \equiv& 
\left(0,\gamma_{rr}^{-1/2},0,0\right), \label{eq_tetrad}
\end{eqnarray}
}

Although the FBA is a priori multi-D, the accretion energy converts to thermal energy in the vicinity of PNS, and eventually spreads all over the PNS surface. This suggests that the asymmetry of neutrino emission becomes milder than that of FBA, \addedc{and numerical simulations also support this assumption \citep{Vartanyan2019}.}
Whether large asymmetries of neutrino emission can be created by non-radial FBA is an interesting question which we defer to future work.
\addedc{We also note that the spherically symmetric conditions artificially suppress the PNS convection. One may wonder if this may cause to underestimate the diffusion component of the neutrino luminosity. According to recent multi-D simulations, however, the PNS convection subsides by $\sim 5$s after bounce \citep{Nagakura2021a}, suggesting that it does not affect FBA neutrinos in the late phase ($t \gtrsim 10$s).}

We employ $512$ radial grid points covering the range $r\in\left[0:100\right]\,\mathrm{km}$.
We note that the resolution is high around the PNS surface (where the minimum mesh width is $\sim30\,\mathrm{m}$). Such a high spatial resolution is mandatory in studying FBA, since the scale heights of matter- and neutrino-radiation field around the PNS are very small.
The number of energy mesh is $20$, covering the range $\epsilon\in\left[0:300\right]\,\mathrm{MeV}$ and the mesh width is logarithmically distributed. The zenith angle $\theta_\nu\in\left[0:\pi\right]$ has $10$ grid points.

We employ the Furusawa-Togashi EOS \citep{Furusawa2017} with some extension. 
It should be mentioned that, in the case with low mass accretion rate, the thermodynamical quantities can be outside of the range covered by the EOS table.
To deal with this issue, we extended the EOS table in a pragmatic way; it is smoothly connected to the gamma-law EOS as the pressure given as $P=(\Gamma-1)\rho\epsilon$, where $\rho$ and $\epsilon$ denote the density and the specific internal energy, respectively. The gamma law index $\Gamma$ is obtained from the edge of the EOS table. 
We found that $\Gamma$ is almost $4/3$ for various input parameters. 
We note, however, that our prescription is rather pragmatic, and it does not have the ability to capture the realistic matter evolution; in particular for shock dynamics. For this reason, we stop the calculation if the shock wave reaches the position where thermodynamical quantities are out of the range of the Furusawa-Togashi EOS.
\addedc{On the other hand, these prescriptions do not compromise the present result, since neutrino emission occurs in high density regions, which are always covered by the Furusawa-Togashi EOS.
}

Regarding neutrino-matter interactions, we incorporate them based on the standard set \citep{Bruenn1985} with some extensions: energy-changing neutrino-electron scattering and the nucleon–nucleon bremsstrahlung.
See \citet{Sumiyoshi2005,Nagakura2017} for the numerical implementations of these reactions.

\subsection{Models}
\label{sec:PNSmodel}
The accurate determination of the neutrino emission from FBA requires resolving the PNS surface where the accretion energy is converted to thermal energy.
We also note that the neutrino opacity hinges on the energy, and the low energy neutrinos can escape from the very high density region ($> 10^{14} {\rm g/cm^3}$), exhibiting that the PNS structure needs to be determined to quantify the neutrino energy spectrum. Hence, we construct the PNS structure by assuming steady state, before running FBA simulations.

We assume an isotropic temperature of $T=2\,\mathrm{MeV}$ and the electron fraction of $Y_e=0.05$ inside the PNS as a reference model, which represents the matter state of the PNS in the cooling phase. For the sake of completeness, the temperature dependence in the neutrino signal is also checked in this study (see Sec.~\ref{subsec:pnstemperadepe}). Given $T$ and $Y_e$, we prepare two different PNS structures by solving TOV equations, by varying the central baryon mass density. 
The first one has the central density of $\rho_c=8.5\times10^{14}\,\mathrm{g}\cdot\mathrm{cm}^{-3}$, leading to the total mass $M_\mathrm{PNS}=1.41M_\odot$.
For the second one, we set $\rho_c=1.2\times10^{15}\,\mathrm{g}\cdot\mathrm{cm}^{-3}$, that leads to $M_\mathrm{PNS}=1.98M_\odot$. 
The spacetime metric obtained from solving the TOV equations is used for the radiation-hydrodynamic simulations, and kept fixed in time.

Below, we describe our FBA model. One thing to note here is that the self-consistent treatment of FBA requires successful CCSN explosion models. It should be noted, however, that detailed features of FBA such as the mass accretion rate, the thermodynamical states, and their time evolution strongly depend on the progenitor, the timing of the shock revival, and the ejecta morphology. In this study, we are not interested in such details of the neutrino signal, but rather in generic features that can be applied to any types of FBA. To this end, we treat FBA in a simple manner capturing the essential features. In our models, we assume a mass inflow from the outer boundary of the computational domain. The accretion rate is one of the control parameters, and we study four cases: $\dot M=10^{-2}, 5 \times 10^{-3}, 2 \times 10^{-3},$ and $10^{-3}\, M_\odot\cdot\mathrm{s}^{-1}$. 
\addedc{
The choice of mass accretion rate is motivated by previous studies of FBA by 
\citep{Chan2018,Moriya2019,Janka2022});. According to their results, strong FBA ($10^{-3} M_{\odot}/s$) can occur in the late phase ($> 10$s) for some progenitors. In this study, we increase the mass accretion rate to see its dependence on the neutrino luminosity. We note that it is necessary to check this dependence by decreasing the mass accretion rate for the sake of completeness, but these simulations are currently not available due to some technical problems associated to EOS tables. Addressing this issue is postponed to a future work.} 
The matter temperature is set as $T=0.5\,\mathrm{MeV}$. 
We note that this setup (cold FBA) leads to a conservative estimation of the neutrino signal by FBA. $Y_e$ is set to be $0.5$. We run each model until the system reaches a quasi-steady state.

For computational reasons, the temperature inside $8\,\mathrm{km}$ is fixed in time.
\addedc{It is well inside the PNS; in fact the matter density is $\rho > 1.5 \times 10^{14} {\rm g/cm^3}$ and its temperature is also low ($\sim 2$MeV), indicating that the boundary condition does not affect the neutrino emission. On the other hand,
}
this prescription is necessary to prevent the PNS from over-cooling. In fact, if the temperature evolution is fully solved, its monotonic decrease with time may cause numerical crashes. As we shall show below, the PNS temperature does not affect the neutrino emission by FBA, indicating that this prescription does not compromise the present result.

\section{Results}
\label{sec:results}

\subsection{Matter distributions}
\label{subsec:matterdistri}
\begin{figure}[htbp]
    \begin{tabular}{c}
      \begin{minipage}[t]{1.0\hsize}
        \centering
        \includegraphics[scale=0.7]{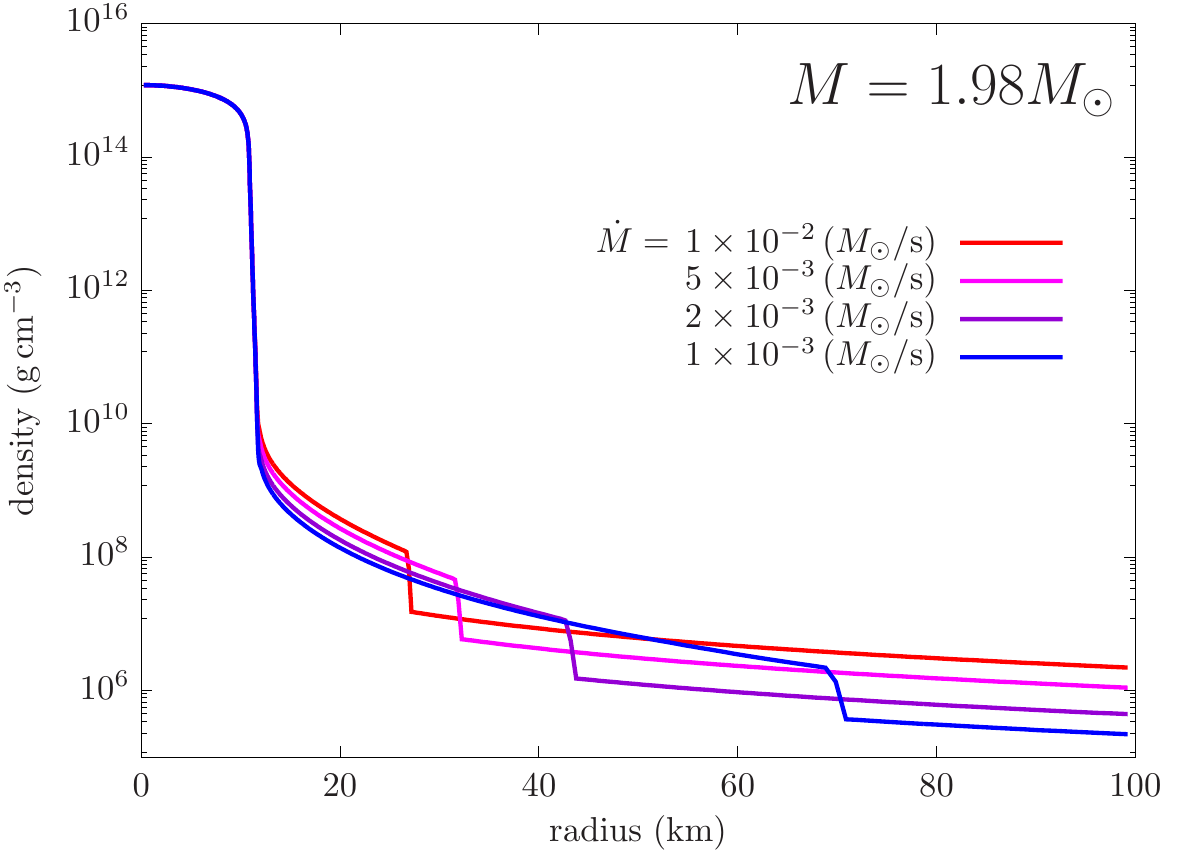}
      \end{minipage} \\
      \begin{minipage}[t]{1.0\hsize}
        \centering
        \includegraphics[scale=0.7]{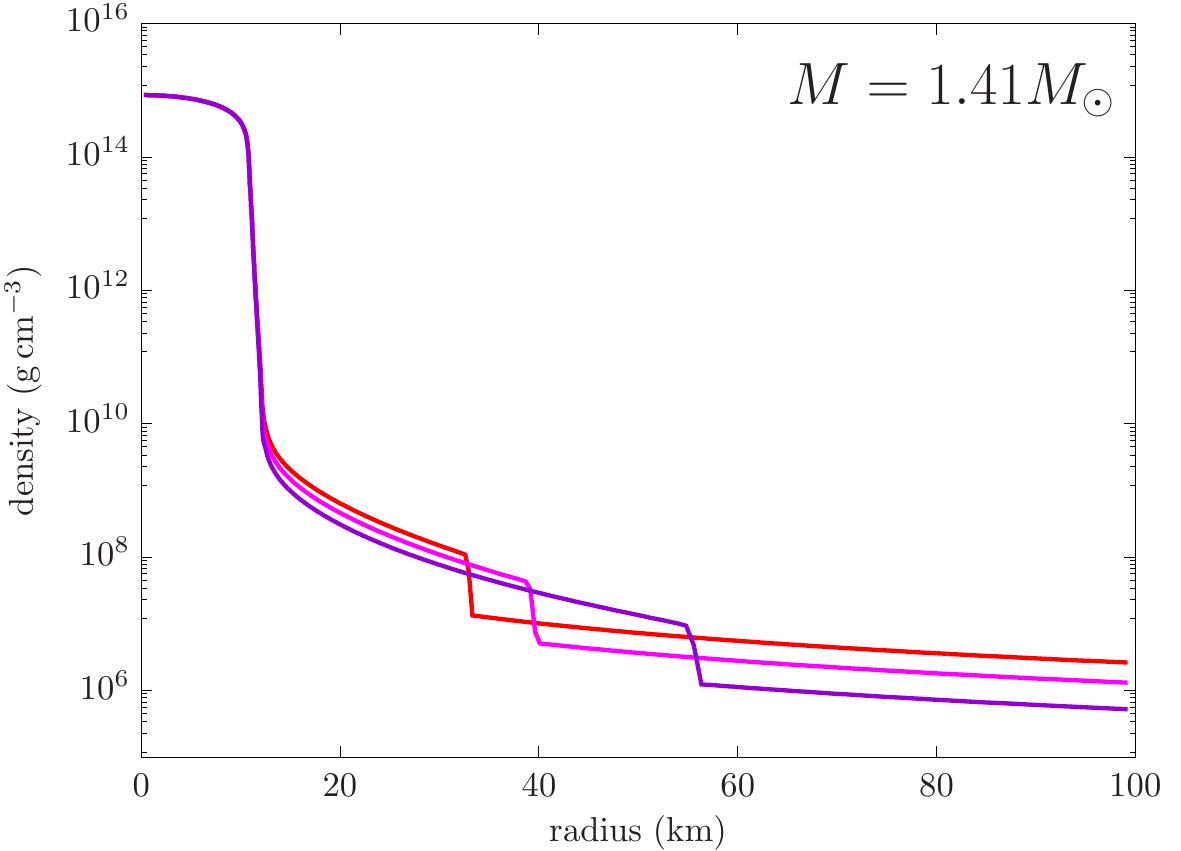}
      \end{minipage}
    \end{tabular}
    \caption{Radial profiles of the density for the $1.98M_\odot$ model (top) and the $1.41M_\odot$ (bottom). The different colors indicate different accretion rates.}
    \label{fig_rho}
\end{figure}

\begin{figure}[htbp]
    \begin{tabular}{c}
      \begin{minipage}[t]{1.0\hsize}
        \centering
        \includegraphics[scale=0.7]{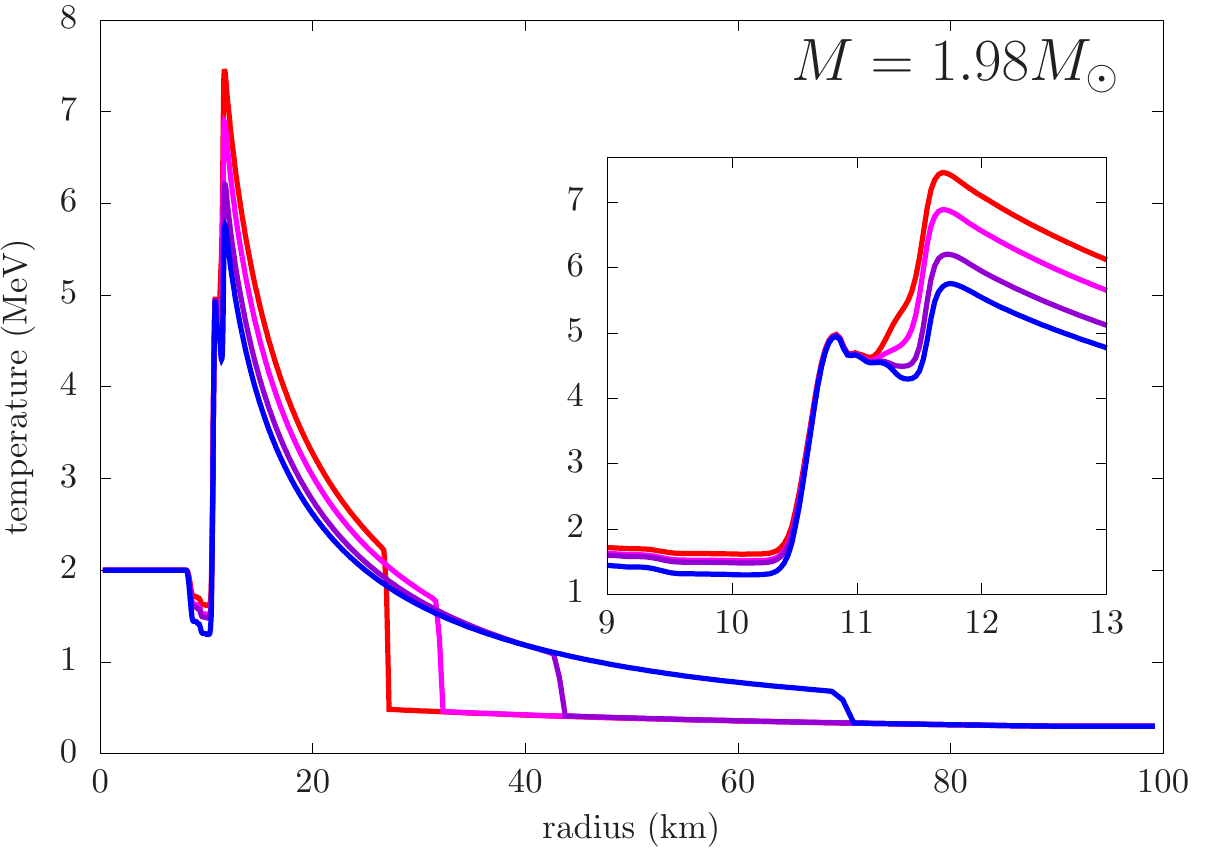}
      \end{minipage} \\
      \begin{minipage}[t]{1.0\hsize}
        \centering
        \includegraphics[scale=0.7]{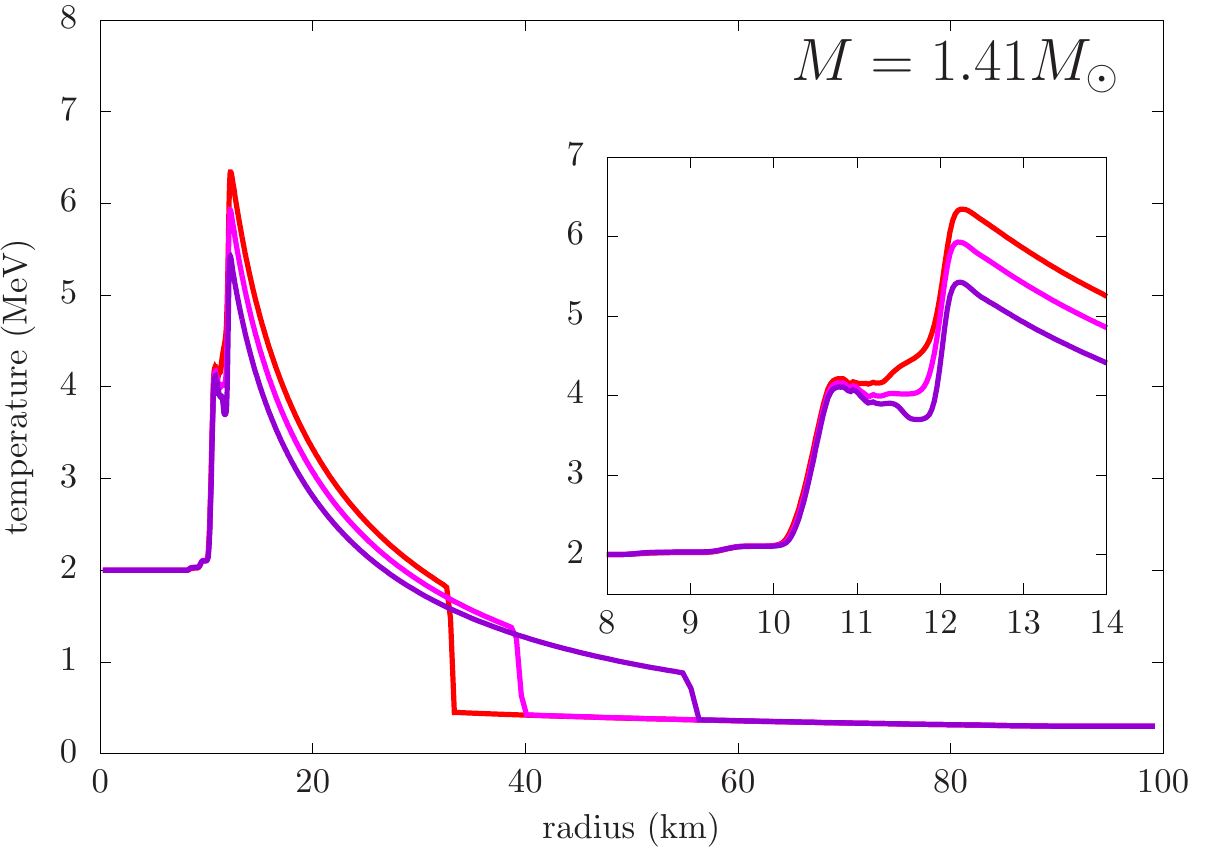}
      \end{minipage}
    \end{tabular}
    \caption{Same as figure \ref{fig_rho}, but for the temperature.}
    \label{fig_temp}
\end{figure}

\begin{figure}[htbp]
    \begin{tabular}{c}
      \begin{minipage}[t]{1.0\hsize}
        \centering
        \includegraphics[scale=0.7]{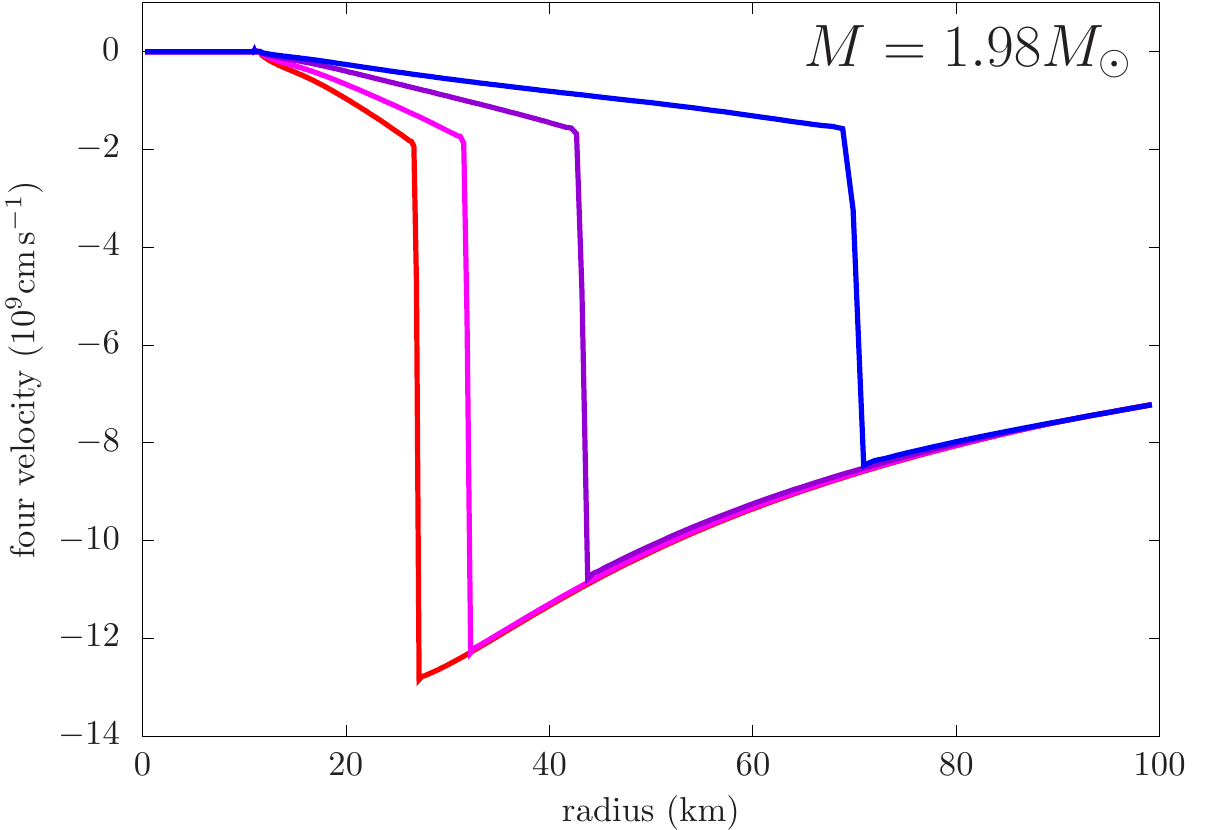}
      \end{minipage} \\
      \begin{minipage}[t]{1.0\hsize}
        \centering
        \includegraphics[scale=0.7]{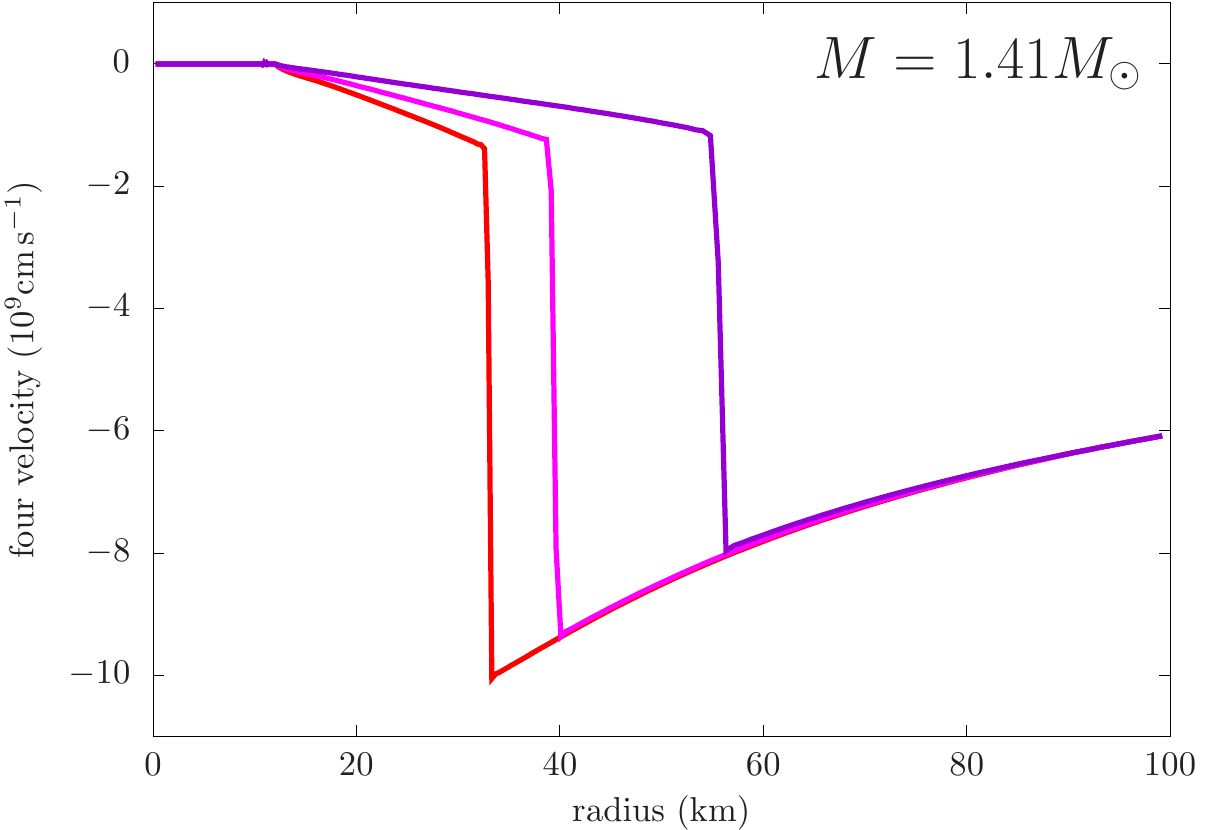}
      \end{minipage}
    \end{tabular}
    \caption{Same as figure \ref{fig_rho}, but for the four-velocity.}
    \label{fig_velo}
\end{figure}

We first focus on the matter distributions of FBA after the system has settled to a quasi-steady state. \addedc{The time it took to reach the steady state varies for different models, in the range of $50\lesssim t \lesssim120\,\mathrm{ms}$. It took longer for lower accretion rate models.}
Figures \ref{fig_rho}, \ref{fig_temp}, and \ref{fig_velo} show the density, temperature, and four-velocity of the fluid.
In these figures, color distinguishes models with different mass accretion rates $\dot M$ and PNS masses $M_\mathrm{PNS}$: four models
($\dot M=1\times10^{-2}$ to $1\times10^{-3}\, M_\odot\cdot\mathrm{s}^{-1}$) for $M_\mathrm{PNS}=1.98M_\odot$, and three models ($\dot M=1\times10^{-2}$ to $2\times10^{-3}\, M_\odot\cdot\mathrm{s}^{-1}$) for $M_\mathrm{PNS}=1.41M_\odot$.

As can be seen in these figures, an accretion shock wave is formed due to FBA. We note that similar phenomena are observed in recent multi-D CCSN simulations (see, e.g., Fig.~9 in \citet{Nagakura2021a}). According to these CCSN models, FBA tends to be cold or lower entropy (otherwise the thermal pressure hampers accretion), and the downflow onto the PNS becomes supersonic. At the surface of the PNS, the fluid needs to be subsonic, implying that a shock wave is inevitably formed. As displayed in these figures, the shock position is larger for the lower accretion rate and the lower PNS mass. This is attributed to the lower ram pressure in the preshock region.

Another notable feature displayed in Fig.~\ref{fig_temp} is that sharp peak profiles in the temperature distribution emerge in the vicinity of the PNS. To see the profile clearly, we magnify the corresponding region ($10 {\rm km} \le r \le 13 {\rm km}$), which is displayed in the same panel.
The sharp increase of temperature is due to the hard wall of the PNS surface, in which the matter density changes by four orders of magnitude over a width $\lesssim 1\,\mathrm{km}$. 
In this region, the kinetic energy of FBA can be efficiently converted into thermal energy, and therefore the temperature of FBA also increases rapidly with decreasing radius. \footnote{A smaller PNS radius means that a larger gravitational energy is converted into thermal energy. Since our focus is the late phase, the PNS radius is thought to have shrunk to a small radius due to cooling (in our setting, $\sim11\,\mathrm{km}$). This situation is actually advantageous for creating a high temperature peak and leads to a larger amount of neutrino emission.}
We also note that the thermal energy of matter is proportional to $\dot{M} {v_{\rm ff}}^2$, where $v_{\rm ff}$ denotes the free-fall velocity where the kinetic energy is dissipated. Since the dissipation region is less sensitive to the PNS mass, $v_{\rm ff}$ is proportional to $M_{\rm PNS}^{0.5}$. This is the rationale behind the higher peak temperature for the higher mass accretion rate and the higher PNS mass (see Fig.~\ref{fig_temp}).

Contrary to the trend which we have discussed so far, the temperature decreases rapidly with decreasing radius at $\lesssim 11.5 {\rm km}$ and $\lesssim 12 {\rm km}$ for $M_\mathrm{PNS}=1.98M_\odot$ and $1.41M_\odot$, respectively (see the magnified figure of Fig.~\ref{fig_temp}). This exhibits that neutrino cooling gives feedback on the matter distribution. On the other hand, the temperature profile is very complicated in the transition layer between the cold PNS envelope and the inner edge of FBA. As we shall show below, weak processes are responsible for the complex radial profile in the temperature distribution. 
It is also interesting to note that the matter profile in the region $10 {\rm km} \lesssim r \lesssim 11 {\rm km}$ does not depend on the mass accretion rate. Although we postpone the detailed investigation to future work, this may be due to a self-regulation mechanism around the PNS surface. Since the matter pressure needs to be connected smoothly across the layer, the fluid element at the PNS surface undergoes shrinking. This implies that the gravitational energy is converted into thermal energy, which also accounts for the increase of neutrino luminosity, in particular for heavy-leptonic neutrinos ($\nu_x$).



\subsection{Neutrino distributions}
\label{subsec:neutrinodistri}
Before going into details, let us first provide the information on species-dependent neutrino spheres. As a reference, we show them in the case of $\dot M=10^{-3}\, M_\odot\cdot\mathrm{s}^{-1}$ and the PNS mass $1.98M_\odot$.
The neutrino spheres for the energy of $23.4\,\mathrm{MeV}$, roughly corresponding to the average energy of neutrinos, are $11.2\,\mathrm{km}$ $10.7\,\mathrm{km}$, $3.66\,\mathrm{km}$ for $\nu_e$, $\bar\nu_e$, and $\nu_x$, respectively. The density at each neutrino sphere is $2.55\times10^{12}\,\mathrm{g}\cdot\mathrm{cm}^{-3}$, $1.88\times10^{14}\,\mathrm{g}\cdot\mathrm{cm}^{-3}$ and 
$1.11\times10^{15}\,\mathrm{g}\cdot\mathrm{cm}^{-3}$, respectively.
This exhibits that the neutrino sphere is located at higher matter density than in the early post-bounce phase (a few hundreds of milliseconds after core bounce); for instance, the neutrino sphere of $\nu_e$ is located at $\sim 10^{11} {\rm g/cm^3}$ in the early post-bounce phase. This difference can be understood as follows. The density gradient becomes so steep in the late phase, indicating that the scale height in this region becomes small. Since the neutrino optical depth is determined not only by the local reaction rate but also by the scale height, the optical depth tends to be smaller in the late phase for the region with the same matter density. It is also worthy to note that these neutrino spheres are located much deeper than the inner boundary adopted in the simulations of \citet{Fryer2009}.

\begin{figure}[htbp]
    \begin{tabular}{c}
      \begin{minipage}[t]{1.0\hsize}
        \centering
        \includegraphics[scale=0.7]{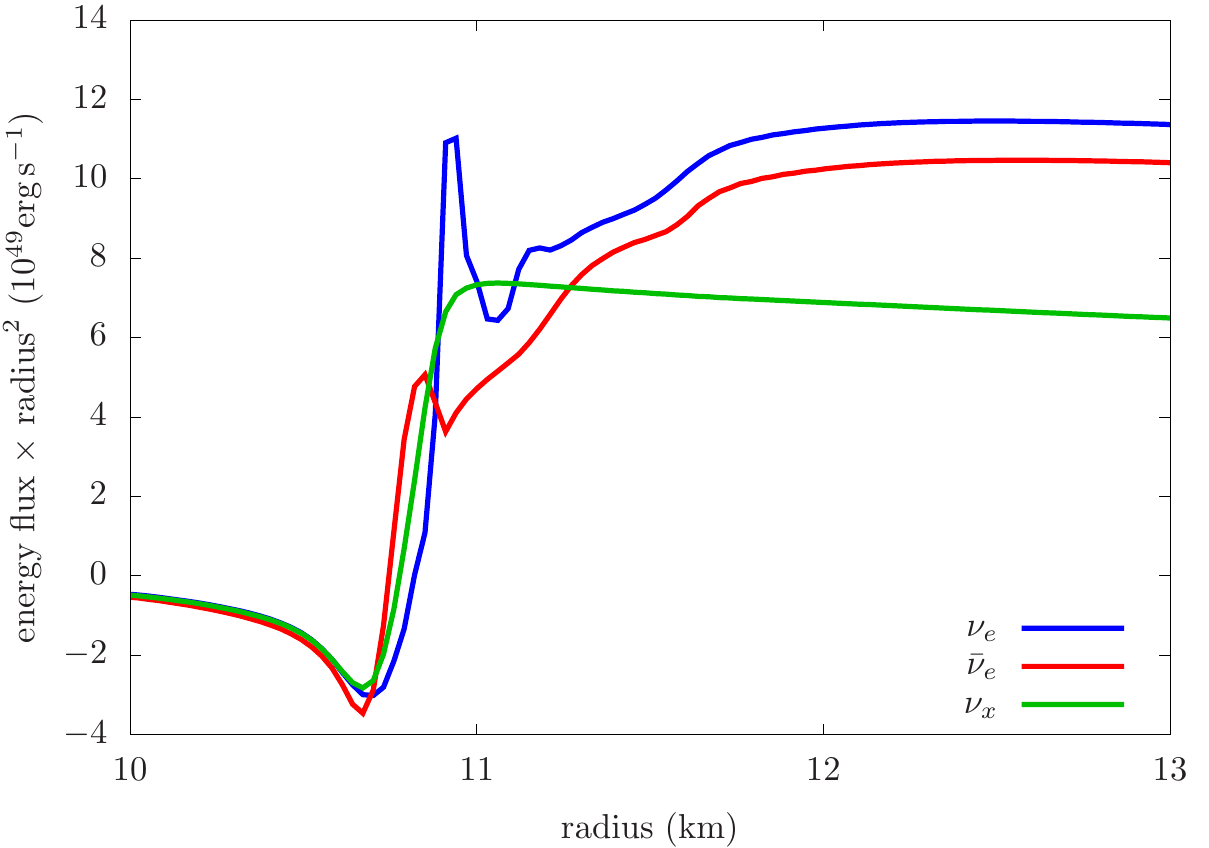}
      \end{minipage} \\
      \begin{minipage}[t]{1.0\hsize}
        \centering
        \includegraphics[scale=0.7]{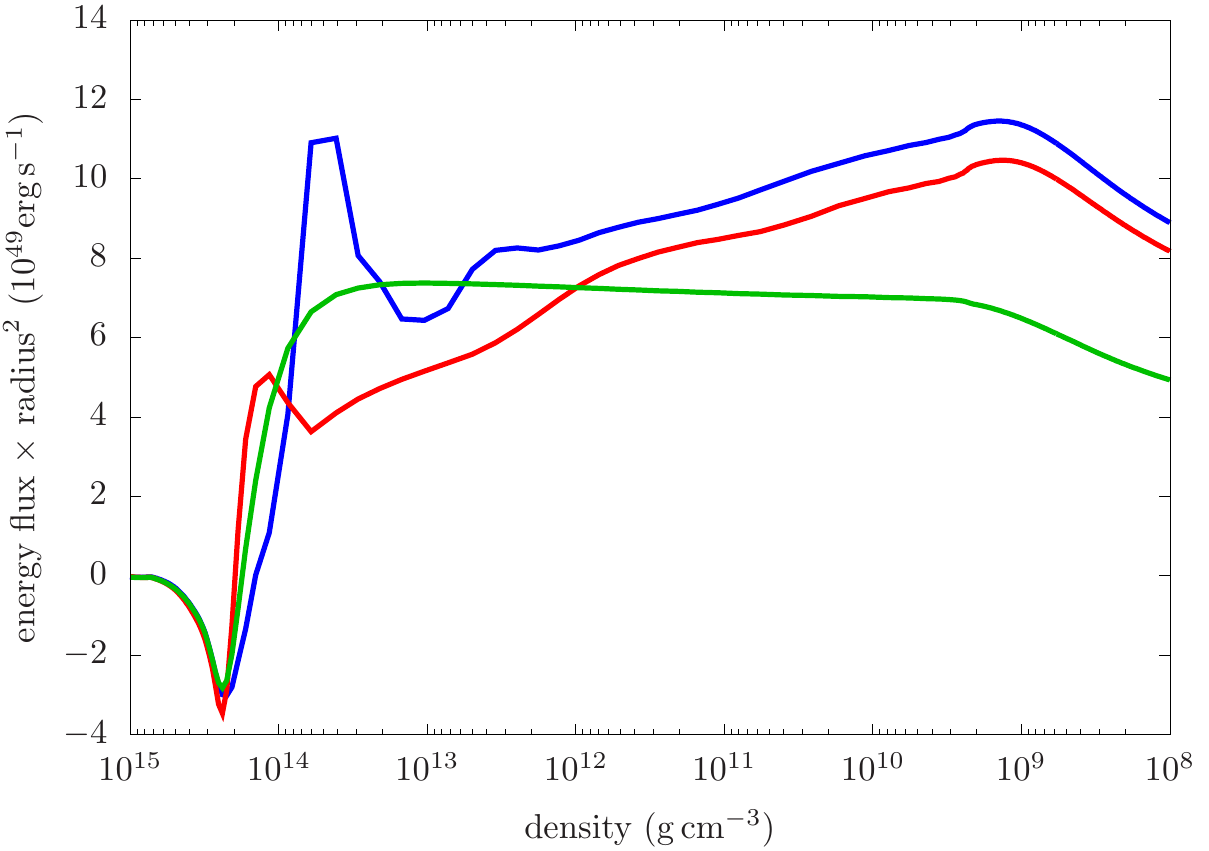}
      \end{minipage}
    \end{tabular}
    \caption{Energy flux times the square of radius for the PNS mass $1.98M_\odot$ and the accretion rate $\dot M=10^{-3}\, M_\odot\cdot\mathrm{s}^{-1}$. The horizontal axis are the radius and density, for top and bottom panels, respectively.}
    \label{fig_flux}
\end{figure}

To delve into the neutrino feedback on matter, we portray the radial profiles of the energy flux ($F_\nu$) of each species of neutrinos in the case of the PNS mass $1.98M_\odot$ and the accretion rate $\dot M=10^{-3}\, M_\odot\cdot\mathrm{s}^{-1}$ (see Fig~\ref{fig_flux}.). In the figure, neutrino fluxes are multiplied by a factor $r^2$. We note that $F_\nu r^2$ is approximately constant in space, if there are no neutrino emission and absorption\footnote{Strictly speaking, $F_\nu r^2$ is not constant in curved spacetime. However, the deviation due to general relativistic (GR) effects is minor and not important for the argument; hence we multiply by a factor $r^2$ without any GR corrections just for simplicity.}. This indicates that the information on neutrino cooling (or heating) is imprinted in the radial profile of $F_\nu r^2$. As shown in the top panel of Fig~\ref{fig_flux}, neutrino fluxes for $\nu_e$ and $\bar{\nu}_e$ increase with radius in the region $11 {\rm km} \le r \le 11.5 {\rm km}$, indicating that these neutrinos are substantially produced there. The fact that $\nu_x$ is approximately constant in space indicates that $\nu_x$ is not produced in this region. It is also worthy to note that $\nu_e$ absorption dominates over emission in the narrow region at $\sim 10.9 {\rm km}$ (see blue line). A similar profile is also observed for $\bar{\nu}_e$ at smaller radius (see the red line).

It is also informative to see the temperature profile as a function of the matter density, which is displayed in the bottom panel of Fig~\ref{fig_flux}. For $\nu_e$ ($\bar{\nu}_e$), strong neutrino production occurs at very high density $5 \times 10^{13} {\rm g/cm^3}
\lesssim \rho \lesssim 2 \times 10^{14} {\rm g/cm^3}$ ($10^{14} {\rm g/cm^3} \lesssim \rho \lesssim 2 \times 10^{14} {\rm g/cm^3}$). When these neutrinos propagate outwards in the lower density environment, neutrino absorption becomes dominant in the region $10^{13} {\rm g/cm^3} \lesssim \rho \lesssim 5 \times 10^{13} {\rm g/cm^3}$ ($5 \times 10^{13} {\rm g/cm^3} \lesssim \rho \lesssim 10^{14} {\rm g/cm^3}$), but neutrino emission again dominates over absorption until $\rho \sim 10^{9} {\rm g/cm^3}$. These non-monotonic profiles of $\nu_e$ and $\bar{\nu}_e$ fluxes are clearly associated with the matter temperature profile, which shall be discussed later. Our result also suggests that there is a substantial amount of diffusion component for both $\nu_e$ and $\bar{\nu}_e$ in their energy fluxes, which are missing components in the simulations of \citet{Fryer2009}. We also find that $\nu_x$ profile is much simpler than others; $\nu_x$ is mainly produced at $5 \times 10^{13} {\rm g/cm^3} \lesssim \rho \lesssim 2 \times 10^{14} {\rm g/cm^3}$ and then they freely escape from the system. This suggests that $\nu_x$ production mainly occurs in such a high density region. It is, hence, mandatory to cover the high density region in numerical simulations to quantify the neutrino signal from FBA.

\begin{figure}[htbp]
    \begin{tabular}{c}
      \begin{minipage}[t]{1.0\hsize}
        \centering
        \includegraphics[scale=0.7]{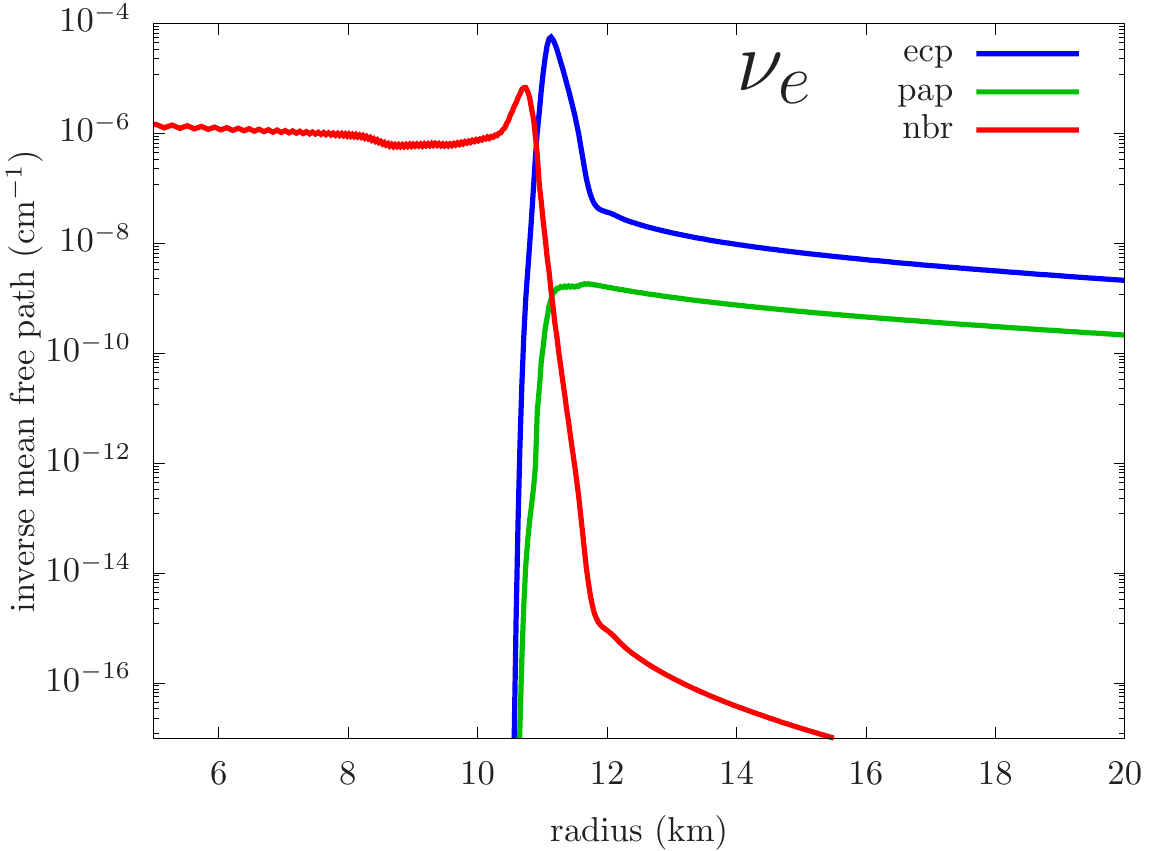}
      \end{minipage} \\
      \begin{minipage}[t]{1.0\hsize}
        \centering
        \includegraphics[scale=0.7]{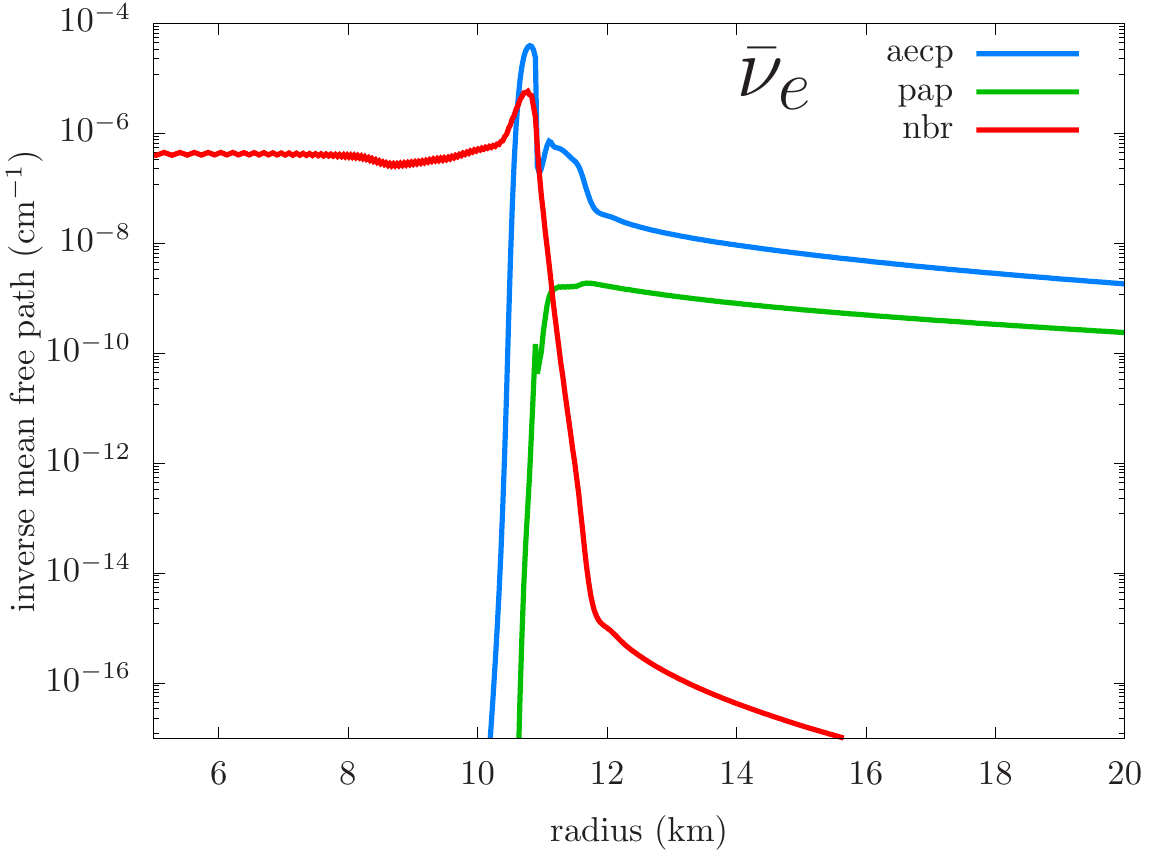}
      \end{minipage} \\
      \begin{minipage}[t]{1.0\hsize}
        \centering
        \includegraphics[scale=0.7]{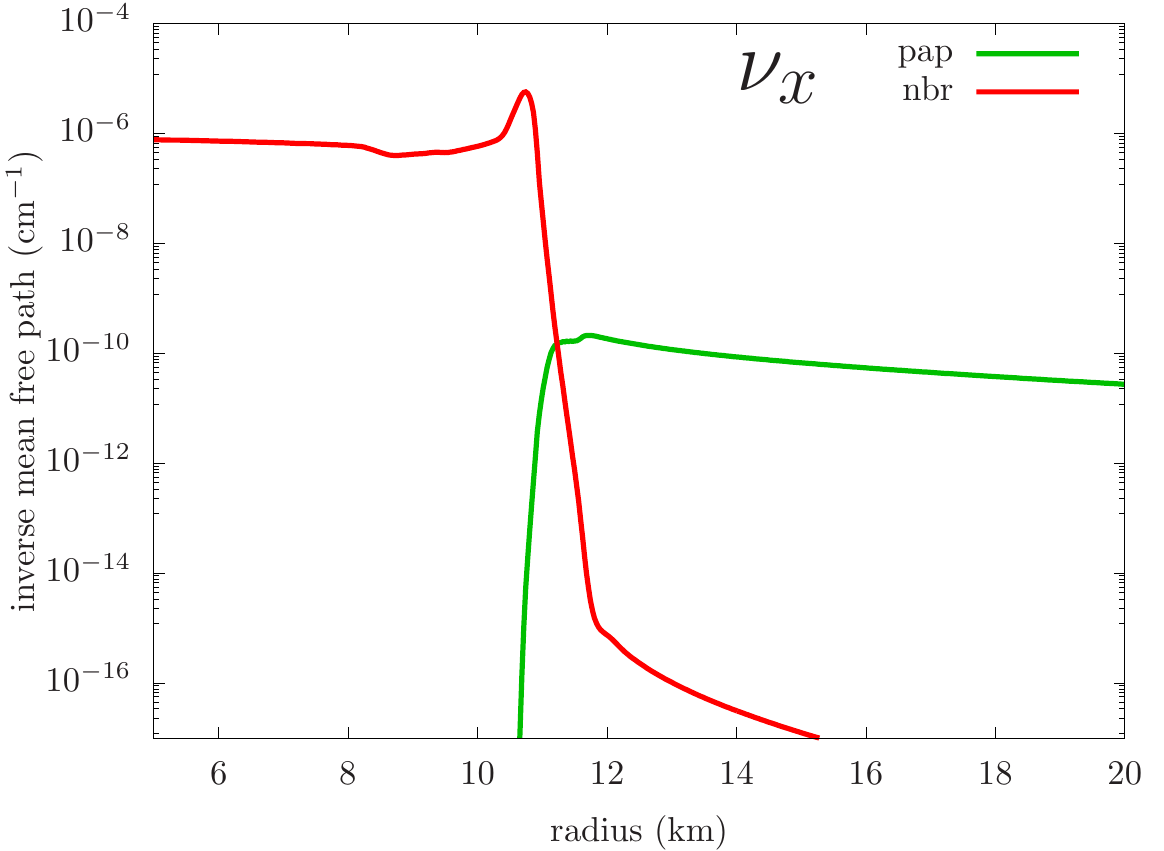}
      \end{minipage} 
    \end{tabular}
    \caption{Radial profile of the inverse mean free path for $\nu_e$ (top), $\bar\nu_e$ (middle) and $\nu_x$ (bottom) with the energy of $23.4\,\mathrm{MeV}$, for the model with the accretion rate $\dot M=10^{-3}\, M_\odot\cdot\mathrm{s}^{-1}$ and the PNS mass $1.98M_\odot$. The abbreviation of the neutrino reactions are as follows: the electron-capture on nucleon (ecp), the positron capture (aecp), the electron–positron process (pap) and the nucleon–nucleon bremsstrahlung (nbr).}
    \label{fig_mfp}
\end{figure}
\begin{figure}[htbp]
    \begin{tabular}{c}
      \begin{minipage}[t]{1.0\hsize}
        \centering
        \includegraphics[scale=0.7]{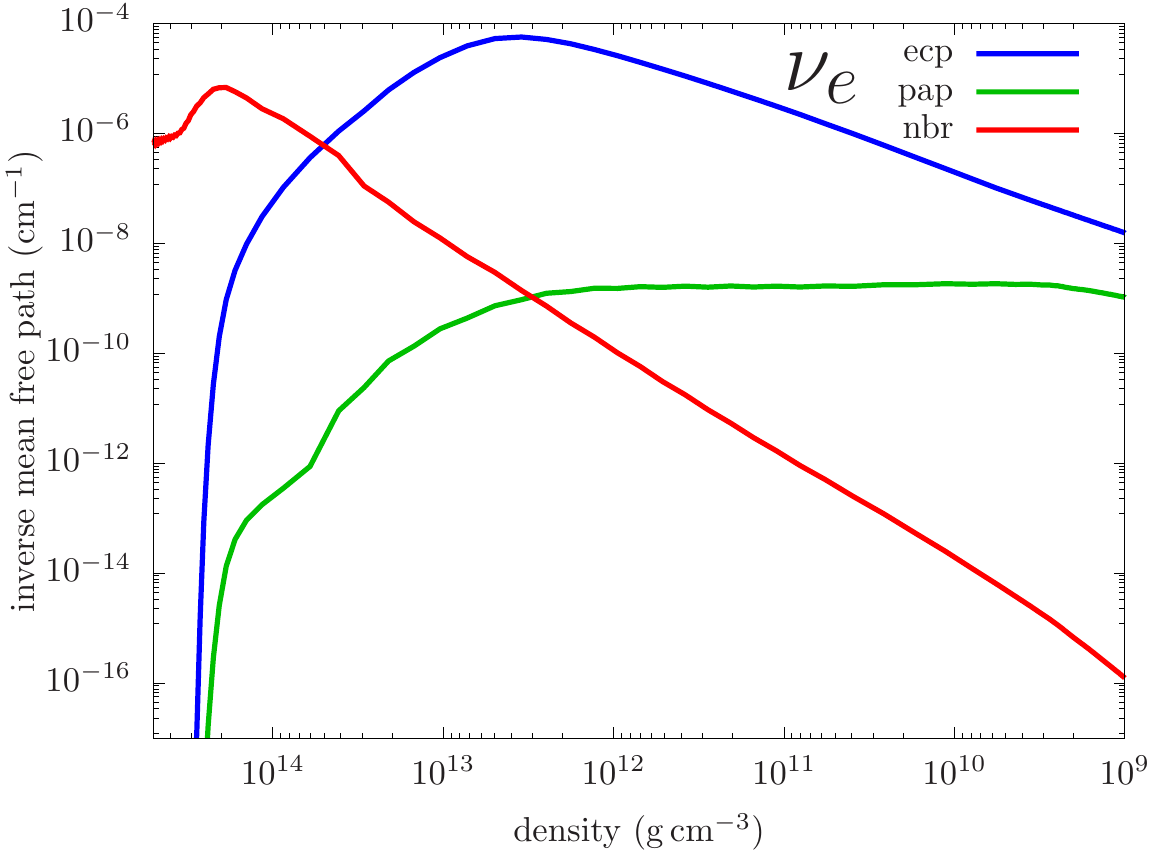}
      \end{minipage} \\
      \begin{minipage}[t]{1.0\hsize}
        \centering
        \includegraphics[scale=0.7]{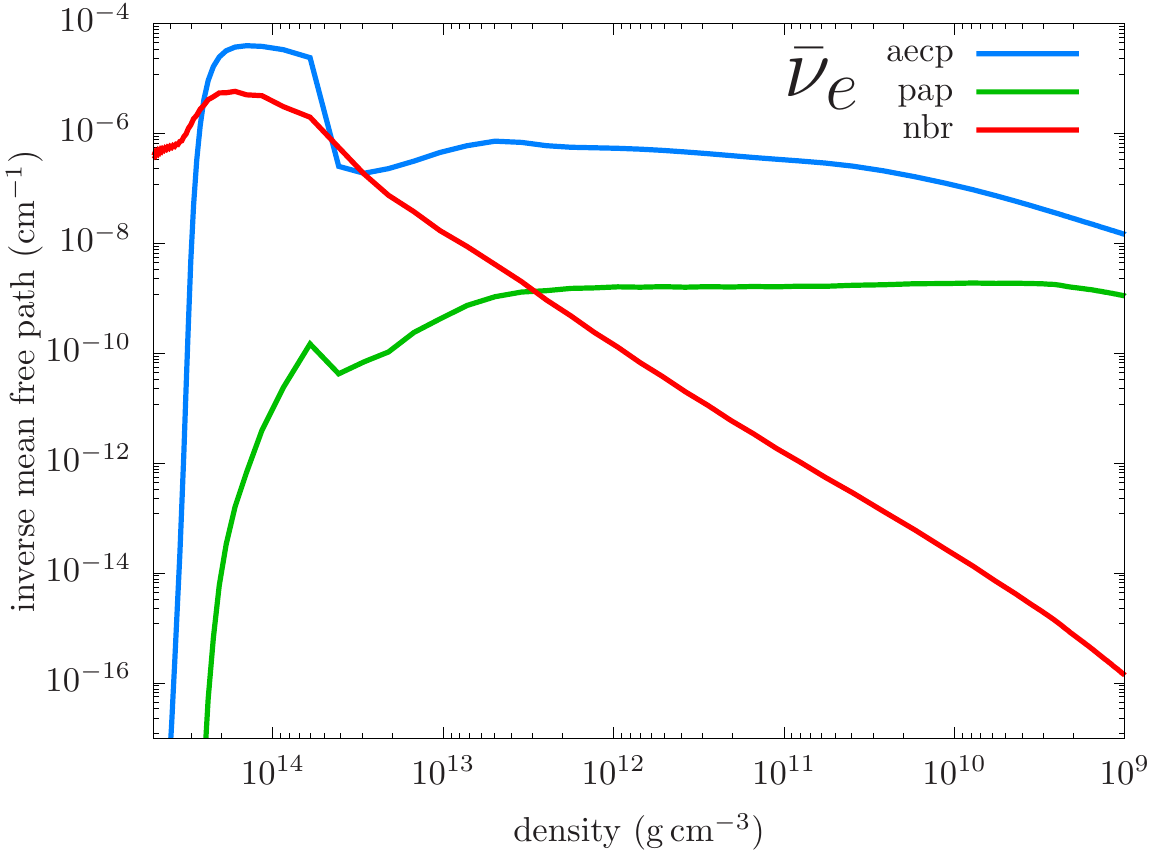}
      \end{minipage} \\
      \begin{minipage}[t]{1.0\hsize}
        \centering
        \includegraphics[scale=0.7]{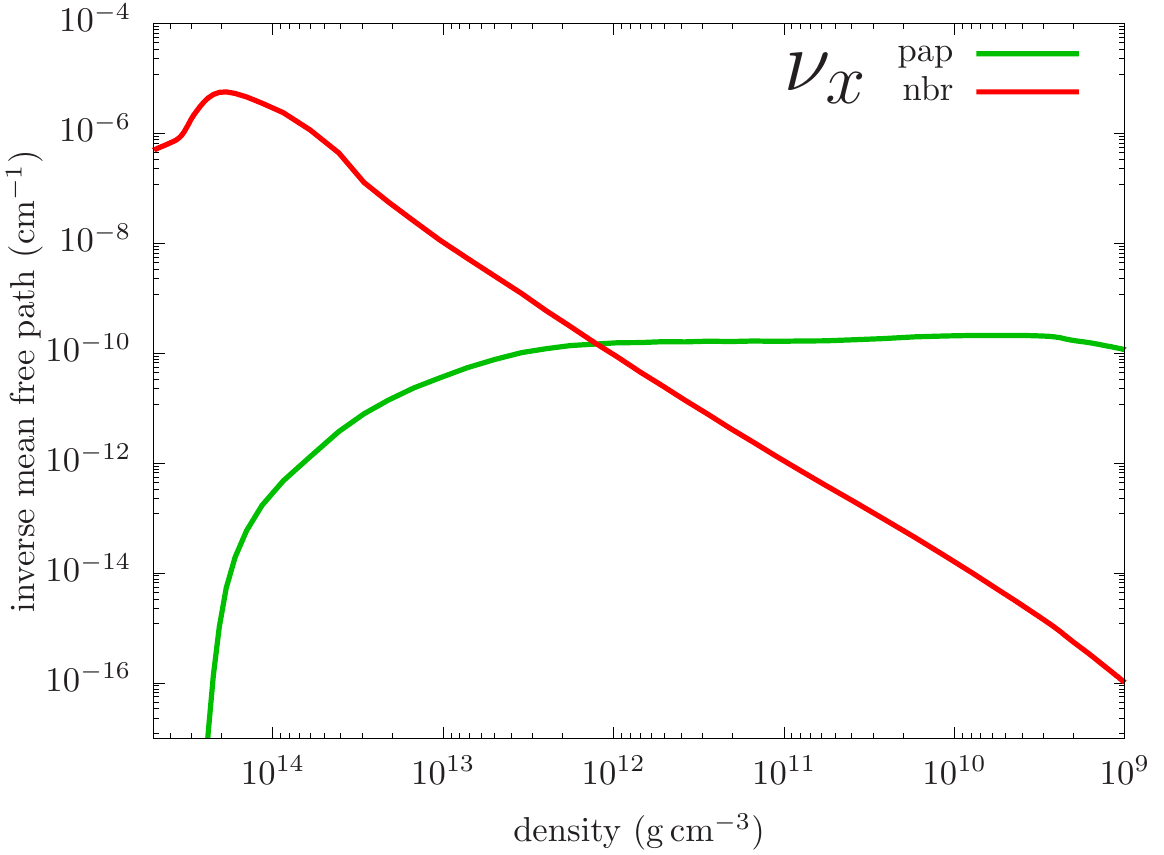}
      \end{minipage} 
    \end{tabular}
    \caption{Same as figure \ref{fig_mfp}, but the horizontal axis is the density.}
    \label{fig_mfp_rho}
\end{figure}
%

To see what weak process accounts for the neutrino emission and absorption, we display the radial profile of inverse mean free path of each weak process in Figs.~\ref{fig_mfp} and \ref{fig_mfp_rho}, for the accretion rate of $\dot M=10^{-3}\, M_\odot\cdot\mathrm{s}^{-1}$ and the PNS mass $1.98M_\odot$ model. These figures show the inverse mean free path as a function of the radius and the matter density, respectively.
Except for the high density region ($\sim 5 \times 10^{13} {\rm g/cm^3}$ for $\nu_e$ and $\sim 2 \times 10^{14} {\rm g/cm^3}$ for $\bar{\nu}_e$), electron-capture on free proton and positron-capture on free neutron dominate the $\nu_e$ and $\bar{\nu}_e$ emission, respectively. We also find that nucleon–nucleon bremsstrahlung becomes dominant in the high density regions, leading to a non-monotonic radial profile of neutrino opacity. This is responsible for the non-monotonic profile for both neutrino fluxes in $\nu_e$ and $\bar{\nu}$. This leads to the complex radial profile of matter temperature. Indeed, the inverse mean free path peaks at 
$11\lesssim r \lesssim12\,\mathrm{km}$ for $\nu_e$ and
$10\lesssim r \lesssim11\,\mathrm{km}$ for $\bar\nu_e$, and these spatial positions are roughly the same as those at
the temperature dips.
This exhibits that temperature dips are caused by the neutrino cooling by $\nu_e$ and $\bar\nu_e$.
For $\nu_x$, on the other hand, the neutrino opacity is dominated by nucleon–nucleon bremsstrahlung at the emission region ($5 \times 10^{13} {\rm g/cm^3} \lesssim \rho \lesssim 2 \times 10^{14} {\rm g/cm^3}$). Although the electron-positron pair production becomes dominant at $\rho \lesssim 10^{12} {\rm g/cm^3}$, the emissivity is very small. In fact, the radial profile of $F_\nu r^2$ for $\nu_x$ is almost constant in space over the low density region (see Fig~\ref{fig_flux}).

\subsection{Neutrino luminosity and mean energy}
\label{subsec:neutrinolumimean}

\begin{figure}[t]
    \begin{tabular}{c}
    \begin{minipage}[t]{0.8\hsize}
        \centering
        \includegraphics[scale=0.7]{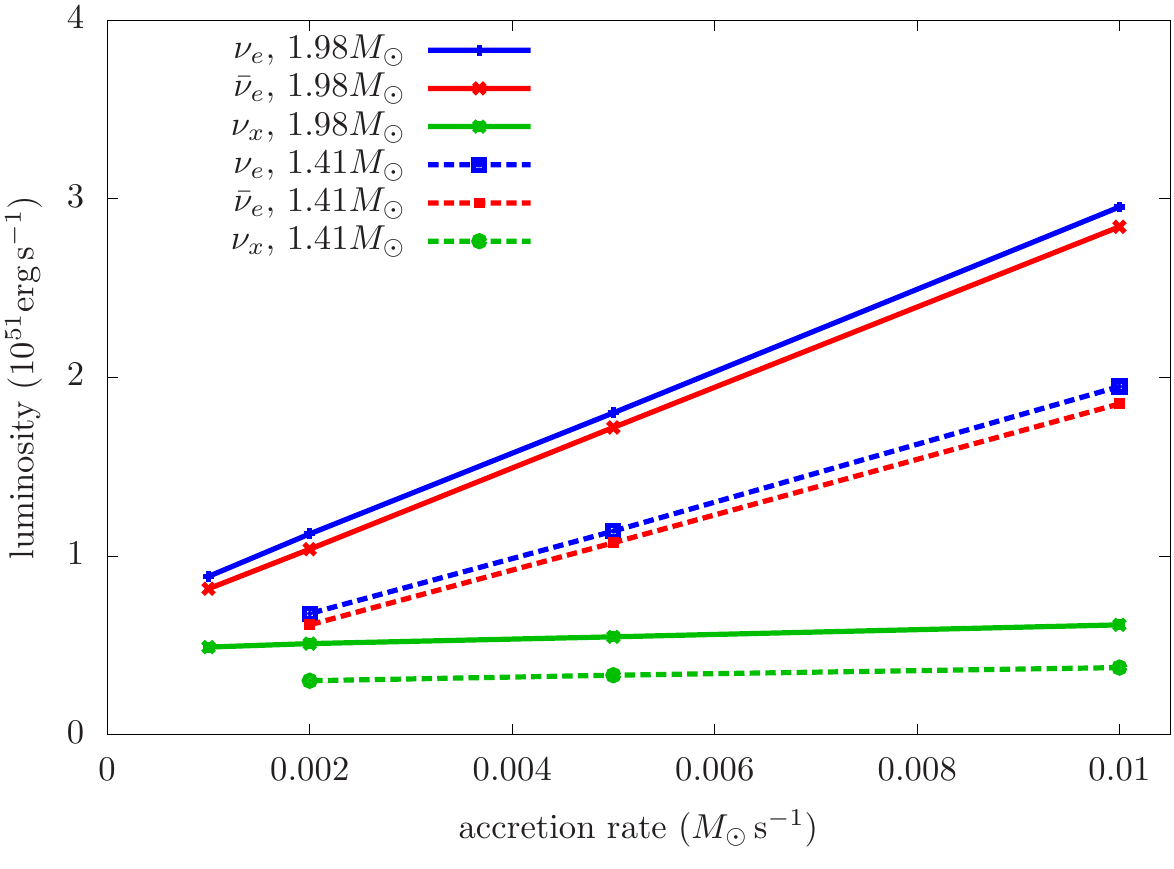}
      \end{minipage} 
    \end{tabular}
  \caption{Energy luminosity versus the accretion rate. Different colors show the different flavors, and the solid lines correspond to the $1.98M_\odot$ model and the broken lines for the $1.41M_\odot$ model.}
  \label{fig_lumi}
\end{figure}

Figure \ref{fig_lumi} summarizes the energy luminosity for all simulated models. 
As shown in the figure, larger accretion rates and larger PNS masses leads to higher luminosities.
The energy luminosity for $\nu_e$ and $\bar\nu_e$ are of the order of $o(10^{51})\,\mathrm{erg}\cdot\mathrm{s}^{-1}$, and $\nu_x$ luminosities are below $10^{51}\,\mathrm{erg}\cdot\mathrm{s}^{-1}$. 
One thing we do notice here is that neutrino luminosities obtained in our simulations are systematically higher than those reported in \citet{Fryer2009}. This is again due to the fact that the simulations of \citet{Fryer2009} did not cover the high density region, which results in underestimating neutrino luminosities. Our result suggests that it is mandatory to include the high density region in theoretical models to quantify the neutrino signal from FBA and to extract physical information from the neutrino signal in real observation (see Sec.~\ref{sec:detectability} for more details). Luminosities for $\nu_x$ hardly depend on the accretion rates. This is because $\nu_x$ are mainly emitted from the inner PNS, as we saw in the radial profiles of the flux. The luminosities of $\nu_e$ and $\bar{\nu}_e$ are also not proportional to the accretion rates and shifted to higher values, due to the same reason. 

\begin{figure}[t]
    \begin{tabular}{c}
    \begin{minipage}[t]{0.8\hsize}
        \centering
        \includegraphics[scale=0.7]{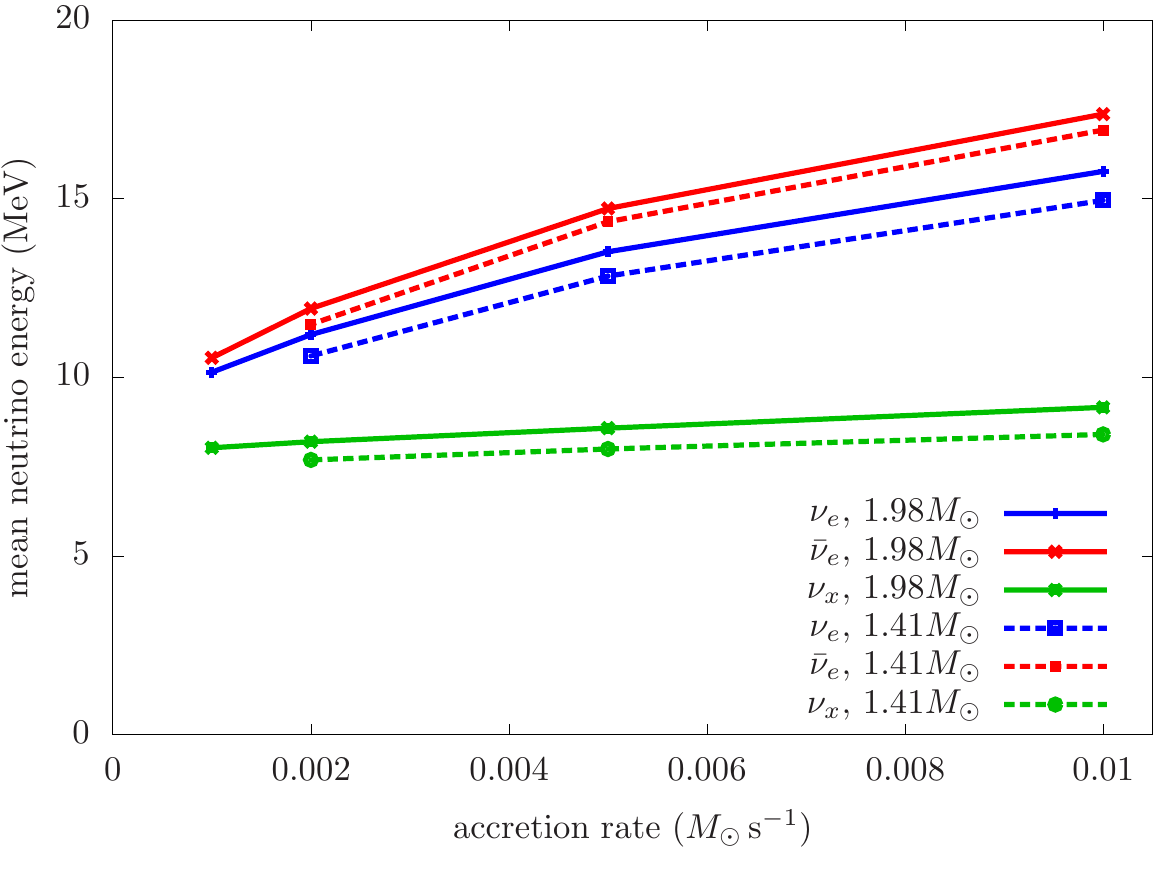}
      \end{minipage} 
    \end{tabular}
  \caption{Same as \ref{fig_lumi}, but for the mean energy of emitted neutrinos.}
  \label{fig_meanene}
\end{figure}

In Fig.~\ref{fig_meanene}, we provide mean energies of the emitted neutrinos. 
The mean energy of $\nu_e$ is $\epsilon\sim13\,\mathrm{MeV}$ for the highest accretion case ($\dot M=10^{-2}\, M_\odot\cdot\mathrm{s}^{-1}$), and it is still $\epsilon\gtrsim10\,\mathrm{MeV}$
for other cases with lower mass accretion rate. The mean energy of $\bar\nu_e$ is always higher than that of $\nu_e$, and reaches a maximum of $\epsilon\sim17\,\mathrm{MeV}$. We also find that, similar to the luminosity, larger accretion rates and PNS masses lead to higher neutrino mean energies. This is due to the higher matter temperature in the neutrino emission region (see Fig.~\ref{fig_temp}).

It is worthy to note that $\nu_x$ has the lowest mean energy among three flavors, which is $\epsilon\sim10\,\mathrm{MeV}$. 
This tendency is clearly different from the canonical hierarchy of neutrino mean energy in CCSNe.
In general, the mean energy of $\nu_x$ is the highest among all flavors of neutrinos in early post-bounce phase, and then the mean energy of all flavors becomes almost identical in the late phase. This exhibits that FBA leads to a qualitatively different neutrino emission from PNS cooling, and that the neutrino detection rate should depend on the neutrino oscillation model.
The low luminosity and low mean energy of $\nu_x$ are attributed to the temperature distribution in the $\nu_x$ emission region.
As shown in Fig.~\ref{fig_flux}, most of the $\nu_x$ are produced in the region of
$5 \times 10^{13} {\rm g/cm^3} \lesssim \rho \lesssim 2 \times 10^{14} {\rm g/cm^3}$. This region corresponds to the transition layer between the PNS surface and the inner edge of FBA. Although the matter temperature sharply increases with radius, it is still very low ($\lesssim 4 {\rm MeV}$). As a result, both the luminosity and the mean energy of $\nu_x$ become much lower than those associated to $\nu_e$ and $\bar{\nu}_e$. We note that the radius of the emission region for $\nu_x$ is smaller than for other flavors. This causes a lower neutrino luminosity, although this effect is minor since the difference of emission region among all flavors of neutrinos is only $\lesssim 1 {\rm km}$.

We remind the readers that our current focus is the late phase ($t\gtrsim10\,\mathrm{s}$), where the typical mean energy of the diffusive neutrino component from PNS is $\epsilon\lesssim10\,\mathrm{MeV}$ \citep{Suwa2019}.
This is much smaller than neutrinos from FBA. As discussed in Sec.~\ref{sec:detectability}, higher luminosities and mean energies of neutrinos are more favorable for neutrino detection.
Our results support the claim in \citet{Fryer2009} that FBA can substantially increase the neutrino event rate, which is quantified in the next section.

\section{Detectability of FBA Neutrinos}
\label{sec:detectability}

\subsection{Detectors and neutrino oscillation models}
\label{subsec:detecSetup}

We evaluate the detectability of the FBA neutrinos by two representative terrestrial neutrino detectors, Super-Kamiokande (Hereafter Super-K) and Deep Underground Neutrino Experiment (DUNE). 
In this estimation, we employ the neutrino cross section data taken from the SNOwGLoBES \citep{snowglobes}.
We ignore any smearing effects caused by the detector response and various noises just for simplicity. 

Super-K is a water-Cherenkov detector using pure water \citep{Fukuda2003} with gadolinium compound loaded recently \citep{Abe2022}.
The main detection channel of Super-K is the inverse-beta interaction 
\begin{equation}
\bar\nu_e+p\rightarrow e^+ + n.
\end{equation}
We assume the fiducial volume of 
$32.5\,\mathrm{kton}$ for the estimation of the event rate. Its update version, Hyper-Kamiokande is also 
under construction \citep{Abe2018}. Its fiducial volume will be $220\,\mathrm{kton}$, and the detection rate can be easily scaled from the result of Super-K.
\addedc{We assume pure water for the evaluation of the event rates. It should be mentioned that the gadolinium-loading in SK plays an important role to decouple the FBA neutrino signal from the background \citep{Li2022,Simpson2019}. Unlike the strong neutrino burst in the early post-bounce phase, the luminosity is lower and the timescale is longer for FBA neutrinos, indicating that the reduction of the background is very important to identify the signal.}

DUNE is a future-planned neutrino detector. It will use liquid argon as the neutrino detector medium.
The main detection channel of DUNE is the neutrino-argon charged-current interaction
\begin{equation}
\nu_e+{}^{40}\mathrm{Ar}\rightarrow e^- + {}^{40}\mathrm{K}^{\ast}.
\end{equation}
We assume a full volume of $40\,\mathrm{kton}$ for the estimation of the event rate. 

For the estimation of the neutrino flux arriving on the earth, we take into account the neutrino oscillation effect in the same way as \citet{Dighe2000,Nagakura2021a}. Neutrino flavors are assumed to convert adiabatically by the Mikheyev–Smirnov–Wolfenstein (MSW) effect. Although it is a simple oscillation model, this provides an essential feature of how detectability of FBA neutrinos depends on flavor conversions.

Following \citet{Nagakura2021a}, the neutrino fluxes arriving on earth $F_e$, $\bar{F}_e$, $F_x$, $\bar{F}_x$ (corresponds to $\nu_e$, $\bar\nu_e$, $\nu_x$, $\bar\nu_x$, respectively) are calculated from the values of the fluxes without neutrino oscillation ($F_e^0$, $\bar{F}_e^0$, $F_x^0$, $\bar{F}_x^0$) as:
\begin{eqnarray}
F_e&=&pF_e^0+(1-p)F_x^0, \\
\bar{F}_e&=&\bar{p}\bar{F}_e^0+(1-\bar{p})\bar{F}_x^0, \\
F_x&=&\frac{1}{2}(1-p)F_e^0+\frac{1}{2}(1+p)F_x^0, \\
\bar{F}_x&=&\frac{1}{2}(1-\bar{p})\bar{F}_e^0+\frac{1}{2}(1+\bar{p})\bar{F}_x^0, \\
\end{eqnarray}
where $p$, $\bar{p}$ are survival probabilities. In the normal-mass hierarchy case, they are defined as
\begin{eqnarray}
p&=&\mathrm{sin}^2\theta_{13}, \\
\bar{p}&=&\mathrm{cos}^2\theta_{12}\mathrm{cos}^2\theta_{13}.
\end{eqnarray}
On the other hand, in the inverted-mass hierarchy case, they are defined as
\begin{eqnarray}
p&=&\mathrm{sin}^2\theta_{12}\mathrm{cos}^2\theta_{13}, \\
\bar{p}&=&\mathrm{sin}^2\theta_{13}. 
\end{eqnarray}
The values of the neutrino mixing angles $\theta_{12}$, $\theta_{13}$ are assumed to be $\mathrm{sin}^2\theta_{12}=2.97\times10^{-1}$ and $\mathrm{sin}^2\theta_{13}=2.15\times10^{-2}$, adopted from \citet{Capozzi2017}. 
We assume $F_x^0=\bar{F}_x^0$ in this study.

\subsection{Neutrino Event Rates}
\begin{figure}[htbp]
    \begin{tabular}{c}
      \begin{minipage}[t]{1.0\hsize}
        \centering
        \includegraphics[scale=0.7]{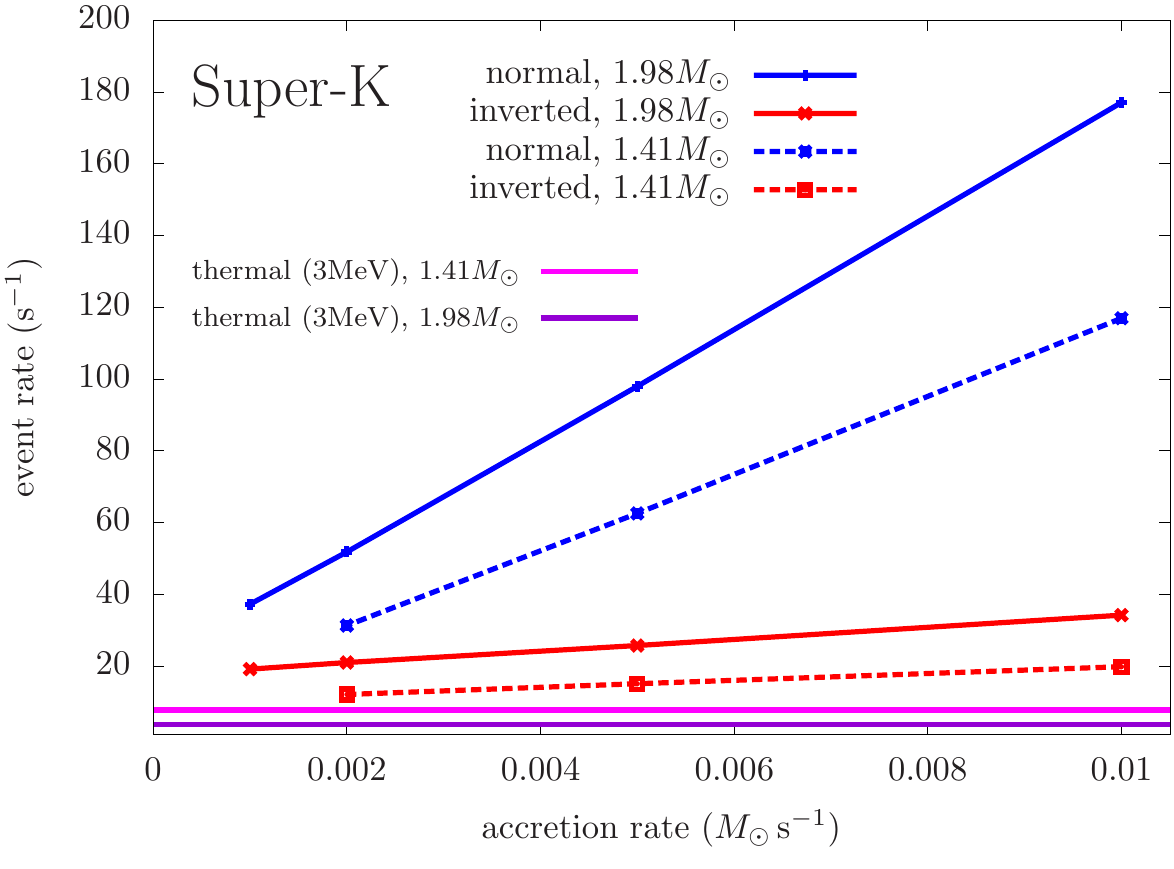}
      \end{minipage} \\
      \begin{minipage}[t]{1.0\hsize}
        \centering
        \includegraphics[scale=0.7]{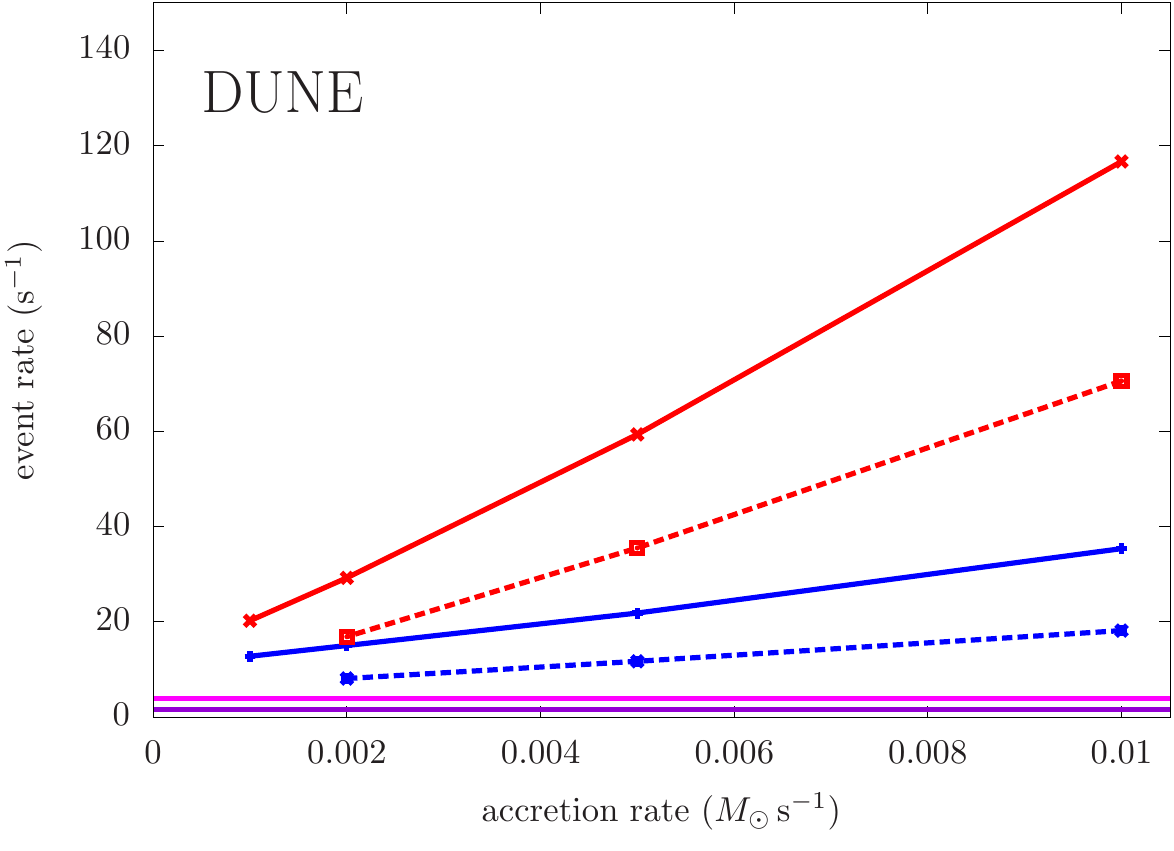}
      \end{minipage}
    \end{tabular}
    \caption{The event rate of neutrinos for the Super-K (top) and DUNE (bottom), assuming the distance of $10\,\mathrm{kpc}$. The horizontal lines denote the event rates assuming thermal emission.}
    \label{fig_event}
\end{figure}

Figure \ref{fig_event} shows the neutrino event rates per unit time, in which we integrate over energy, while the energy-dependent feature is discussed later. The distance is assumed to be $10\,\mathrm{kpc}$. 
As shown in Fig.~\ref{fig_event}, the event rate clearly depends on the mass hierarchy, where the difference is more than double. 
In the case with the normal(inverted)-mass hierarchy, $p$ ($\bar{p}$) becomes small, indicating that neutrinos (anti-neutrinos) undergo large flavor conversions. As shown in Sec.~\ref{subsec:neutrinolumimean}, both the energy luminosities and the average energies of $\nu_e$ and $\bar{\nu}_e$ are higher than those of $\nu_x$ at the source, indicating that the large flavor conversion results in reducing the $\nu_e$ and $\bar{\nu}_e$ number flux. As a result, the number of event rate at Super-K and DUNE becomes lower in the case of inverted-mass hierarchy and normal one, respectively. Hence, simultaneous observation of FBA neutrinos with Super-K and DUNE will provide a strong constraint on neutrino mass hierarchy. 

\addedc{The estimated event rate is found to be $o(10)\,\mathrm{s}^{-1}$ for the accretion rate of $\dot M\sim10^{-3}\, M_\odot\cdot\mathrm{s}^{-1}$. This result also suggests that if we detect a large number of neutrinos in the very late phase, the detection will be an evidence for the occurrence of FBA neutrinos. 
It is also worthy to note that similar accretion rates were found in previous studies \citet{Chan2018,Moriya2019,Janka2022} in the late phase.}

We also find that the dependence of the event rate on the mass accretion rate hinges on the neutrino oscillation model. In the case of normal (inverted) mass hierarchy, the detection rate at Super-K (DUNE) becomes remarkably higher for higher mass accretion rates, whereas it is less sensitive to the accretion rate in the case of inverted (normal) one. This trend can also be understood through the species-dependent feature of neutrino emission at the CCSN source. As shown in Figs.~\ref{fig_lumi}~and~\ref{fig_meanene}, both the luminosity and the average energy of $\nu_x$ weakly depend on the mass accretion rate, and therefore the large flavor conversion makes the detection count at each detector less sensitive to the mass accretion rate. Nevertheless, the number of event count at each detector is remarkably higher than that emitted from the PNS. As a reference, we show the case for the neutrino signal without FBA but only with isothermal PNS of $T= 3 {\rm MeV}$ in Fig.~\ref{fig_event} (see below for the details).
This figure illustrates that the detection rate of neutrinos from FBA is remarkably higher than in the case with thermal neutrinos from the PNS. It is also worthy to note that flavor dependent features would be resolved by using other reaction channels or joint analysis with other detectors \citep[see, e.g.][]{Beacom2002,Dasgupta2011,Nagakura2021b}, that would provide a key information to distinguish the neutrinos powered by FBA from those radiated only from the inner region of the PNS.

\begin{figure}[t]
    \begin{center}
    \begin{tabular}{cc}
      \begin{minipage}[t]{1.0\hsize}
        \centering
        \includegraphics[scale=0.7]{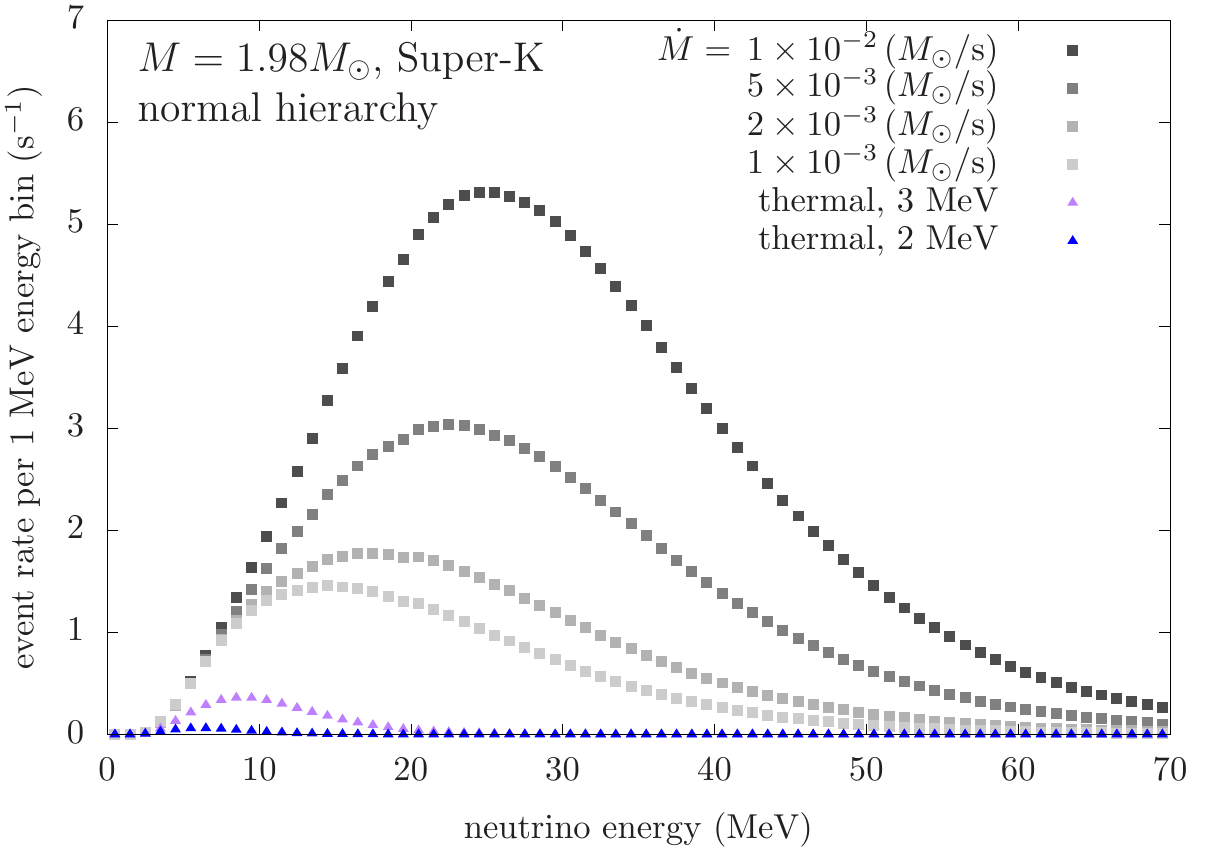}
      \end{minipage}
      \begin{minipage}[t]{1.0\hsize}
        \centering
        \includegraphics[scale=0.7]{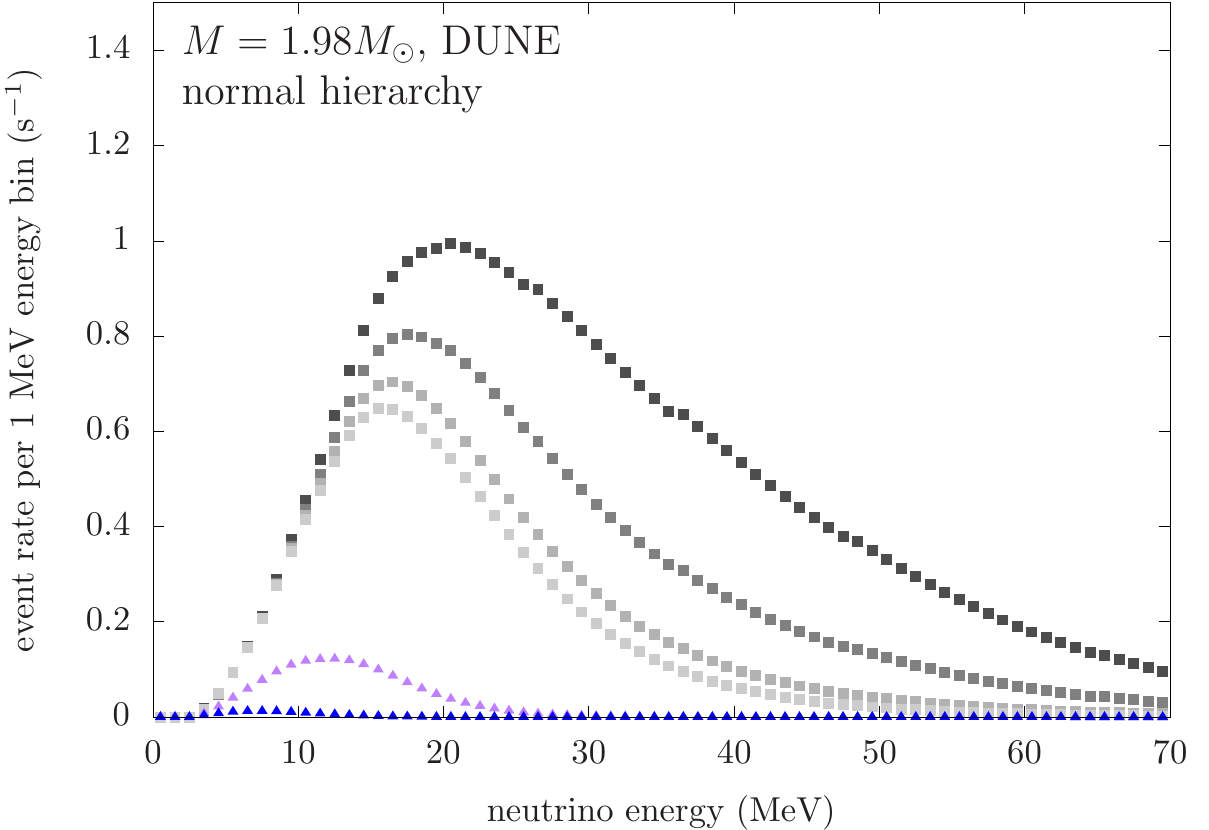}
      \end{minipage} \\
      \begin{minipage}[t]{1.0\hsize}
        \centering
        \includegraphics[scale=0.7]{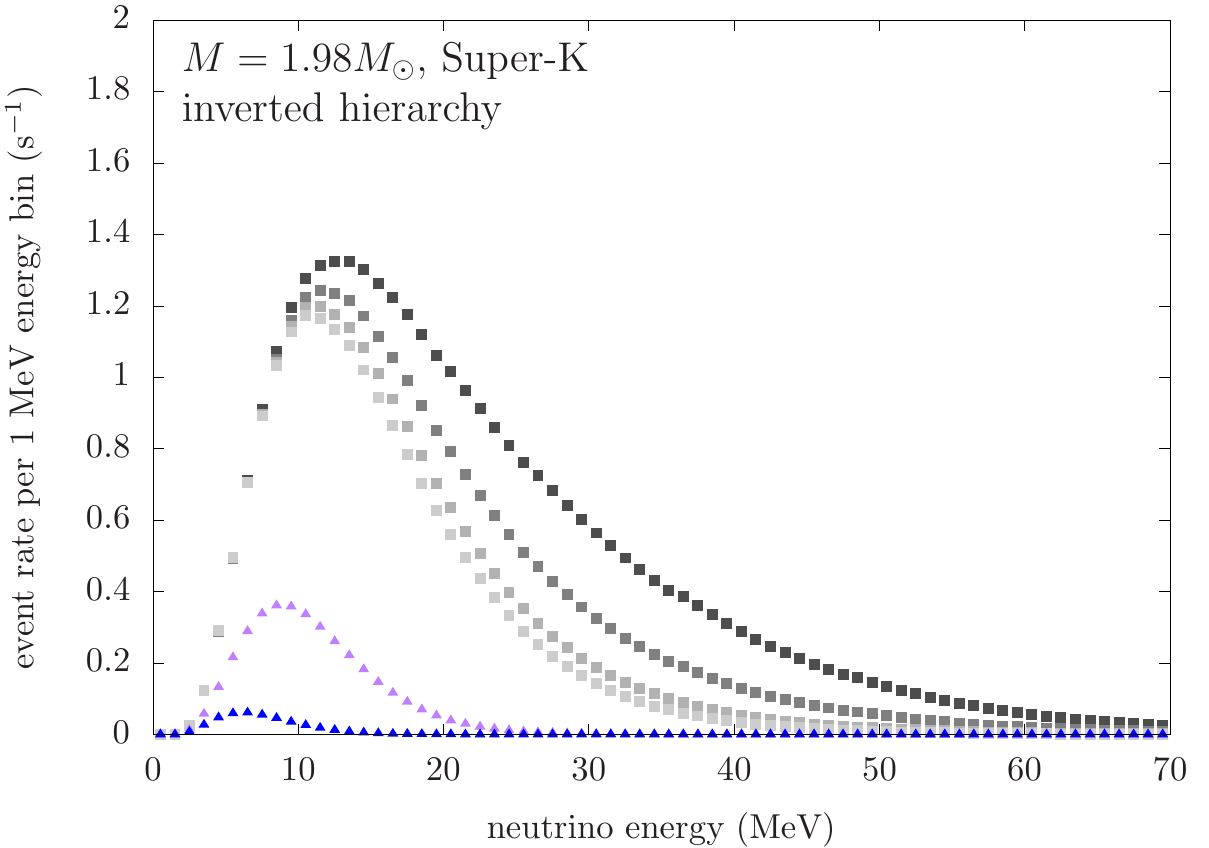}
      \end{minipage} 
      \begin{minipage}[t]{1.0\hsize}
        \centering
        \includegraphics[scale=0.7]{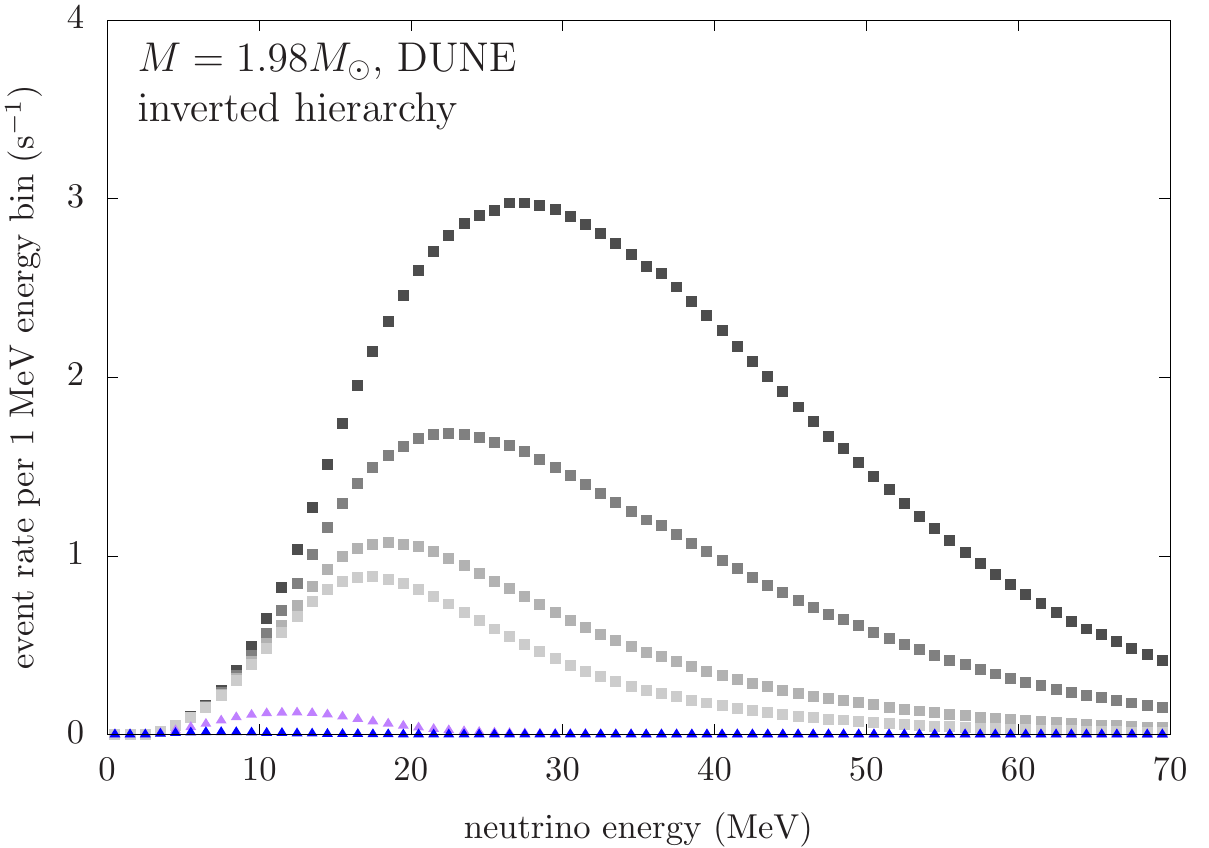}
      \end{minipage} 
    \end{tabular}
    \centering
    \caption{The event rate per $1\,\mathrm{MeV}$ energy bin for $M_\mathrm{PNS}=1.98M_\odot$ model, assuming the distance of $10\,\mathrm{kpc}$. The top and bottom panel corresponds to normal and inverted mass hierarchy, respectively. The left panels are for Super-K and right panels are for DUNE. The gray plots corresponds to the numerical results, and the purple and blue plots corresponds to thermal values.}
    \label{fig_event_perbin_198}
    \end{center}
\end{figure}

\begin{figure}[ht]
    \begin{center}
    \begin{tabular}{cc}
      \begin{minipage}[t]{1.0\hsize}
        \centering
        \includegraphics[scale=0.7]{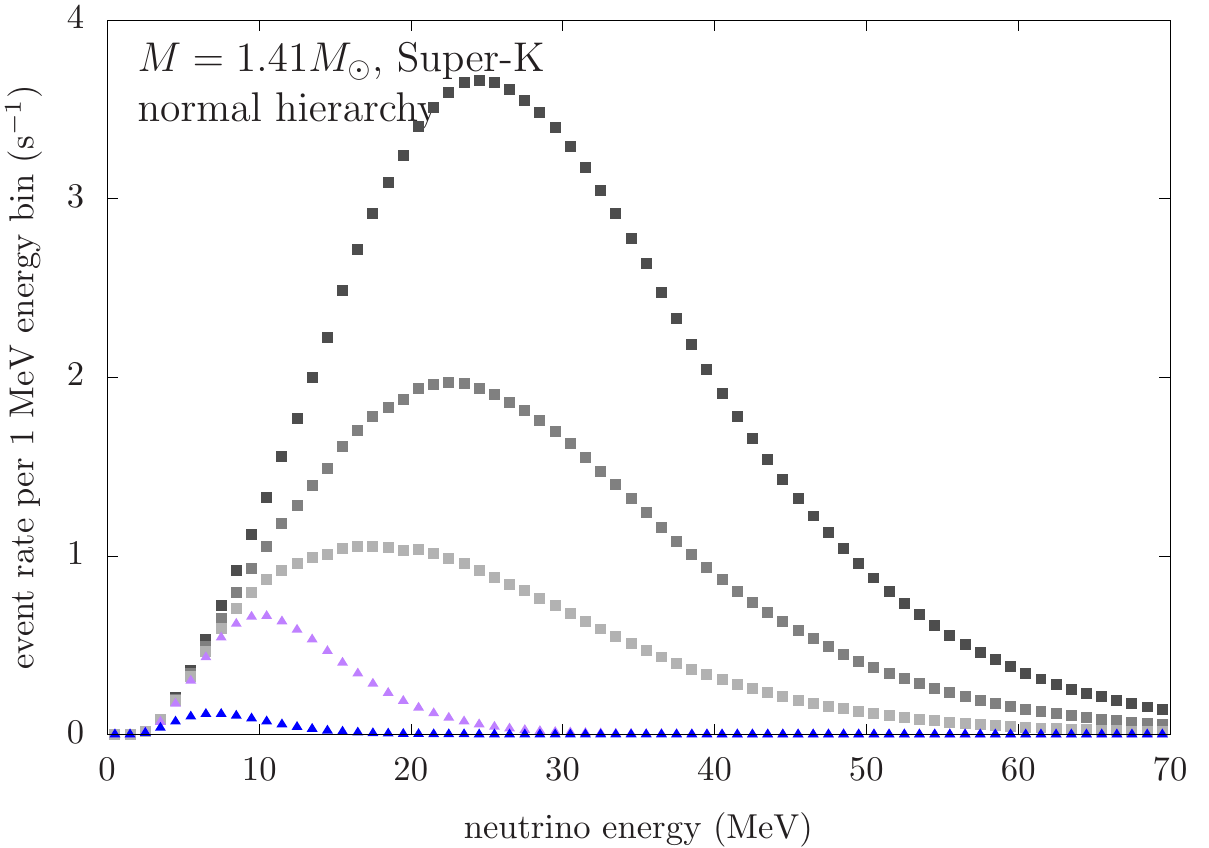}
      \end{minipage} 
      \begin{minipage}[t]{1.0\hsize}
        \centering
        \includegraphics[scale=0.7]{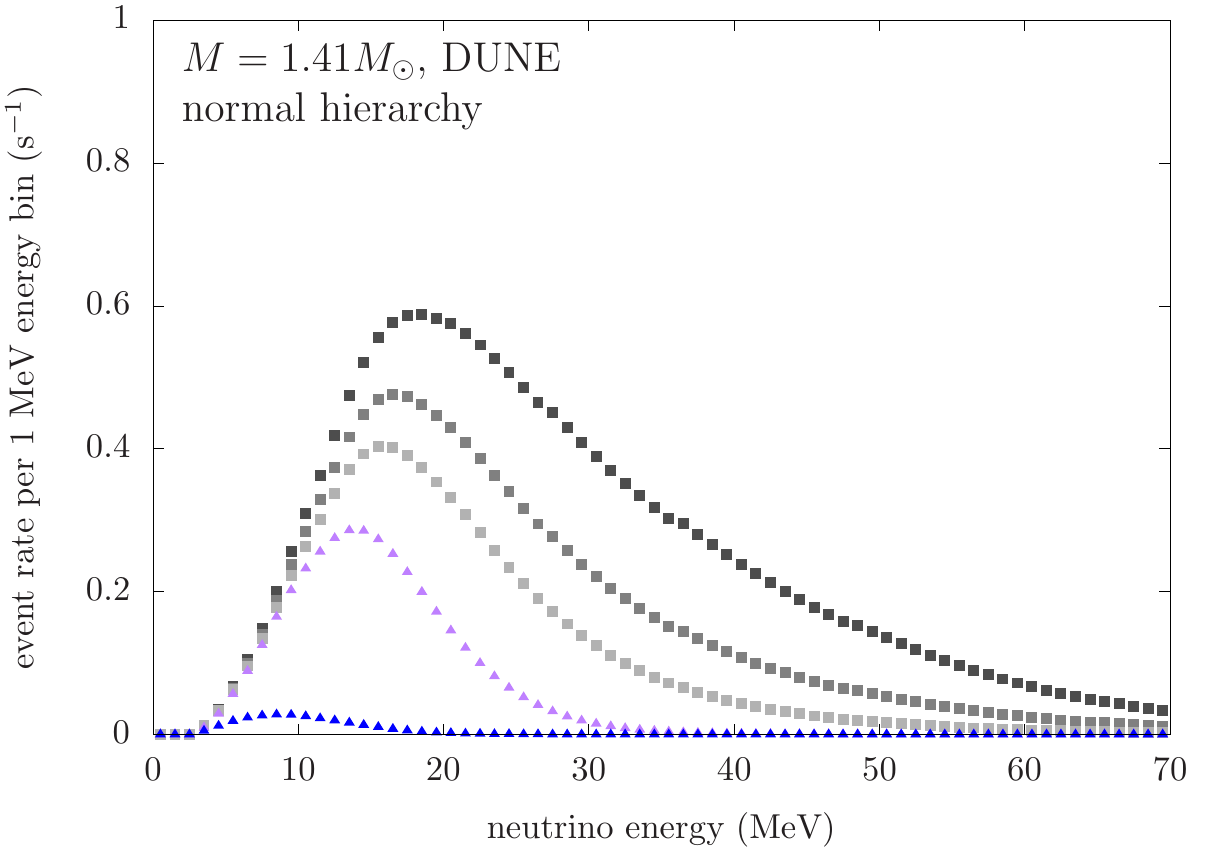}
      \end{minipage} \\
      \begin{minipage}[t]{1.0\hsize}
        \centering
        \includegraphics[scale=0.7]{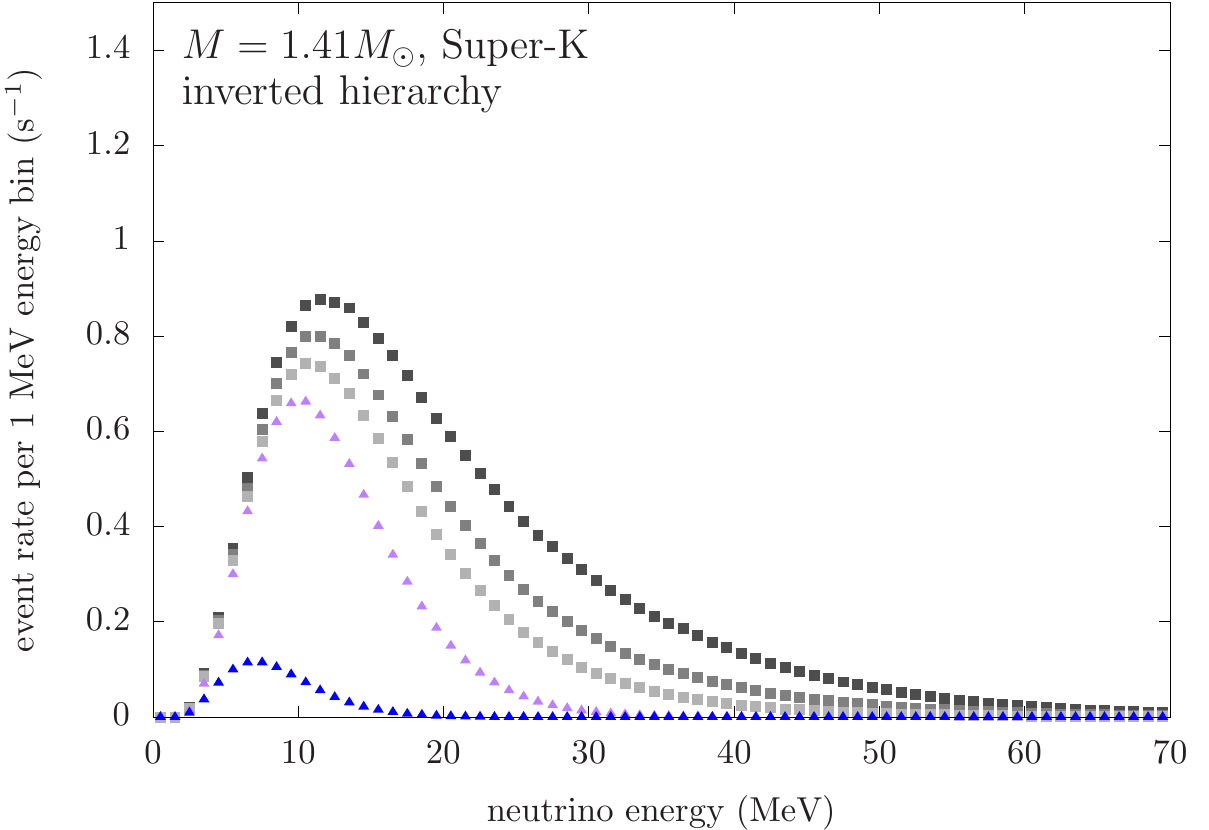}
      \end{minipage} 
      \begin{minipage}[t]{1.0\hsize}
        \centering
        \includegraphics[scale=0.7]{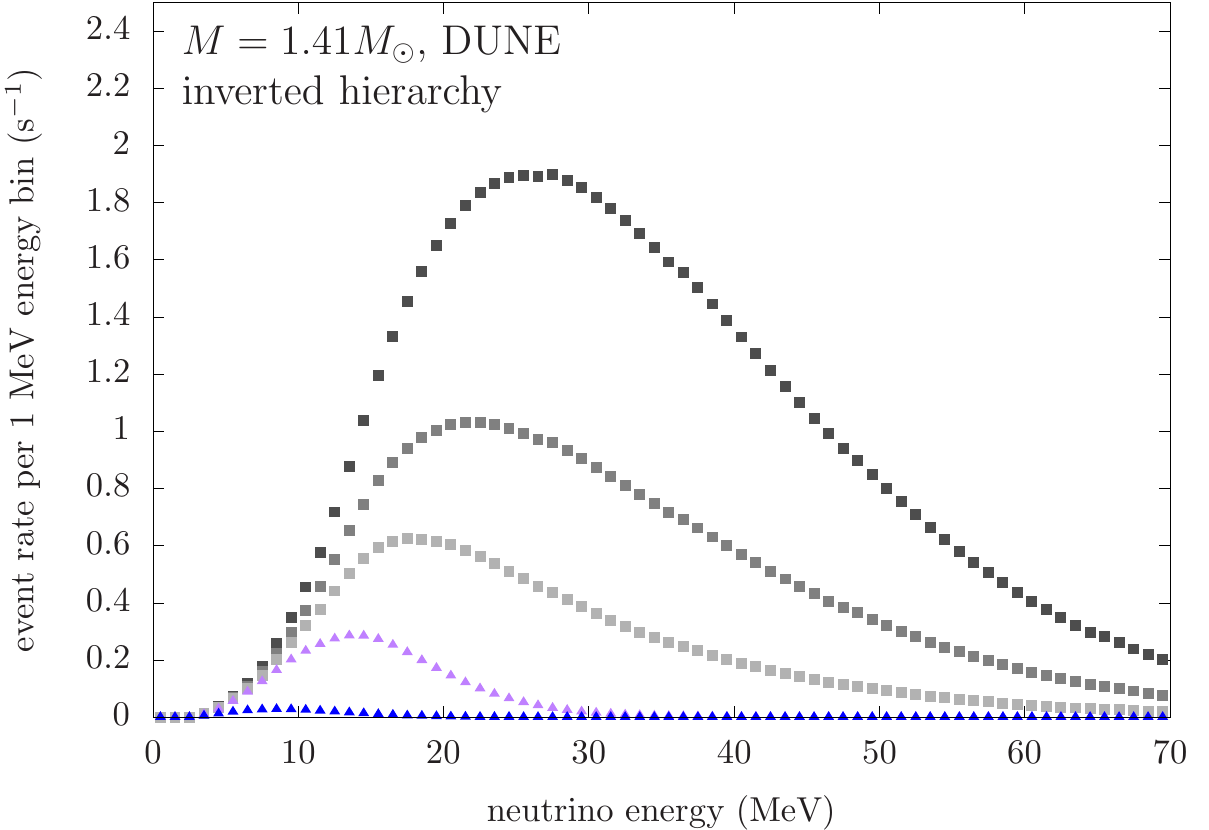}
      \end{minipage}
    \end{tabular}
    \centering
    \caption{Same as figure \ref{fig_event_perbin_198}, but for PNS mass $1.41M_\odot$ model.}
    \label{fig_event_perbin_141}
    \end{center}
\end{figure}

In Figs. \ref{fig_event_perbin_198} and \ref{fig_event_perbin_141}, we show the energy-dependent event rate, in which the energy bin is set to be $1\,\mathrm{MeV}$.
As references, we also make a plot for the case with the purely thermal emission (Fermi-Dirac distribution with zero chemical potential) with a PNS temperature of $2$ and $3 \mathrm{MeV}$. The emission radius is assumed to be $11$ km, and we also take into account the gravitational redshift in this estimation.
\addedc{For all simulated models, the event rates are orders of magnitude larger than the background event rates (see the latest experimental data of Super-K in \citet{HaradaM2023}.}
These figures illustrate that a large neutrino emission can be expected in the case of higher mass accretion rate. Another notable feature found in these figures is the high energy tail in each spectrum. Even in the case with the low mass accretion rate ($\dot M=10^{-3}\, M_\odot\cdot\mathrm{s}^{-1}$), neutrinos with $\gtrsim 30 {\rm MeV}$ may be observed. 
\addedc{
It should be noted that these high energy neutrinos cannot be detected in the late phase ($t \gg 10$s) by thermal emission of PNS, unless the source is extremely close \citep{Nakazato2022}. If we detect them in real observations in the late phase, these neutrinos would be generated by FBA.}
We note that Figs.~\ref{fig_event_perbin_198}~and~\ref{fig_event_perbin_141} show the energy event rate per second, indicating that the actual event count may be a factor of $>10$ larger than this value (since we are currently considering in the phase of $> 10 {\rm s}$ after core bounce).


%
\clearpage
\onecolumngrid
\twocolumngrid
\clearpage

\onecolumngrid

\twocolumngrid
\clearpage

\subsection{PNS Temperature Dependence}
\label{subsec:pnstemperadepe}
It is interesting to see how the neutrino signal from FBA depends on the PNS temperature. The increase of the PNS temperature would lead to higher neutrino emission inside the PNS, which potentially alters the neutrino signal. In this test, we employ the same numerical setup as that used in our model with the PNS mass of $1.98M_\odot$ and the accretion rate of $\dot M=10^{-3}\, M_\odot\cdot\mathrm{s}^{-1}$ except for the PNS temperature. We consider two cases: $3\,\mathrm{MeV}$ and $4\,\mathrm{MeV}$.
We note that $T=4\,\mathrm{MeV}$ is too hot for PNS in the late phase which we consider in this paper ($>10$s after core bounce), but the result is still informative.

\begin{figure}[htbp]
    \begin{tabular}{c}
      \begin{minipage}[t]{1.0\hsize}
        \centering
        \includegraphics[scale=0.7]{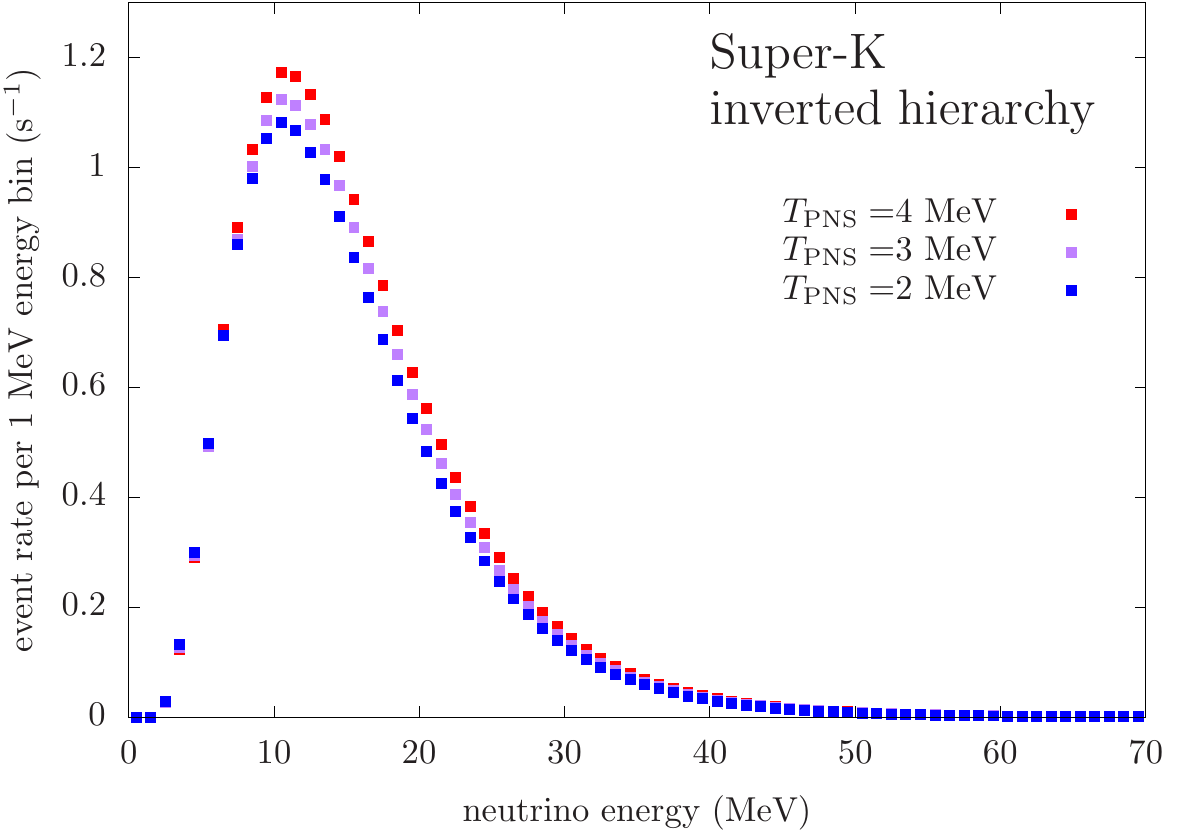}
      \end{minipage} \\
      \begin{minipage}[t]{1.0\hsize}
        \centering
        \includegraphics[scale=0.7]{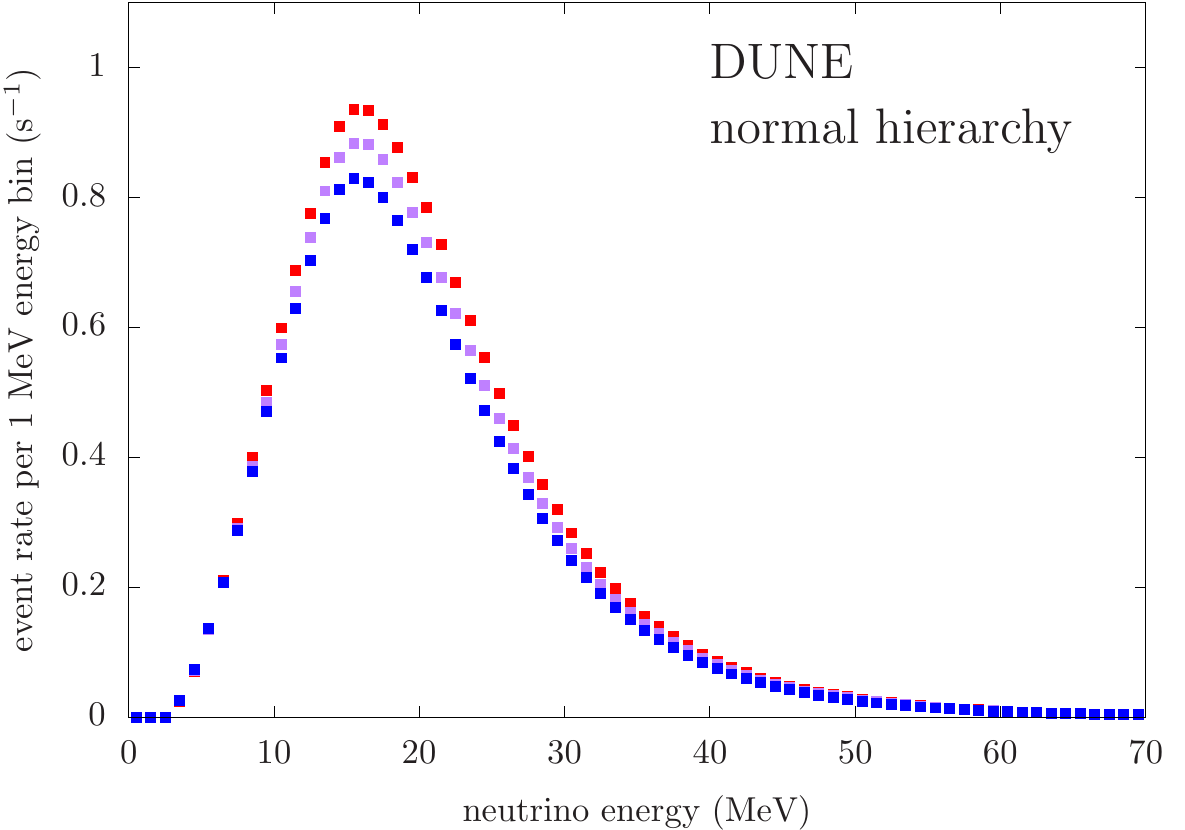}
      \end{minipage}
    \end{tabular}
    \caption{The event rate per $1\,\mathrm{MeV}$ energy bin for different PNS temperatures; $2\,\mathrm{MeV}$ (blue), $3\,\mathrm{MeV}$ (purple) and $4\,\mathrm{MeV}$ (red), assuming the distance of $10\,\mathrm{kpc}$. Top panel is for Super-K and bottom panel for DUNE. Mass hierarchy is inverted (top) and normal (bottom). PNS mass is $M_\mathrm{PNS}=1.98M_\odot$ and the accretion rate is $\dot M=10^{-3}\, M_\odot\cdot\mathrm{s}^{-1}$.}
    \label{fig_2_3_4MeV}
\end{figure}

In Fig.~\ref{fig_2_3_4MeV}, we show the energy spectrum of the neutrino event rate at Super-K and DUNE in the case of normal- and inverted mass hierarchy, respectively. We note that each oscillation model corresponds to the case having the lower number of event rate than the other mass hierarchy. As shown in these figures, even in these pessimistic cases, the PNS temperature does not affect the neutrino event rate. This result supports the claim that neutrinos from FBA overwhelm the thermal neutrinos from the PNS, unless they are extremely hot ($T \gg 4\,\mathrm{MeV}$).

\section{Summary and Conclusion}
\label{sec:summary}
In this paper we investigated neutrino emission from fallback mass accretion (FBA) onto PNS in the late phase of CCSN ($> 10$s) by using general relativistic neutrino radiation-hydrodynamic simulations with full Boltzmann neutrino transport. In our numerical simulations, we covered the very high density region ($> 10^{14} {\rm g/cm^3}$) where we set a quasi-steady PNS structure as initial conditions. We changed the mass accretion rate in a parametric manner, and ran each simulation until the system settled to a quasi-steady state.

We found that a higher accretion rate and a higher PNS mass leads to a higher 
temperature in the transition layer from the PNS surface to the inner edge of FBA, where most of the neutrinos are radiated. As a result, both luminosities and mean energies of neutrinos tend to be higher with increasing mass accretion rates. On the other hand, the sensitivity of neutrino emission to the mass accretion rate hinge on neutrino species. Although $\nu_e$ and $\bar{\nu}_e$ emission strongly vary with the mass accretion rate, $\nu_x$ is less sensitive. This is due to the fact that $\nu_x$ is produced in the highest density region ($\rho \gtrsim 5 \times 10^{13} {\rm g/cm^3}$), indicating that the impact of FBA on the temperature distribution tends to be weak. Nevertheless, both the luminosity and the mean energy of $\nu_x$ are remarkably higher than those estimated by standard PNS cooling models.

The present study supports the claim by \citet{Fryer2009} that FBA can substantially change the neutrino emission in the late phase of CCSN. On the other hand, we also find that most of the neutrinos by FBA are produced in the high density region which the simulations of \citet{Fryer2009} did not cover. As a result, the neutrino luminosities in his estimation are underestimated by a factor of $\gtrsim 5$, and this systematic error has a non-negligible effect to extract physical information from neutrino signal in real observations. We also find that the dominant weak processes for neutrino emission depends on species: electron-capture by free proton, positron-capture by free neutron, and nucleon–nucleon bremsstrahlung for $\nu_e$, $\bar{\nu}_e$, and $\nu_x$, respectively. Although the electron-positron pair can be a dominant emission process for $\nu_x$ in the low density region, the emissivity is too low to change the neutrino flux.

Based on the numerical results, we estimate the expected event rate for Super-K and DUNE with the adiabatic MSW oscillation model.
One thing we need to stress is that neutrino emission from FBA has a rich flavor-dependent structure, indicating that the neutrino observation should depend on the neutrino oscillation model. Indeed, the difference of event rate between normal- and inverted mass hierarchy at each detector becomes more than double. In short, the detection rate tends to be smaller if the flavor conversion is strong. This is attributed to the fact that $\nu_x$ luminosity and mean energy are systematically lower than those of other species.
Nevertheless, the event rate is the order of $o(100)\,\mathrm{s}^{-1}$ for the optical case with the highest accretion rate in both detectors, and still $o(10)\,\mathrm{s}^{-1}$ for the least optimal setting, which is much larger than the canonical PNS cooling model.
We also provide energy-dependent features in the neutrino signal.
We find that the peak energy of neutrino detection is remarkably higher than the thermal emission of PNS with $ \le 3 {\rm MeV}$. Our result suggests that high energy neutrinos ($\gtrsim 30 {\rm MeV}$) may be observed in the late phase, which will be evidence that neutrinos are emitted by FBA.

As a final remark, we point out a couple of limitations in our study. First, we assumed spherical symmetry. 
In the multi-D case, the accretion shock wave may be unstable to non-radial perturbations \citep{Blondin2003,Yamasaki2005,Yamasaki2006,Yamasaki2007,Foglizzo2007}, \addedc{and FBA is usually accompanied by turbulence \citep{Vartanyan2022}}, which potentially leads to temporal variations in the neutrino signal. On the other hand, it would be hard to resolve the temporal variation by the current- and even future-planned neutrino detectors, unless the CCSN source is very close \citep[see, e.g.,][]{Nagakura2021c}. This is because the neutrino luminosity is very low in the late phase, and the temporal variation would be smeared out by noise. It should be mentioned, however, that the thermodynamical properties in the post-shock flow may be influenced by the shock instability, which may change the neutrino signal. 
We postpone this detailed study to future work.
Second, the number of models simulated in this study is limited due to the computational cost. It should be stressed that high spatial resolution is required to resolve both matter and neutrino distributions around the surface of the PNS, implying that the timestep is severely limited by the Courant condition. 
To prepare for future observations, however, we need a systematic study by covering wider ranges of PNS masses and mass accretion rates than those studied in the present paper. The EOS dependence is also worthy of investigation.
Numerical simulations are not suitable to carry out such a systematic study, and therefore we are planning to take a semi-analytic approach to address this issue.
\addedc{If we cover full parameter space, we may be able to infer the EOS parameters or accretion rates from the future neutrino detection. An analysis pipeline based on a Bayesian approach has been already developed for thermal neutrino detection \citep{HaradaA2023}.}
The results with similar approach will be reported elsewhere.

\begin{acknowledgments}
We thank Hirotada Okawa, Kohsuke Sumiyoshi, Yudai Suwa, Akira Harada, Shun Furusawa and Shoichi Yamada for fruitful discussions. Numerical computations were carried out on Cray XC50 at Center for Computational Astrophysics, National Astronomical Observatory of Japan (NAOJ). 
This research used the K and Fugaku supercomputers provided by RIKEN, the FX10 provided by Tokyo University, the FX100 provided by Nagoya University, the Grand Chariot provided by Hokkaido University, and Oakforest-PACS provided by JCAHPC through the HPCI System Research Project (Project ID: hp130025, 140211, 150225, 150262, 160071, 160211, 170031, 170230, 170304, 180111, 180179, 180239, 190100, 190160, 200102, 200124, 210050, 210051, 210164, 220047, 220173, 220047, 220223 and 230033). This work is supported by 
Grant-in-Aid for Scientific Research
(19K03837, 20H01905)
and 
Grant-in-Aid for Scientific Research on Innovative areas 
"Gravitational wave physics and astronomy:Genesis" 
(17H06357, 17H06365) and ”Unraveling the History of the Universe and Matter Evolution with Underground Physics” (19H05802 and 19H05811)
from the Ministry of Education, Culture, Sports, Science and Technology (MEXT), Japan.
R. A. is supported by JSPS Grant-in-Aid for JSPS Fellows (Grant No. 22J10298) from MEXT. 
H.N is supported by Grant-inAid for Scientific Research (23K03468).
For providing high performance computing resources, Computing Research Center, KEK, JLDG on SINET of NII, Research Center for Nuclear Physics, Osaka University, Yukawa Institute of Theoretical Physics, Kyoto University, Nagoya University, and Information Technology Center, University of Tokyo are acknowledged. This work was supported by 
MEXT as "Program for Promoting Researches on the Supercomputer Fugaku" 
(Toward a unified view of the universe: from large scale structures to planets, JPMXP1020200109) and the Particle, Nuclear and Astro Physics Simulation Program (Nos. 2020-004, 2021-004, 2022-003) of Institute of Particle and Nuclear Studies, High Energy Accelerator Research Organization (KEK).
\end{acknowledgments}






\bibliography{sample631}{}

\begin{thebibliography}{}
\expandafter\ifx\csname natexlab\endcsname\relax\def\natexlab#1{#1}\fi
\providecommand{\url}[1]{\href{#1}{#1}}
\providecommand{\dodoi}[1]{doi:~\href{http://doi.org/#1}{\nolinkurl{#1}}}
\providecommand{\doeprint}[1]{\href{http://ascl.net/#1}{\nolinkurl{http://ascl.net/#1}}}
\providecommand{\doarXiv}[1]{\href{https://arxiv.org/abs/#1}{\nolinkurl{https://arxiv.org/abs/#1}}}

\bibitem[{{Abbasi} {et~al.}(2011){Abbasi}, {Abdou}, {Abu-Zayyad}, {Ackermann},
  {Adams}, {Aguilar}, {Ahlers}, {Allen}, {Altmann}, {Andeen}, {Auffenberg},
  {Bai}, {Baker}, {Barwick}, {Baum}, {Bay}, {Bazo Alba}, {Beattie}, {Beatty},
  {Bechet}, {Becker}, {Becker}, {Benabderrahmane}, {Benzvi}, {Berdermann},
  {Berghaus}, {Berley}, {Bernardini}, {Bertrand}, {Besson}, {Bindig}, {Bissok},
  {Blaufuss}, {Blumenthal}, {Boersma}, {Bohm}, {Bose}, {B{\"o}ser}, {Botner},
  {Brown}, {Buitink}, {Caballero-Mora}, {Carson}, {Chirkin}, {Christy},
  {Clevermann}, {Cohen}, {Colnard}, {Cowen}, {Cruz Silva}, {D'Agostino},
  {Danninger}, {Daughhetee}, {Davis}, {de Clercq}, {Degner}, {Demir{\"o}rs},
  {Descamps}, {Desiati}, {de Vries-Uiterweerd}, {Deyoung},
  {D{\'\i}az-V{\'e}lez}, {Dierckxsens}, {Dreyer}, {Dumm}, {Dunkman}, {Eisch},
  {Ellsworth}, {Engdeg{\r{a}}rd}, {Euler}, {Evenson}, {Fadiran}, {Fazely},
  {Fedynitch}, {Feintzeig}, {Feusels}, {Filimonov}, {Finley}, {Fischer-Wasels},
  {Fox}, {Franckowiak}, {Franke}, {Gaisser}, {Gallagher}, {Gerhardt},
  {Gladstone}, {Gl{\"u}senkamp}, {Goldschmidt}, {Goodman}, {G{\'o}ra}, {Grant},
  {Griesel}, {Gro{\ss}}, {Grullon}, {Gurtner}, {Ha}, {Haj Ismail}, {Hallgren},
  {Halzen}, {Han}, {Hanson}, {Heinen}, {Helbing}, {Hellauer}, {Hickford},
  {Hill}, {Hoffman}, {Hoffmann}, {Homeier}, {Hoshina}, {Huelsnitz},
  {H{\"u}l{\ss}}, {Hulth}, {Hultqvist}, {Hussain}, {Ishihara}, {Jakobi},
  {Jacobsen}, {Japaridze}, {Johansson}, {Kampert}, {Kappes}, {Karg}, {Karle},
  {Kenny}, {Kiryluk}, {Kislat}, {Klein}, {K{\"o}hne}, {Kohnen}, {Kolanoski},
  {K{\"o}pke}, {Kopper}, {Koskinen}, {Kowalski}, {Kowarik}, {Krasberg},
  {Kroll}, {Kurahashi}, {Kuwabara}, {Labare}, {Laihem}, {Landsman}, {Larson},
  {Lauer}, {L{\"u}nemann}, {Madsen}, {Marotta}, {Maruyama}, {Mase}, {Matis},
  {Meagher}, {Merck}, {M{\'e}sz{\'a}ros}, {Meures}, {Miarecki}, {Middell},
  {Milke}, {Miller}, {Montaruli}, {Morse}, {Movit}, {Nahnhauer}, {Nam},
  {Naumann}, {Nygren}, {Odrowski}, {Olivas}, {Olivo}, {O'Murchadha}, {Panknin},
  {Paul}, {P{\'e}rez de Los Heros}, {Petrovic}, {Piegsa}, {Pieloth}, {Porrata},
  {Posselt}, {Price}, {Przybylski}, {Rawlins}, {Redl}, {Resconi}, {Rhode},
  {Ribordy}, {Richard}, {Richman}, {Rodrigues}, {Rothmaier}, {Rott}, {Ruhe},
  {Rutledge}, {Ruzybayev}, {Ryckbosch}, {Sander}, {Santander}, {Sarkar},
  {Schatto}, {Schmidt}, {Sch{\"o}nwald}, {Schukraft}, {Schulte}, {Schultes},
  {Schulz}, {Schunck}, {Seckel}, {Semburg}, {Seo}, {Sestayo}, {Seunarine},
  {Silvestri}, {Singh}, {Slipak}, {Spiczak}, {Spiering}, {Stamatikos},
  {Stanev}, {Stezelberger}, {Stokstad}, {St{\"o}{\ss}l}, {Strahler},
  {Str{\"o}m}, {St{\"u}er}, {Sullivan}, {Swillens}, {Taavola}, {Taboada},
  {Tamburro}, {Tepe}, {Ter-Antonyan}, {Tilav}, {Toale}, {Toscano}, {Tosi}, {van
  Eijndhoven}, {Vandenbroucke}, {van Overloop}, {van Santen}, {Vehring},
  {Voge}, {Walck}, {Waldenmaier}, {Wallraff}, {Walter}, {Weaver}, {Wendt},
  {Westerhoff}, {Whitehorn}, {Wiebe}, {Wiebusch}, {Williams}, {Wischnewski},
  {Wissing}, {Wolf}, {Wood}, {Woschnagg}, {Xu}, {Xu}, {Xu}, {Yanez}, {Yodh},
  {Yoshida}, {Zarzhitsky}, {Zoll}, \& {IceCube Collaboration}}]{Abbasi2011}
{Abbasi}, R., {Abdou}, Y., {Abu-Zayyad}, T., {et~al.} 2011, \aap, 535, A109,
  \dodoi{10.1051/0004-6361/201117810}

\bibitem[{{Abe} {et~al.}(2022){Abe}, {Bronner}, {Hayato}, {Hiraide}, {Ikeda},
  {Imaizumi}, {Kameda}, {Kanemura}, {Kataoka}, {Miki}, {Miura}, {Moriyama},
  {Nagao}, {Nakahata}, {Nakayama}, {Okada}, {Okamoto}, {Orii}, {Pronost},
  {Sekiya}, {Shiozawa}, {Sonoda}, {Suzuki}, {Takeda}, {Takemoto}, {Takenaka},
  {Tanaka}, {Watanabe}, {Yano}, {Han}, {Kajita}, {Okumura}, {Tashiro}, {Xia},
  {Megias}, {Bravo-Bergu{\~n}o}, {Labarga}, {Marti}, {Zaldivar}, {Pointon},
  {Blaszczyk}, {Kearns}, {Raaf}, {Stone}, {Wan}, {Wester}, {Bian},
  {Griskevich}, {Kropp}, {Locke}, {Mine}, {Smy}, {Sobel}, {Takhistov}, {Hill},
  {Kim}, {Lim}, {Park}, {Bodur}, {Scholberg}, {Walter}, {Bernard}, {Coffani},
  {Drapier}, {Hedri}, {Giampaolo}, {Gonin}, {Mueller}, {Paganini}, {Quilain},
  {Ishizuka}, {Nakamura}, {Jang}, {Learned}, {Anthony}, {Martin}, {Scott},
  {Sztuc}, {Uchida}, {Cao}, {Berardi}, {Catanesi}, {Radicioni}, {Calabria},
  {Machado}, {De Rosa}, {Collazuol}, {Iacob}, {Lamoureux}, {Mattiazzi},
  {Ospina}, {Ludovici}, {Maekawa}, {Nishimura}, {Friend}, {Hasegawa}, {Ishida},
  {Kobayashi}, {Jakkapu}, {Matsubara}, {Nakadaira}, {Nakamura}, {Oyama},
  {Sakashita}, {Sekiguchi}, {Tsukamoto}, {Boschi}, {Gao}, {Di Lodovico},
  {Migenda}, {Taani}, {Zsoldos}, {Kotsar}, {Nakano}, {Ozaki}, {Shiozawa},
  {Suzuki}, {Takeuchi}, {Yamamoto}, {Ali}, {Ashida}, {Feng}, {Hirota},
  {Kikawa}, {Mori}, {Nakaya}, {Wendell}, {Yasutome}, {Fernandez}, {McCauley},
  {Mehta}, {Tsui}, {Fukuda}, {Itow}, {Menjo}, {Niwa}, {Sato}, {Tsukada},
  {Lagoda}, {Lakshmi}, {Mijakowski}, {Zalipska}, {Jiang}, {Jung}, {Vilela},
  {Wilking}, {Yanagisawa}, {Hagiwara}, {Harada}, {Horai}, {Ishino}, {Ito},
  {Kitagawa}, {Koshio}, {Ma}, {Piplani}, {Sakai}, {Barr}, {Barrow}, {Cook},
  {Goldsack}, {Samani}, {Wark}, {Nova}, {Yang}, {Jenkins}, {Malek}, {McElwee},
  {Stone}, {Thiesse}, {Thompson}, {Okazawa}, {Kim}, {Seo}, {Yu}, {Ichikawa},
  {Nakamura}, {Nishijima}, {Koshiba}, {Iwamoto}, {Nakajima}, {Ogawa},
  {Yokoyama}, {Martens}, {Vagins}, {Kuze}, {Izumiyama}, {Yoshida}, {Inomoto},
  {Ishitsuka}, {Ito}, {Kinoshita}, {Matsumoto}, {Ohta}, {Shinoki}, {Suganuma},
  {Martin}, {Tanaka}, {Towstego}, {Akutsu}, {Hartz}, {Konaka}, {de Perio},
  {Prouse}, {Chen}, {Xu}, {Posiadala-Zezula}, {Hadley}, {O'Flaherty},
  {Richards}, {Jamieson}, {Walker}, {Minamino}, {Okamoto}, {Pintaudi}, {Sano},
  {Sasaki}, \& {Super-Kamiokande Collaboration}}]{Abe2022}
{Abe}, K., {Bronner}, C., {Hayato}, Y., {et~al.} 2022, Nuclear Instruments and
  Methods in Physics Research A, 1027, 166248,
  \dodoi{10.1016/j.nima.2021.166248}

\bibitem[{{Abi} {et~al.}(2021){Abi}, {Acciarri}, {Acero}, {Adamov}, {Adams},
  {Adinolfi}, {Ahmad}, {Ahmed}, {Alion}, {Alonso Monsalve}, {Alt}, {Anderson},
  {Andreopoulos}, {Andrews}, {Andrianala}, {Andringa}, {Ankowski}, {Antonova},
  {Antusch}, {Aranda-Fernandez}, {Ariga}, {Arnold}, {Arroyave}, {Asaadi},
  {Aurisano}, {Aushev}, {Autiero}, {Azfar}, {Back}, {Back}, {Backhouse},
  {Baesso}, {Bagby}, {Bajou}, {Balasubramanian}, {Baldi}, {Bambah}, {Barao},
  {Barenboim}, {Barker}, {Barkhouse}, {Barnes}, {Barr}, {Barranco Monarca},
  {Barros}, {Barrow}, {Bashyal}, {Basque}, {Bay}, {Alba}, {Beacom},
  {Bechetoille}, {Behera}, {Bellantoni}, {Bellettini}, {Bellini},
  {Beltramello}, {Belver}, {Benekos}, {Bento Neves}, {Berger}, {Berkman},
  {Bernardini}, {Berner}, {Berns}, {Bertolucci}, {Betancourt}, {Bezawada},
  {Bhattacharjee}, {Bhuyan}, {Biagi}, {Bian}, {Biassoni}, {Biery}, {Bilki},
  {Bishai}, {Bitadze}, {Blake}, {Blanco Siffert}, {Blaszczyk}, {Blazey},
  {Blucher}, {Boissevain}, {Bolognesi}, {Bolton}, {Bonesini}, {Bongrand},
  {Bonini}, {Booth}, {Booth}, {Bordoni}, {Borkum}, {Boschi}, {Bostan}, {Bour},
  {Boyd}, {Boyden}, {Bracinik}, {Braga}, {Brailsford}, {Brandt}, {Bremer},
  {Brew}, {Brianne}, {Brice}, {Brizzolari}, {Bromberg}, {Brooijmans}, {Brooke},
  {Bross}, {Brunetti}, {Buchanan}, {Budd}, {Caiulo}, {Calafiura}, {Calcutt},
  {Calin}, {Calvez}, {Calvo}, {Camilleri}, {Caminata}, {Campanelli},
  {Caratelli}, {Carini}, {Carlus}, {Carniti}, {Caro Terrazas}, {Carranza},
  {Castillo}, {Castromonte}, {Cattadori}, {Cavalier}, {Cavanna}, {Centro},
  {Cerati}, {Cervelli}, {Cervera Villanueva}, {Chalifour}, {Chang},
  {Chardonnet}, {Chatterjee}, {Chattopadhyay}, {Chaves}, {Chen}, {Chen},
  {Chen}, {Cherdack}, {Chi}, {Childress}, {Chiriacescu}, {Cho}, {Choubey},
  {Christensen}, {Christian}, {Christodoulou}, {Church}, {Clarke}, {Coan},
  {Cocco}, {Coelho}, {Conley}, {Conrad}, {Convery}, {Corwin}, {Cotte},
  {Cremaldi}, {Cremonesi}, {Crespo-Anad{\'o}n}, {Cristaldo}, {Cross}, {Cuesta},
  {Cui}, {Cussans}, {Dabrowski}, {da Motta}, {Da Silva Peres}, {David},
  {David}, {Davies}, {Davini}, {Dawson}, {De}, {De Almeida}, {Debbins}, {De
  Bonis}, {Decowski}, {de Gouv{\^e}a}, {De Holanda}, {De Icaza Astiz},
  {Deisting}, {De Jong}, {Delbart}, {Delepine}, {Delgado}, {Dell-Acqua}, {De
  Lurgio}, {de Mello Neto}, {DeMuth}, {Dennis}, {Densham}, {Deptuch}, {De
  Roeck}, {De Romeri}, {De Vries}, {Dharmapalan}, {Dias}, {Diaz}, {D{\'\i}az},
  {Di Domizio}, {Di Giulio}, {Ding}, {Di Noto}, {Distefano}, {Diurba}, {Diwan},
  {Djurcic}, {Dokania}, {Dolinski}, {Domine}, {Douglas}, {Drielsma},
  {Duchesneau}, {Duffy}, {Dunne}, {Durkin}, {Duyang}, {Dvornikov}, {Dwyer},
  {Dyshkant}, {Eads}, {Edmunds}, {Eisch}, {Emery}, {Ereditato}, {Escobar},
  {Escudero Sanchez}, {Evans}, {Ewart}, {Ezeribe}, {Fahey}, {Falcone},
  {Farnese}, {Farzan}, {Felix}, {Fernandez-Martinez}, {Fernandez Menendez},
  {Ferraro}, {Fields}, {Filkins}, {Filthaut}, {Fitzpatrick}, {Flanagan},
  {Fleming}, {Flight}, {Fowler}, {Fox}, {Franc}, {Francis}, {Franco},
  {Freeman}, {Freestone}, {Fried}, {Friedland}, {Fuess}, {Furic}, {Furmanski},
  {Gago}, {Gallagher}, {Gallego-Ros}, {Gallice}, {Galymov}, {Gamberini},
  {Gamble}, {Gandhi}, {Gandrajula}, {Gao}, {Garcia-Gamez}, {Garc{\'\i}a-Peris},
  {Gardiner}, {Gastler}, {Ge}, {Gelli}, {Gendotti}, {Gent},
  {Ghorbani-Moghaddam}, {Gibin}, {Gil-Botella}, {Girerd}, {Giri}, {Gnani},
  {Gogota}, {Gold}, {Gollapinni}, {Gollwitzer}, {Gomes}, {Gomez Bermeo}, {Gomez
  Fajardo}, {Gonnella}, {Gonzalez-Cuevas}, {Goodman}, {Goodwin}, {Goswami},
  {Gotti}, {Goudzovski}, {Grace}, {Graham}, {Gramellini}, {Gran}, {Granados},
  {Grant}, {Grant}, {Gratieri}, {Green}, {Green}, {Greenler}, {Greenwood},
  {Greer}, {Griffith}, {Groh}, {Grudzinski}, {Grzelak}, {Gu}, {Guarino},
  {Guenette}, {Guglielmi}, {Guo}, {Guthikonda}, {Gutierrez}, {Guzowski},
  {Guzzo}, {Gwon}, {Habig}, {Hackenburg}, {Hadavand}, {Haenni}, {Hahn},
  {Haigh}, {Haiston}, {Hamernik}, {Hamilton}, {Han}, {Harder}, {Harris},
  {Hartnell}, {Hasegawa}, {Hatcher}, {Hazen}, {Heavey}, {Heeger}, {Heise},
  {Hennessy}, {Henry}, {Hernandez Morquecho}, {Herner}, {Hertel}, {Hesam},
  {Hewes}, {Higuera}, {Hill}, {Hillier}, {Himmel}, {Hoff}, {Hohl}, {Holin},
  {Hoppe}, {Horton-Smith}, {Hostert}, {Hourlier}, {Howard}, {Howell}, {Huang},
  {Huang}, {Hugon}, {Iles}, {Ilic}, {Iliescu}, {Illingworth}, {Ioannisian},
  {Itay}, {Izmaylov}, {James}, {Jargowsky}, {Jediny}, {Jes{\`u}s-Valls}, {Ji},
  {Jiang}, {Jim{\'e}nez}, {Jipa}, {Joglekar}, {Johnson}, {Johnson}, {Jones},
  {Jones}, {Jung}, {Junk}, {Jwa}, {Kabirnezhad}, {Kaboth}, {Kadenko}, {Kamiya},
  {Karagiorgi}, {Karcher}, {Karolak}, {Karyotakis}, {Kasai}, {Kasetti},
  {Kashur}, {Kazaryan}, {Kearns}, {Keener}, {Kelly}, {Kemp}, {Ketchum},
  {Kettell}, {Khabibullin}, {Khotjantsev}, {Khvedelidze}, {Kim}, {King},
  {Kirby}, {Kirby}, {Klein}, {Koehler}, {Koerner}, {Kohn}, {Koller},
  {Kordosky}, {Kosc}, {Kose}, {Kosteleck{\'y}}, {Kothekar}, {Krennrich},
  {Kreslo}, {Kudenko}, {Kudryavtsev}, {Kulagin}, {Kumar}, {Kumar}, {Kuruppu},
  {Kus}, {Kutter}, {Lambert}, {Lande}, {Lane}, {Lang}, {Langford}, {Lasorak},
  {Last}, {Lastoria}, {Laundrie}, {Lawrence}, {Lazanu}, {LaZur}, {Le},
  {Learned}, {LeBrun}, {Lehmann Miotto}, {Lehnert}, {Leigui de Oliveira},
  {Leitner}, {Leyton}, {Li}, {Li}, {Li}, {Li}, {Li}, {Liao}, {Lin}, {Lin},
  {Lister}, {Littlejohn}, {Liu}, {Lockwitz}, {Loew}, {Lokajicek}, {Lomidze},
  {Long}, {Loo}, {Lorca}, {Lord}, {LoSecco}, {Louis}, {Luk}, {Luo}, {Lurkin},
  {Lux}, {Luzio}, {MacFarland}, {Machado}, {Machado}, {Macias}, {Macier},
  {Maddalena}, {Madigan}, {Magill}, {Mahn}, {Maio}, {Major}, {Maloney},
  {Mandrioli}, {Maneira}, {Manenti}, {Manly}, {Mann}, {Manolopoulos}, {Manrique
  Plata}, {Marchionni}, {Marciano}, {Marfatia}, {Mariani}, {Maricic},
  {Marinho}, {Marino}, {Marshak}, {Marshall}, {Marshall}, {Marteau},
  {Martin-Albo}, {Martinez}, {Martinez Caicedo}, {Martynenko}, {Mason},
  {Mastbaum}, {Masud}, {Matsuno}, {Matthews}, {Mauger}, {Mauri},
  {Mavrokoridis}, {Mazza}, {Mazzacane}, {Mazzucato}, {McCluskey}, {McConkey},
  {McFarland}, {McGrew}, {McNab}, {Mefodiev}, {Mehta}, {Melas}, {Mellinato},
  {Mena}, {Menary}, {Mendez}, {Menegolli}, {Meng}, {Messier}, {Metcalf},
  {Mewes}, {Meyer}, {Miao}, {Michna}, {Miedema}, {Migenda}, {Milincic},
  {Miller}, {Mills}, {Milne}, {Mineev}, {Miranda}, {Miryala}, {Mishra},
  {Mishra}, {Mislivec}, {Mladenov}, {Mocioiu}, {Moffat}, {Moggi}, {Mohanta},
  {Mohayai}, {Mokhov}, {Molina}, {Molina Bueno}, {Montanari}, {Montanari},
  {Montanari}, {Montano Zetina}, {Moon}, {Mooney}, {Moor}, {Moreno}, {Morgan},
  {Morris}, {Mossey}, {Motuk}, {Moura}, {Mousseau}, {Mu}, {Mualem}, {Mueller},
  {Muether}, {Mufson}, {Muheim}, {Muir}, {Mulhearn}, {Muramatsu}, {Murphy},
  {Musser}, {Nachtman}, {Nagu}, {Nalbandyan}, {Nandakumar}, {Naples}, {Narita},
  {Navas-Nicol{\'a}s}, {Nayak}, {Nebot-Guinot}, {Necib}, {Negishi}, {Nelson},
  {Nesbit}, {Nessi}, {Newbold}, {Newcomer}, {Newhart}, {Nichol}, {Niner},
  {Nishimura}, {Norman}, {Norrick}, {Northrop}, {Novella}, {Nowak}, {Oberling},
  {Olivares Del Campo}, {Olivier}, {Onel}, {Onishchuk}, {Ott}, {Pagani},
  {Pakvasa}, {Palamara}, {Palestini}, {Paley}, {Pallavicini}, {Palomares},
  {Pantic}, {Paolone}, {Papadimitriou}, {Papaleo}, {Papanestis},
  {Paramesvaran}, {Parke}, {Parsa}, {Parvu}, {Pascoli}, {Pasqualini},
  {Pasternak}, {Pater}, {Patrick}, {Patrizii}, {Patterson}, {Patton}, {Patzak},
  {Paudel}, {Paulos}, {Paulucci}, {Pavlovic}, {Pawloski}, {Payne}, {Pec},
  {Peeters}, {Penichot}, {Pennacchio}, {Penzo}, {Peres}, {Perry}, {Pershey},
  {Pessina}, {Petrillo}, {Petta}, {Petti}, {Piastra}, {Pickering},
  {Pietropaolo}, {Pillow}, {Pinzino}, {Plunkett}, {Poling}, {Pons},
  {Poonthottathil}, {Pordes}, {Potekhin}, {Potenza}, {Potukuchi}, {Pozimski},
  {Pozzato}, {Prakash}, {Prakash}, {Prince}, {Prior}, {Pugnere}, {Qi}, {Qian},
  {Raaf}, {Raboanary}, {Radeka}, {Rademacker}, {Radics}, {Rafique}, {Raguzin},
  {Rai}, {Rajaoalisoa}, {Rakhno}, {Rakotondramanana}, {Rakotondravohitra},
  {Ramachers}, {Rameika}, {Ramirez Delgado}, {Ramson}, {Rappoldi}, {Raselli},
  {Ratoff}, {Ravat}, {Razafinime}, {Real}, {Rebel}, {Redondo},
  {Reggiani-Guzzo}, {Rehak}, {Reichenbacher}, {Reitzner}, {Renshaw}, {Rescia},
  {Resnati}, {Reynolds}, {Riccobene}, {Rice}, {Rielage}, {Rigaut}, {Rivera},
  {Rochester}, {Roda}, {Rodrigues}, {Rodriguez Alonso}, {Rodriguez Rondon},
  {Roeth}, {Rogers}, {Rosauro-Alcaraz}, {Rossella}, {Rout}, {Roy}, {Rubbia},
  {Rubbia}, {Russell}, {Russell}, {Ruterbories}, {Saakyan}, {Sacerdoti},
  {Safford}, {Sahu}, {Sala}, {Samios}, {Sanchez}, {Sanders}, {Sankey},
  {Santana}, {Santos-Maldonado}, {Saoulidou}, {Sapienza}, {Sarasty},
  {Sarcevic}, {Savage}, {Savinov}, {Scaramelli}, {Scarff}, {Scarpelli},
  {Schaffer}, {Schellman}, {Schlabach}, {Schmitz}, {Scholberg}, {Schukraft},
  {Segreto}, {Sensenig}, {Seong}, {Sergi}, {Sergiampietri}, {Sgalaberna},
  {Shaevitz}, {Shafaq}, {Shamma}, {Sharma}, {Sharma}, {Shaw},
  {Shepherd-Themistocleous}, {Shin}, {Shooltz}, {Shrock}, {Simard}, {Simos},
  {Sinclair}, {Sinev}, {Singh}, {Singh}, {Singh}, {Sipos}, {Sippach}, {Sirri},
  {Sitraka}, {Siyeon}, {Smargianaki}, {Smith}, {Smith}, {Smith}, {Smolik},
  {Smy}, {Snopok}, {Soares Nunes}, {Sobel}, {Soderberg}, {Solano Salinas},
  {S{\"o}ldner-Rembold}, {Solomey}, {Solovov}, {Sondheim}, {Sorel},
  {Soto-Oton}, {Sousa}, {Soustruznik}, {Spagliardi}, {Spanu}, {Spitz},
  {Spooner}, {Spurgeon}, {Staley}, {Stancari}, {Stanco}, {Steiner}, {Stewart},
  {Stillwell}, {Stock}, {Stocker}, {Stokes}, {Strait}, {Strauss}, {Striganov},
  {Stuart}, {Summers}, {Surdo}, {Susic}, {Suter}, {Sutera}, {Svoboda},
  {Szczerbinska}, {Szelc}, {Talaga}, {Tanaka}, {Tapia Oregui}, {Tapper},
  {Tariq}, {Tatar}, {Tayloe}, {Teklu}, {Tenti}, {Terao}, {Ternes}, {Terranova},
  {Testera}, {Thea}, {Thompson}, {Thorn}, {Timm}, {Tonazzo}, {Torti},
  {T{\'o}rtola}, {Tortorici}, {Totani}, {Toups}, {Touramanis}, {Trevor},
  {Trzaska}, {Tsai}, {Tsamalaidze}, {Tsang}, {Tsverava}, {Tufanli}, {Tull},
  {Tyley}, {Tzanov}, {Uchida}, {Urheim}, {Usher}, {Vagins}, {Vahle},
  {Valdiviesso}, {Valencia}, {Vallari}, {Valle}, {Vallecorsa}, {Van Berg}, {Van
  de Water}, {Vanegas Forero}, {Varanini}, {Vargas}, {Varner}, {Vasel},
  {Vasseur}, {Vaziri}, {Ventura}, {Verdugo}, {Vergani}, {Vermeulen},
  {Verzocchi}, {Vieira de Souza}, {Vignoli}, {Vilela}, {Viren}, {Vrba},
  {Wachala}, {Waldron}, {Wallbank}, {Wang}, {Wang}, {Wang}, {Wang},
  {Warburton}, {Warner}, {Wascko}, {Waters}, {Watson}, {Weatherly}, {Weber},
  {Weber}, {Wei}, {Weinstein}, {Wenman}, {Wetstein}, {While}, {White},
  {Whitehead}, {Whittington}, {Wilking}, {Wilkinson}, {Williams}, {Wilson},
  {Wilson}, {Wolcott}, {Wongjirad}, {Wood}, {Wood}, {Worcester}, {Worcester},
  {Wret}, {Wu}, {Wu}, {Xiao}, {Yang}, {Yang}, {Yershov}, {Yonehara}, {Young},
  {Yu}, {Yu}, {Zaki}, {Zalesak}, {Zambelli}, {Zamorano}, {Zani}, {Zazueta},
  {Zeller}, {Zennamo}, {Zeug}, {Zhang}, {Zhao}, {Zhivun}, {Zhu}, {Zimmerman},
  {Zito}, {Zucchelli}, {Zuklin}, {Zutshi}, \& {Zwaska}}]{Abi2021}
{Abi}, B., {Acciarri}, R., {Acero}, M.~A., {et~al.} 2021, European Physical
  Journal C, 81, 423, \dodoi{10.1140/epjc/s10052-021-09166-w}

\bibitem[{Akaho {et~al.}(2021)Akaho, Harada, Nagakura, Sumiyoshi, Iwakami,
  Okawa, Furusawa, Matsufuru, \& Yamada}]{Akaho2021}
Akaho, R., Harada, A., Nagakura, H., {et~al.} 2021, The Astrophysical Journal,
  909, 210, \dodoi{10.3847/1538-4357/abe1bf}

\bibitem[{{Akaho} {et~al.}(2023){Akaho}, {Harada}, {Nagakura}, {Iwakami},
  {Okawa}, {Furusawa}, {Matsufuru}, {Sumiyoshi}, \& {Yamada}}]{Akaho2023}
{Akaho}, R., {Harada}, A., {Nagakura}, H., {et~al.} 2023, \apj, 944, 60,
  \dodoi{10.3847/1538-4357/acad76}

\bibitem[{{An} {et~al.}(2016){An}, {An}, {An}, {Antonelli}, {Baussan},
  {Beacom}, {Bezrukov}, {Blyth}, {Brugnera}, {Buizza Avanzini}, {Busto},
  {Cabrera}, {Cai}, {Cai}, {Cammi}, {Cao}, {Cao}, {Chang}, {Chen}, {Chen},
  {Chen}, {Chiesa}, {Clemenza}, {Clerbaux}, {Conrad}, {D'Angelo}, {De Kerret},
  {Deng}, {Deng}, {Ding}, {Djurcic}, {Dornic}, {Dracos}, {Drapier}, {Dusini},
  {Dye}, {Enqvist}, {Fan}, {Fang}, {Favart}, {Ford}, {G{\"o}ger-Neff}, {Gan},
  {Garfagnini}, {Giammarchi}, {Gonchar}, {Gong}, {Gong}, {Gonin}, {Grassi},
  {Grewing}, {Guan}, {Guarino}, {Guo}, {Guo}, {Guo}, {Hagner}, {Han}, {He},
  {Heng}, {Hsiung}, {Hu}, {Hu}, {Hu}, {Huang}, {Huang}, {Huo}, {Ioannisian},
  {Jeitler}, {Ji}, {Jiang}, {Jollet}, {Kang}, {Karagounis}, {Kazarian},
  {Krumshteyn}, {Kruth}, {Kuusiniemi}, {Lachenmaier}, {Leitner}, {Li}, {Li},
  {Li}, {Li}, {Li}, {Li}, {Li}, {Li}, {Li}, {Liang}, {Lin}, {Lin}, {Lin},
  {Ling}, {Lippi}, {Liu}, {Liu}, {Liu}, {Liu}, {Liu}, {Liu}, {Liu}, {Liu},
  {Liu}, {Lombardi}, {Long}, {Lu}, {Lu}, {Lu}, {Lu}, {Lubsandorzhiev},
  {Ludhova}, {Luo}, {Lyashuk}, {M{\"o}llenberg}, {Ma}, {Mantovani}, {Mao},
  {Mari}, {McDonough}, {Meng}, {Meregaglia}, {Meroni}, {Mezzetto}, {Miramonti},
  {Mueller}, {Naumov}, {Oberauer}, {Ochoa-Ricoux}, {Olshevskiy}, {Ortica},
  {Paoloni}, {Peng}, {Peng}, {Previtali}, {Qi}, {Qian}, {Qian}, {Qian}, {Qin},
  {Raffelt}, {Ranucci}, {Ricci}, {Robens}, {Romani}, {Ruan}, {Ruan},
  {Salamanna}, {Shaevitz}, {Sinev}, {Sirignano}, {Sisti}, {Smirnov}, {Soiron},
  {Stahl}, {Stanco}, {Steinmann}, {Sun}, {Sun}, {Taichenachev}, {Tang},
  {Tkachev}, {Trzaska}, {van Waasen}, {Volpe}, {Vorobel}, {Votano}, {Wang},
  {Wang}, {Wang}, {Wang}, {Wang}, {Wang}, {Wang}, {Wang}, {Wang}, {Wang},
  {Wang}, {Wang}, {Wang}, {Wang}, {Wei}, {Wen}, {Wiebusch}, {Wonsak}, {Wu},
  {Wulz}, {Wurm}, {Xi}, {Xia}, {Xie}, {Xing}, {Xu}, {Yan}, {Yang}, {Yang},
  {Yang}, {Yang}, {Yang}, {Yao}, {Yegin}, {Yermia}, {You}, {Yu}, {Yu}, {Yu},
  {Zavatarelli}, {Zhan}, {Zhang}, {Zhang}, {Zhang}, {Zhang}, {Zhang}, {Zhang},
  {Zhang}, {Zhao}, {Zheng}, {Zhong}, {Zhou}, {Zhou}, {Zhou}, {Zhou}, {Zhou},
  {Zhou}, {Zhou}, {Zhou}, {Zhou}, \& {Zou}}]{An2016}
{An}, F., {An}, G., {An}, Q., {et~al.} 2016, Journal of Physics G Nuclear
  Physics, 43, 030401, \dodoi{10.1088/0954-3899/43/3/030401}

\bibitem[{{Asakura} {et~al.}(2016){Asakura}, {Gando}, {Gando}, {Hachiya},
  {Hayashida}, {Ikeda}, {Inoue}, {Ishidoshiro}, {Ishikawa}, {Ishio}, {Koga},
  {Matsuda}, {Mitsui}, {Motoki}, {Nakamura}, {Obara}, {Oura}, {Shimizu},
  {Shirahata}, {Shirai}, {Suzuki}, {Tachibana}, {Tamae}, {Ueshima}, {Watanabe},
  {Xu}, {Kozlov}, {Takemoto}, {Yoshida}, {Fushimi}, {Piepke}, {Banks},
  {Berger}, {Fujikawa}, {O'Donnell}, {Learned}, {Maricic}, {Matsuno}, {Sakai},
  {Winslow}, {Efremenko}, {Karwowski}, {Markoff}, {Tornow}, {Detwiler},
  {Enomoto}, {Decowski}, \& {KamLAND Collaboration}}]{Asakura2016}
{Asakura}, K., {Gando}, A., {Gando}, Y., {et~al.} 2016, \apj, 818, 91,
  \dodoi{10.3847/0004-637X/818/1/91}

\bibitem[{{Barr{\`e}re} {et~al.}(2022){Barr{\`e}re}, {Guilet}, {Reboul-Salze},
  {Raynaud}, \& {Janka}}]{Barrere2022}
{Barr{\`e}re}, P., {Guilet}, J., {Reboul-Salze}, A., {Raynaud}, R., \& {Janka},
  H.~T. 2022, \aap, 668, A79, \dodoi{10.1051/0004-6361/202244172}

\bibitem[{{Beacom} {et~al.}(2002){Beacom}, {Farr}, \& {Vogel}}]{Beacom2002}
{Beacom}, J.~F., {Farr}, W.~M., \& {Vogel}, P. 2002, \prd, 66, 033001,
  \dodoi{10.1103/PhysRevD.66.033001}

\bibitem[{{Bionta} {et~al.}(1987){Bionta}, {Blewitt}, {Bratton}, {Casper},
  {Ciocio}, {Claus}, {Cortez}, {Crouch}, {Dye}, {Errede}, {Foster}, {Gajewski},
  {Ganezer}, {Goldhaber}, {Haines}, {Jones}, {Kielczewska}, {Kropp}, {Learned},
  {Losecco}, {Matthews}, {Miller}, {Mudan}, {Park}, {Price}, {Reines},
  {Schultz}, {Seidel}, {Shumard}, {Sinclair}, {Sobel}, {Stone}, {Sulak},
  {Svoboda}, {Thornton}, {van der Velde}, \& {Wuest}}]{Bionta1987}
{Bionta}, R.~M., {Blewitt}, G., {Bratton}, C.~B., {et~al.} 1987, \prl, 58,
  1494, \dodoi{10.1103/PhysRevLett.58.1494}

\bibitem[{{Blondin} {et~al.}(2003){Blondin}, {Mezzacappa}, \&
  {DeMarino}}]{Blondin2003}
{Blondin}, J.~M., {Mezzacappa}, A., \& {DeMarino}, C. 2003, \apj, 584, 971,
  \dodoi{10.1086/345812}

\bibitem[{{Bollig} {et~al.}(2021){Bollig}, {Yadav}, {Kresse}, {Janka},
  {M{\"u}ller}, \& {Heger}}]{Bollig2021}
{Bollig}, R., {Yadav}, N., {Kresse}, D., {et~al.} 2021, \apj, 915, 28,
  \dodoi{10.3847/1538-4357/abf82e}

\bibitem[{{Bruenn}(1985)}]{Bruenn1985}
{Bruenn}, S.~W. 1985, \apjs, 58, 771, \dodoi{10.1086/191056}

\bibitem[{{Burrows} \& {Lattimer}(1986)}]{Burrows1986}
{Burrows}, A., \& {Lattimer}, J.~M. 1986, \apj, 307, 178,
  \dodoi{10.1086/164405}

\bibitem[{{Burrows} \& {Vartanyan}(2021)}]{Burrows2021}
{Burrows}, A., \& {Vartanyan}, D. 2021, \nat, 589, 29,
  \dodoi{10.1038/s41586-020-03059-w}

\bibitem[{{Capozzi} {et~al.}(2017){Capozzi}, {Di Valentino}, {Lisi}, {Marrone},
  {Melchiorri}, \& {Palazzo}}]{Capozzi2017}
{Capozzi}, F., {Di Valentino}, E., {Lisi}, E., {et~al.} 2017, \prd, 95, 096014,
  \dodoi{10.1103/PhysRevD.95.096014}

\bibitem[{{Chan} {et~al.}(2018){Chan}, {M{\"u}ller}, {Heger}, {Pakmor}, \&
  {Springel}}]{Chan2018}
{Chan}, C., {M{\"u}ller}, B., {Heger}, A., {Pakmor}, R., \& {Springel}, V.
  2018, \apjl, 852, L19, \dodoi{10.3847/2041-8213/aaa28c}

\bibitem[{{Chevalier}(1989)}]{Chevalier1989}
{Chevalier}, R.~A. 1989, \apj, 346, 847, \dodoi{10.1086/168066}

\bibitem[{{Coleman} \& {Burrows}(2022)}]{Coleman2022}
{Coleman}, M. S.~B., \& {Burrows}, A. 2022, \mnras, 517, 3938,
  \dodoi{10.1093/mnras/stac2573}

\bibitem[{{Colgate}(1971)}]{Colgate1971}
{Colgate}, S.~A. 1971, \apj, 163, 221, \dodoi{10.1086/150760}

\bibitem[{{Dasgupta} \& {Beacom}(2011)}]{Dasgupta2011}
{Dasgupta}, B., \& {Beacom}, J.~F. 2011, \prd, 83, 113006,
  \dodoi{10.1103/PhysRevD.83.113006}

\bibitem[{{Dexter} \& {Kasen}(2013)}]{Dexter2013}
{Dexter}, J., \& {Kasen}, D. 2013, \apj, 772, 30,
  \dodoi{10.1088/0004-637X/772/1/30}

\bibitem[{{Dighe} \& {Smirnov}(2000)}]{Dighe2000}
{Dighe}, A.~S., \& {Smirnov}, A.~Y. 2000, \prd, 62, 033007,
  \dodoi{10.1103/PhysRevD.62.033007}

\bibitem[{{Fischer} {et~al.}(2020){Fischer}, {Guo}, {Dzhioev},
  {Mart{\'\i}nez-Pinedo}, {Wu}, {Lohs}, \& {Qian}}]{Fischer2020}
{Fischer}, T., {Guo}, G., {Dzhioev}, A.~A., {et~al.} 2020, \prc, 101, 025804,
  \dodoi{10.1103/PhysRevC.101.025804}

\bibitem[{{Fischer} {et~al.}(2012){Fischer}, {Mart{\'\i}nez-Pinedo}, {Hempel},
  \& {Liebend{\"o}rfer}}]{Fischer2012}
{Fischer}, T., {Mart{\'\i}nez-Pinedo}, G., {Hempel}, M., \& {Liebend{\"o}rfer},
  M. 2012, \prd, 85, 083003, \dodoi{10.1103/PhysRevD.85.083003}

\bibitem[{{Fischer} {et~al.}(2010){Fischer}, {Whitehouse}, {Mezzacappa},
  {Thielemann}, \& {Liebend{\"o}rfer}}]{Fischer2010}
{Fischer}, T., {Whitehouse}, S.~C., {Mezzacappa}, A., {Thielemann}, F.~K., \&
  {Liebend{\"o}rfer}, M. 2010, \aap, 517, A80,
  \dodoi{10.1051/0004-6361/200913106}

\bibitem[{{Foglizzo} {et~al.}(2007){Foglizzo}, {Galletti}, {Scheck}, \&
  {Janka}}]{Foglizzo2007}
{Foglizzo}, T., {Galletti}, P., {Scheck}, L., \& {Janka}, H.~T. 2007, \apj,
  654, 1006, \dodoi{10.1086/509612}

\bibitem[{{Fryer}(2009)}]{Fryer2009}
{Fryer}, C.~L. 2009, \apj, 699, 409, \dodoi{10.1088/0004-637X/699/1/409}

\bibitem[{{Fryxell} {et~al.}(1991){Fryxell}, {Mueller}, \&
  {Arnett}}]{Fryxell1991}
{Fryxell}, B., {Mueller}, E., \& {Arnett}, D. 1991, \apj, 367, 619,
  \dodoi{10.1086/169657}

\bibitem[{{Fukuda} {et~al.}(2003){Fukuda}, {Fukuda}, {Hayakawa}, {Ichihara},
  {Ishitsuka}, {Itow}, {Kajita}, {Kameda}, {Kaneyuki}, {Kasuga}, {Kobayashi},
  {Kobayashi}, {Koshio}, {Miura}, {Moriyama}, {Nakahata}, {Nakayama}, {Namba},
  {Obayashi}, {Okada}, {Oketa}, {Okumura}, {Oyabu}, {Sakurai}, {Shiozawa},
  {Suzuki}, {Takeuchi}, {Toshito}, {Totsuka}, {Yamada}, {Desai}, {Earl},
  {Hong}, {Kearns}, {Masuzawa}, {Messier}, {Stone}, {Sulak}, {Walter}, {Wang},
  {Scholberg}, {Barszczak}, {Casper}, {Liu}, {Gajewski}, {Halverson}, {Hsu},
  {Kropp}, {Mine}, {Price}, {Reines}, {Smy}, {Sobel}, {Vagins}, {Ganezer},
  {Keig}, {Ellsworth}, {Tasaka}, {Flanagan}, {Kibayashi}, {Learned}, {Matsuno},
  {Stenger}, {Hayato}, {Ishii}, {Ichikawa}, {Kanzaki}, {Kobayashi}, {Maruyama},
  {Nakamura}, {Oyama}, {Sakai}, {Sakuda}, {Sasaki}, {Echigo}, {Iwashita},
  {Kohama}, {Suzuki}, {Hasegawa}, {Inagaki}, {Kato}, {Maesaka}, {Nakaya},
  {Nishikawa}, {Yamamoto}, {Haines}, {Kim}, {Sanford}, {Svoboda}, {Blaufuss},
  {Chen}, {Conner}, {Goodman}, {Guillian}, {Sullivan}, {Turcan}, {Habig},
  {Ackerman}, {Goebel}, {Hill}, {Jung}, {Kato}, {Kerr}, {Malek}, {Martens},
  {Mauger}, {McGrew}, {Sharkey}, {Viren}, {Yanagisawa}, {Doki}, {Inaba}, {Ito},
  {Kirisawa}, {Kitaguchi}, {Mitsuda}, {Miyano}, {Saji}, {Takahata},
  {Takahashi}, {Higuchi}, {Kajiyama}, {Kusano}, {Nagashima}, {Nitta}, {Takita},
  {Yamaguchi}, {Yoshida}, {Kim}, {Kim}, {Yoo}, {Okazawa}, {Etoh}, {Fujita},
  {Gando}, {Hasegawa}, {Hasegawa}, {Hatakeyama}, {Inoue}, {Ishihara},
  {Iwamoto}, {Koga}, {Nishiyama}, {Ogawa}, {Shirai}, {Suzuki}, {Takayama},
  {Tsushima}, {Koshiba}, {Ichikawa}, {Hashimoto}, {Hatakeyama}, {Koike},
  {Horiuchi}, {Nemoto}, {Nishijima}, {Takeda}, {Fujiyasu}, {Futagami},
  {Ishino}, {Kanaya}, {Morii}, {Nishihama}, {Nishimura}, {Suzuki}, {Watanabe},
  {Kielczewska}, {Golebiewska}, {Berns}, {Boyd}, {Doyle}, {George}, {Stachyra},
  {Wai}, {Wilkes}, {Young}, {Kobayashi}, \& {Super-Kamiokande
  Collaboration}}]{Fukuda2003}
{Fukuda}, S., {Fukuda}, Y., {Hayakawa}, T., {et~al.} 2003, Nuclear Instruments
  and Methods in Physics Research A, 501, 418,
  \dodoi{10.1016/S0168-9002(03)00425-X}

\bibitem[{Furusawa {et~al.}(2017)Furusawa, Togashi, Nagakura, Sumiyoshi,
  Yamada, Suzuki, \& Takano}]{Furusawa2017}
Furusawa, S., Togashi, H., Nagakura, H., {et~al.} 2017, Journal of Physics G:
  Nuclear and Particle Physics, 44, 094001, \dodoi{10.1088/1361-6471/aa7f35}

\bibitem[{{Harada} {et~al.}(2023{\natexlab{a}}){Harada}, {Suwa}, {Harada},
  {Koshio}, {Mori}, {Nakanishi}, {Nakazato}, {Sumiyoshi}, \&
  {Wendell}}]{HaradaA2023}
{Harada}, A., {Suwa}, Y., {Harada}, M., {et~al.} 2023{\natexlab{a}}, \apj, 954,
  52, \dodoi{10.3847/1538-4357/ace52e}

\bibitem[{{Harada} {et~al.}(2023{\natexlab{b}}){Harada}, {Abe}, {Bronner},
  {Hayato}, {Hiraide}, {Hosokawa}, {Ieki}, {Ikeda}, {Kameda}, {Kanemura},
  {Kaneshima}, {Kashiwagi}, {Kataoka}, {Miki}, {Mine}, {Miura}, {Moriyama},
  {Nakano}, {Nakahata}, {Nakayama}, {Noguchi}, {Okamoto}, {Sato}, {Sekiya},
  {Shiba}, {Shimizu}, {Shiozawa}, {Sonoda}, {Suzuki}, {Takeda}, {Takemoto},
  {Takenaka}, {Tanaka}, {Watanabe}, {Yano}, {Han}, {Kajita}, {Okumura},
  {Tashiro}, {Tomiya}, {Wang}, {Yoshida}, {Megias}, {Fernandez}, {Labarga},
  {Ospina}, {Zaldivar}, {Pointon}, {Kearns}, {Raaf}, {Wan}, {Wester}, {Bian},
  {Griskevich}, {Locke}, {Smy}, {Sobel}, {Takhistov}, {Yankelevich}, {Hill},
  {Lee}, {Moon}, {Park}, {Bodur}, {Scholberg}, {Walter}, {Beauch{\^e}ne},
  {Drapier}, {Giampaolo}, {Mueller}, {Santos}, {Paganini}, {Quilain},
  {Ishizuka}, {Nakamura}, {Jang}, {Learned}, {Choi}, {Iovine}, {Cao},
  {Anthony}, {Martin}, {Scott}, {Sztuc}, {Uchida}, {Berardi}, {Catanesi},
  {Radicioni}, {Calabria}, {Langella}, {Machado}, {De Rosa}, {Collazuol},
  {Iacob}, {Lamoureux}, {Mattiazzi}, {Ludovici}, {Gonin}, {Pronost},
  {Fujisawa}, {Maekawa}, {Nishimura}, {Okazaki}, {Akutsu}, {Friend},
  {Hasegawa}, {Ishida}, {Kobayashi}, {Jakkapu}, {Matsubara}, {Nakadaira},
  {Nakamura}, {Oyama}, {Sakashita}, {Sekiguchi}, {Tsukamoto}, {Bhuiyan},
  {Burton}, {Di Lodovico}, {Gao}, {Goldsack}, {Katori}, {Migenda}, {Xie},
  {Zsoldos}, {Kotsar}, {Ozaki}, {Suzuki}, {Takagi}, {Takeuchi}, {Feng}, {Feng},
  {Hu}, {Hu}, {Kikawa}, {Mori}, {Nakaya}, {Wendell}, {Yasutome}, {Jenkins},
  {McCauley}, {Mehta}, {Tarrant}, {Fukuda}, {Itow}, {Menjo}, {Ninomiya},
  {Lagoda}, {Lakshmi}, {Mandal}, {Mijakowski}, {Prabhu}, {Zalipska}, {Jia},
  {Jiang}, {Jung}, {Wilking}, {Yanagisawa}, {Hino}, {Ishino}, {Kitagawa},
  {Koshio}, {Nakanishi}, {Sakai}, {Tada}, {Tano}, {Barr}, {Barrow}, {Cook},
  {Samani}, {Wark}, {Holin}, {Nova}, {Yang}, {Yang}, {Yoo}, {Fannon}, {Kneale},
  {Malek}, {McElwee}, {Thiesse}, {Thompson}, {Wilson}, {Okazawa}, {Kim},
  {Kwon}, {Seo}, {Yu}, {Ichikawa}, {Nakamura}, {Tairafune}, {Nishijima},
  {Nakagiri}, {Nakajima}, {Shima}, {Taniuchi}, {Watanabe}, {Yokoyama}, {de
  Perio}, {Martens}, {Tsui}, {Vagins}, {Xia}, {Kuze}, {Izumiyama}, {Matsumoto},
  {Ishitsuka}, {Ito}, {Kinoshita}, {Matsumoto}, {Ommura}, {Shigeta}, {Shinoki},
  {Suganuma}, {Yamauchi}, {Martin}, {Tanaka}, {Towstego}, {Gaur},
  {Gousy-Leblanc}, {Hartz}, {Konaka}, {Li}, {Prouse}, {Chen}, {Xu}, {Zhang},
  {Posiadala-Zezula}, {Boyd}, {Edwards}, {Hadley}, {Nicholson}, {O'Flaherty},
  {Richards}, {Ali}, {Jamieson}, {Marti}, {Minamino}, {Pintaudi}, {Sano},
  {Suzuki}, {Wada}, \& {Super-Kamiokande Collaboration}}]{HaradaM2023}
{Harada}, M., {Abe}, K., {Bronner}, C., {et~al.} 2023{\natexlab{b}}, \apjl,
  951, L27, \dodoi{10.3847/2041-8213/acdc9e}

\bibitem[{{Herant} {et~al.}(1994){Herant}, {Benz}, {Hix}, {Fryer}, \&
  {Colgate}}]{Herant1994}
{Herant}, M., {Benz}, W., {Hix}, W.~R., {Fryer}, C.~L., \& {Colgate}, S.~A.
  1994, \apj, 435, 339, \dodoi{10.1086/174817}

\bibitem[{{Hirata} {et~al.}(1987){Hirata}, {Kajita}, {Koshiba}, {Nakahata},
  {Oyama}, {Sato}, {Suzuki}, {Takita}, {Totsuka}, {Kifune}, {Suda},
  {Takahashi}, {Tanimori}, {Miyano}, {Yamada}, {Beier}, {Feldscher}, {Kim},
  {Mann}, {Newcomer}, {van}, {Zhang}, \& {Cortez}}]{Hirata1987}
{Hirata}, K., {Kajita}, T., {Koshiba}, M., {et~al.} 1987, \prl, 58, 1490,
  \dodoi{10.1103/PhysRevLett.58.1490}

\bibitem[{{Horiuchi} \& {Kneller}(2018)}]{Horiuchi2018}
{Horiuchi}, S., \& {Kneller}, J.~P. 2018, Journal of Physics G Nuclear Physics,
  45, 043002, \dodoi{10.1088/1361-6471/aaa90a}

\bibitem[{{Hyper-Kamiokande Proto-Collaboration}
  {et~al.}(2018){Hyper-Kamiokande Proto-Collaboration}, {:}, {Abe}, {Abe},
  {Aihara}, {Aimi}, {Akutsu}, {Andreopoulos}, {Anghel}, {Anthony}, {Antonova},
  {Ashida}, {Aushev}, {Barbi}, {Barker}, {Barr}, {Beltrame}, {Berardi},
  {Bergevin}, {Berkman}, {Berns}, {Berry}, {Bhadra}, {Bravo-Bergu{\~n}o},
  {Blaszczyk}, {Blondel}, {Bolognesi}, {Boyd}, {Bravar}, {Bronner}, {Buizza
  Avanzini}, {Cafagna}, {Cole}, {Calland}, {Cao}, {Cartwright}, {Catanesi},
  {Checchia}, {Chen-Wishart}, {Choi}, {Choi}, {Coleman}, {Collazuol}, {Cowan},
  {Cremonesi}, {Dealtry}, {De Rosa}, {Densham}, {Dewhurst}, {Drakopoulou}, {Di
  Lodovico}, {Drapier}, {Dumarchez}, {Dunne}, {Dziewiecki}, {Emery}, {Esmaili},
  {Evangelisti}, {Fernandez-Martinez}, {Feusels}, {Finch}, {Fiorentini},
  {Fiorillo}, {Fitton}, {Frankiewicz}, {Friend}, {Fujii}, {Fukuda}, {Fukuda},
  {Ganezer}, {Giganti}, {Gonin}, {Grant}, {Gumplinger}, {Hadley}, {Hartfiel},
  {Hartz}, {Hayato}, {Hayrapetyan}, {Hill}, {Hirota}, {Horiuchi}, {Ichikawa},
  {Iijima}, {Ikeda}, {Imber}, {Inoue}, {Insler}, {Intonti}, {Ioannisian},
  {Ishida}, {Ishino}, {Ishitsuka}, {Itow}, {Iwamoto}, {Izmaylov}, {Jamieson},
  {Jang}, {Jang}, {Jeon}, {Jiang}, {Jonsson}, {Joo}, {Kaboth}, {Kachulis},
  {Kajita}, {Kameda}, {Kataoka}, {Katori}, {Kayrapetyan}, {Kearns},
  {Khabibullin}, {Khotjantsev}, {Kim}, {Kim}, {Kim}, {Kim}, {King},
  {Kishimoto}, {Kobayashi}, {Koga}, {Konaka}, {Kormos}, {Koshio}, {Korzenev},
  {Kowalik}, {Kropp}, {Kudenko}, {Kurjata}, {Kutter}, {Kuze}, {Labarga},
  {Lagoda}, {Lasorak}, {Laveder}, {Lawe}, {Learned}, {Lim}, {Lindner},
  {Litchfield}, {Longhin}, {Loverre}, {Lou}, {Ludovici}, {Ma}, {Magaletti},
  {Mahn}, {Malek}, {Maret}, {Mariani}, {Martens}, {Marti}, {Martin}, {Marzec},
  {Matsuno}, {Mazzucato}, {McCarthy}, {McCauley}, {McFarland}, {McGrew},
  {Mefodiev}, {Mermod}, {Metelko}, {Mezzetto}, {Migenda}, {Mijakowski},
  {Minakata}, {Minamino}, {Mine}, {Mineev}, {Mitra}, {Miura}, {Mochizuki},
  {Monroe}, {Moon}, {Moriyama}, {Mueller}, {Muheim}, {Murase}, {Muto},
  {Nakahata}, {Nakajima}, {Nakamura}, {Nakaya}, {Nakayama}, {Nantais},
  {Needham}, {Nicholls}, {Nishimura}, {Noah}, {Nova}, {Nowak}, {Nunokawa},
  {Obayashi}, {O'Keeffe}, {Okajima}, {Okumura}, {Onishchuk}, {O'Sullivan},
  {O'Sullivan}, {Ovsiannikova}, {Owen}, {Oyama}, {Pac}, {Palladino},
  {Palomino}, {Paolone}, {Parker}, {Parsa}, {Payne}, {Perkin}, {Pidcott},
  {Pinzon Guerra}, {Playfer}, {Popov}, {Posiadala-Zezula}, {Poutissou},
  {Pritchard}, {Prouse}, {Pronost}, {Przewlocki}, {Quilain}, {Radicioni},
  {Ratoff}, {Retiere}, {Riccio}, {Richards}, {Rondio}, {Rose}, {Rott},
  {Rountree}, {Ruggeri}, {Rychter}, {Sacco}, {Sakuda}, {Sanchez},
  {Scantamburlo}, {Scott}, {Sedgwick}, {Seiya}, {Sekiguchi}, {Sekiya}, {Seo},
  {Sgalaberna}, {Shah}, {Shaikhiev}, {Shimizu}, {Shiozawa}, {Shitov}, {Short},
  {Simpson}, {Sinnis}, {Smy}, {Snow}, {Sobczyk}, {Sobel}, {Sonoda}, {Spina},
  {Stewart}, {Stone}, {Suda}, {Suwa}, {Suzuki}, {Suzuki}, {Svoboda}, {Taani},
  {Tacik}, {Takeda}, {Takenaka}, {Taketa}, {Takeuchi}, {Takhistov}, {Tanaka},
  {Tanaka}, {Tanaka}, {Terri}, {Thiesse}, {Thompson}, {Thorpe}, {Tobayama},
  {Touramanis}, {Towstego}, {Tsukamoto}, {Tsui}, {Tzanov}, {Uchida}, {Vagins},
  {Vasseur}, {Vilela}, {Vogelaar}, {Walding}, {Walker}, {Ward}, {Wark},
  {Wascko}, {Weber}, {Wendell}, {Wilkes}, {Wilking}, {Wilson}, {Xin},
  {Yamamoto}, {Yanagisawa}, {Yano}, {Yen}, {Yershov}, {Yeum}, {Yokoyama},
  {Yoshida}, {Yu}, {Yu}, {Zalipska}, {Zaremba}, {Ziembicki}, {Zito}, \&
  {Zsoldos}}]{Abe2018}
{Hyper-Kamiokande Proto-Collaboration}, {:}, {Abe}, K., {et~al.} 2018, arXiv
  e-prints, arXiv:1805.04163.
\newblock \doarXiv{1805.04163}

\bibitem[{{Ikeda} {et~al.}(2007){Ikeda}, {Takeda}, {Fukuda}, {Vagins}, {Abe},
  {Iida}, {Ishihara}, {Kameda}, {Koshio}, {Minamino}, {Mitsuda}, {Miura},
  {Moriyama}, {Nakahata}, {Obayashi}, {Ogawa}, {Sekiya}, {Shiozawa}, {Suzuki},
  {Takeuchi}, {Ueshima}, {Watanabe}, {Yamada}, {Higuchi}, {Ishihara},
  {Ishitsuka}, {Kajita}, {Kaneyuki}, {Mitsuka}, {Nakayama}, {Nishino},
  {Okumura}, {Saji}, {Takenaga}, {Clark}, {Desai}, {Dufour}, {Kearns},
  {Likhoded}, {Litos}, {Raaf}, {Stone}, {Sulak}, {Wang}, {Goldhaber}, {Casper},
  {Cravens}, {Dunmore}, {Kropp}, {Liu}, {Mine}, {Regis}, {Smy}, {Sobel},
  {Ganezer}, {Hill}, {Keig}, {Jang}, {Kim}, {Lim}, {Scholberg}, {Tanimoto},
  {Walter}, {Wendell}, {Ellsworth}, {Tasaka}, {Guillian}, {Learned}, {Matsuno},
  {Messier}, {Hayato}, {Ichikawa}, {Ishida}, {Ishii}, {Iwashita}, {Kobayashi},
  {Nakadaira}, {Nakamura}, {Nitta}, {Oyama}, {Totsuka}, {Suzuki}, {Hasegawa},
  {Hiraide}, {Maesaka}, {Nakaya}, {Nishikawa}, {Sasaki}, {Yamamoto},
  {Yokoyama}, {Haines}, {Dazeley}, {Hatakeyama}, {Svoboda}, {Sullivan},
  {Turcan}, {Habig}, {Sato}, {Itow}, {Koike}, {Tanaka}, {Jung}, {Kato},
  {Kobayashi}, {Malek}, {McGrew}, {Sarrat}, {Terri}, {Yanagisawa}, {Tamura},
  {Idehara}, {Sakuda}, {Sugihara}, {Kuno}, {Yoshida}, {Kim}, {Yang}, {Yoo},
  {Ishizuka}, {Okazawa}, {Choi}, {Seo}, {Gando}, {Hasegawa}, {Inoue}, {Furuse},
  {Ishii}, {Nishijima}, {Ishino}, {Watanabe}, {Koshiba}, {Chen}, {Deng}, {Liu},
  {Kielczewska}, {Zalipska}, {Berns}, {Gran}, {Shiraishi}, {Stachyra},
  {Thrane}, {Washburn}, {Wilkes}, \& {Super-KAMIOKANDE
  Collaboration}}]{Ikeda2007}
{Ikeda}, M., {Takeda}, A., {Fukuda}, Y., {et~al.} 2007, \apj, 669, 519,
  \dodoi{10.1086/521547}

\bibitem[{{Janka} {et~al.}(2022){Janka}, {Wongwathanarat}, \&
  {Kramer}}]{Janka2022}
{Janka}, H.-T., {Wongwathanarat}, A., \& {Kramer}, M. 2022, \apj, 926, 9,
  \dodoi{10.3847/1538-4357/ac403c}

\bibitem[{{Johnston} {et~al.}(2005){Johnston}, {Hobbs}, {Vigeland}, {Kramer},
  {Weisberg}, \& {Lyne}}]{Johnson2005}
{Johnston}, S., {Hobbs}, G., {Vigeland}, S., {et~al.} 2005, \mnras, 364, 1397,
  \dodoi{10.1111/j.1365-2966.2005.09669.x}

\bibitem[{{Johnston} {et~al.}(2007){Johnston}, {Kramer}, {Karastergiou},
  {Hobbs}, {Ord}, \& {Wallman}}]{Johnson2007}
{Johnston}, S., {Kramer}, M., {Karastergiou}, A., {et~al.} 2007, \mnras, 381,
  1625, \dodoi{10.1111/j.1365-2966.2007.12352.x}

\bibitem[{{Lang} {et~al.}(2016){Lang}, {McCabe}, {Reichard}, {Selvi}, \&
  {Tamborra}}]{Lang2016}
{Lang}, R.~F., {McCabe}, C., {Reichard}, S., {Selvi}, M., \& {Tamborra}, I.
  2016, \prd, 94, 103009, \dodoi{10.1103/PhysRevD.94.103009}

\bibitem[{{Li} {et~al.}(2021){Li}, {Roberts}, \& {Beacom}}]{Li2021}
{Li}, S.~W., {Roberts}, L.~F., \& {Beacom}, J.~F. 2021, \prd, 103, 023016,
  \dodoi{10.1103/PhysRevD.103.023016}

\bibitem[{{Li} {et~al.}(2022){Li}, {Vagins}, \& {Wurm}}]{Li2022}
{Li}, Y.-F., {Vagins}, M., \& {Wurm}, M. 2022, Universe, 8, 181,
  \dodoi{10.3390/universe8030181}

\bibitem[{{MacFadyen} \& {Woosley}(1999)}]{MacFadyen1999}
{MacFadyen}, A.~I., \& {Woosley}, S.~E. 1999, \apj, 524, 262,
  \dodoi{10.1086/307790}

\bibitem[{{MacFadyen} {et~al.}(2001){MacFadyen}, {Woosley}, \&
  {Heger}}]{MacFadyen2001}
{MacFadyen}, A.~I., {Woosley}, S.~E., \& {Heger}, A. 2001, \apj, 550, 410,
  \dodoi{10.1086/319698}

\bibitem[{{Mart{\'\i}nez-Pinedo} {et~al.}(2012){Mart{\'\i}nez-Pinedo},
  {Fischer}, {Lohs}, \& {Huther}}]{Martinez2012}
{Mart{\'\i}nez-Pinedo}, G., {Fischer}, T., {Lohs}, A., \& {Huther}, L. 2012,
  \prl, 109, 251104, \dodoi{10.1103/PhysRevLett.109.251104}

\bibitem[{{Moriya} {et~al.}(2010){Moriya}, {Tominaga}, {Tanaka}, {Nomoto},
  {Sauer}, {Mazzali}, {Maeda}, \& {Suzuki}}]{Moriya2010}
{Moriya}, T., {Tominaga}, N., {Tanaka}, M., {et~al.} 2010, \apj, 719, 1445,
  \dodoi{10.1088/0004-637X/719/2/1445}

\bibitem[{{Moriya} {et~al.}(2019){Moriya}, {M{\"u}ller}, {Chan}, {Heger}, \&
  {Blinnikov}}]{Moriya2019}
{Moriya}, T.~J., {M{\"u}ller}, B., {Chan}, C., {Heger}, A., \& {Blinnikov},
  S.~I. 2019, \apj, 880, 21, \dodoi{10.3847/1538-4357/ab2643}

\bibitem[{{Moriya} {et~al.}(2018){Moriya}, {Terreran}, \&
  {Blinnikov}}]{Moriya2018}
{Moriya}, T.~J., {Terreran}, G., \& {Blinnikov}, S.~I. 2018, \mnras, 475, L11,
  \dodoi{10.1093/mnrasl/slx200}

\bibitem[{{Nagakura}(2021)}]{Nagakura2021b}
{Nagakura}, H. 2021, \mnras, 500, 319, \dodoi{10.1093/mnras/staa3287}

\bibitem[{{Nagakura} {et~al.}(2020){Nagakura}, {Burrows}, {Radice}, \&
  {Vartanyan}}]{Nagakura2020}
{Nagakura}, H., {Burrows}, A., {Radice}, D., \& {Vartanyan}, D. 2020, \mnras,
  492, 5764, \dodoi{10.1093/mnras/staa261}

\bibitem[{{Nagakura} {et~al.}(2021{\natexlab{a}}){Nagakura}, {Burrows}, \&
  {Vartanyan}}]{Nagakura2021a}
{Nagakura}, H., {Burrows}, A., \& {Vartanyan}, D. 2021{\natexlab{a}}, \mnras,
  506, 1462, \dodoi{10.1093/mnras/stab1785}

\bibitem[{{Nagakura} {et~al.}(2021{\natexlab{b}}){Nagakura}, {Burrows},
  {Vartanyan}, \& {Radice}}]{Nagakura2021c}
{Nagakura}, H., {Burrows}, A., {Vartanyan}, D., \& {Radice}, D.
  2021{\natexlab{b}}, \mnras, 500, 696, \dodoi{10.1093/mnras/staa2691}

\bibitem[{Nagakura {et~al.}(2017)Nagakura, Iwakami, Furusawa, Sumiyoshi,
  Yamada, Matsufuru, \& Imakura}]{Nagakura2017}
Nagakura, H., Iwakami, W., Furusawa, S., {et~al.} 2017, The Astrophysical
  Journal Supplement Series, 229, 42, \dodoi{10.3847/1538-4365/aa69ea}

\bibitem[{Nagakura {et~al.}(2014)Nagakura, Sumiyoshi, \& Yamada}]{Nagakura2014}
Nagakura, H., Sumiyoshi, K., \& Yamada, S. 2014, The Astrophysical Journal
  Supplement Series, 214, 16, \dodoi{10.1088/0067-0049/214/2/16}

\bibitem[{Nagakura {et~al.}(2019)Nagakura, Sumiyoshi, \& Yamada}]{Nagakura2019}
---. 2019, The Astrophysical Journal, 878, 160,
  \dodoi{10.3847/1538-4357/ab2189}

\bibitem[{{Nakazato} {et~al.}(2013){Nakazato}, {Sumiyoshi}, {Suzuki}, {Totani},
  {Umeda}, \& {Yamada}}]{Nakazato2013}
{Nakazato}, K., {Sumiyoshi}, K., {Suzuki}, H., {et~al.} 2013, \apjs, 205, 2,
  \dodoi{10.1088/0067-0049/205/1/2}

\bibitem[{Nakazato \& Suzuki(2019)}]{Nakazato2019}
Nakazato, K., \& Suzuki, H. 2019, The Astrophysical Journal, 878, 25,
  \dodoi{10.3847/1538-4357/ab1d4b}

\bibitem[{{Nakazato} \& {Suzuki}(2020)}]{Nakazato2020}
{Nakazato}, K., \& {Suzuki}, H. 2020, \apj, 891, 156,
  \dodoi{10.3847/1538-4357/ab7456}

\bibitem[{{Nakazato} {et~al.}(2022){Nakazato}, {Nakanishi}, {Harada}, {Koshio},
  {Suwa}, {Sumiyoshi}, {Harada}, {Mori}, \& {Wendell}}]{Nakazato2022}
{Nakazato}, K., {Nakanishi}, F., {Harada}, M., {et~al.} 2022, \apj, 925, 98,
  \dodoi{10.3847/1538-4357/ac3ae2}

\bibitem[{{Ng} \& {Romani}(2007)}]{Ng2007}
{Ng}, C.~Y., \& {Romani}, R.~W. 2007, \apj, 660, 1357, \dodoi{10.1086/513597}

\bibitem[{{Pascal} {et~al.}(2022){Pascal}, {Novak}, \& {Oertel}}]{Pascal2022}
{Pascal}, A., {Novak}, J., \& {Oertel}, M. 2022, \mnras, 511, 356,
  \dodoi{10.1093/mnras/stac016}

\bibitem[{{Perna} {et~al.}(2014){Perna}, {Duffell}, {Cantiello}, \&
  {MacFadyen}}]{Perna2014}
{Perna}, R., {Duffell}, P., {Cantiello}, M., \& {MacFadyen}, A.~I. 2014, \apj,
  781, 119, \dodoi{10.1088/0004-637X/781/2/119}

\bibitem[{{Roberts}(2012)}]{Roberts2012b}
{Roberts}, L.~F. 2012, \apj, 755, 126, \dodoi{10.1088/0004-637X/755/2/126}

\bibitem[{{Roberts} {et~al.}(2012){Roberts}, {Shen}, {Cirigliano}, {Pons},
  {Reddy}, \& {Woosley}}]{Roberts2012a}
{Roberts}, L.~F., {Shen}, G., {Cirigliano}, V., {et~al.} 2012, \prl, 108,
  061103, \dodoi{10.1103/PhysRevLett.108.061103}

\bibitem[{{Ronchi} {et~al.}(2022){Ronchi}, {Rea}, {Graber}, \&
  {Hurley-Walker}}]{Ronchi2022}
{Ronchi}, M., {Rea}, N., {Graber}, V., \& {Hurley-Walker}, N. 2022, \apj, 934,
  184, \dodoi{10.3847/1538-4357/ac7cec}

\bibitem[{{Shibata} {et~al.}(2014){Shibata}, {Nagakura}, {Sekiguchi}, \&
  {Yamada}}]{Shibata2014}
{Shibata}, M., {Nagakura}, H., {Sekiguchi}, Y., \& {Yamada}, S. 2014, \prd, 89,
  084073, \dodoi{10.1103/PhysRevD.89.084073}

\bibitem[{{Simpson} {et~al.}(2019){Simpson}, {Abe}, {Bronner}, {Hayato},
  {Ikeda}, {Ito}, {Iyogi}, {Kameda}, {Kataoka}, {Kato}, {Kishimoto}, {Marti},
  {Miura}, {Moriyama}, {Mochizuki}, {Nakahata}, {Nakajima}, {Nakayama},
  {Okada}, {Okamoto}, {Orii}, {Pronost}, {Sekiya}, {Shiozawa}, {Sonoda},
  {Takeda}, {Takenaka}, {Tanaka}, {Yano}, {Akutsu}, {Kajita}, {Okumura},
  {Wang}, {Xia}, {Bravo-Bergu{\~n}o}, {Labarga}, {Fernandez}, {Blaszczyk},
  {Kachulis}, {Kearns}, {Raaf}, {Stone}, {Wan}, {Wester}, {Sussman}, {Berkman},
  {Bian}, {Griskevich}, {Kropp}, {Locke}, {Mine}, {Smy}, {Sobel}, {Takhistov},
  {Weatherly}, {Ganezer}, {Hill}, {Kim}, {Lim}, {Park}, {Bodur}, {Scholberg},
  {Walter}, {Coffani}, {Drapier}, {Gonin}, {Imber}, {Mueller}, {Paganini},
  {Ishizuka}, {Nakamura}, {Jang}, {Choi}, {Learned}, {Matsuno}, {Litchfield},
  {Sztuc}, {Uchida}, {Wascko}, {Berardi}, {Calabria}, {Catanesi}, {Intonti},
  {Radicioni}, {De Rosa}, {Collazuol}, {Iacob}, {Ludovici}, {Nishimura}, {Cao},
  {Friend}, {Hasegawa}, {Ishida}, {Kobayashi}, {Nakadaira}, {Nakamura},
  {Oyama}, {Sakashita}, {Sekiguchi}, {Tsukamoto}, {Abe}, {Hasegawa}, {Isobe},
  {Miyabe}, {Nakano}, {Shiozawa}, {Sugimoto}, {Suzuki}, {Takeuchi}, {Ali},
  {Ashida}, {Hayashino}, {Hirota}, {Jiang}, {Kikawa}, {Mori}, {Nakamura},
  {Nakaya}, {Wendell}, {Anthony}, {McCauley}, {Pritchard}, {Tsui}, {Fukuda},
  {Itow}, {Murrase}, {Niwa}, {Taani}, {Tsukada}, {Mijakowski}, {Frankiewicz},
  {Jung}, {Li}, {Palomino}, {Santucci}, {Vilela}, {Wilking}, {Yanagisawa},
  {Fukuda}, {Harada}, {Hagiwara}, {Horai}, {Ishino}, {Ito}, {Koshio}, {Sakuda},
  {Takahira}, {Xu}, {Kuno}, {Cook}, {Wark}, {Di Lodovico}, {Molina Sedgwick},
  {Richards}, {Zsoldos}, {Kim}, {Tacik}, {Thiesse}, {Thompson}, {Okazawa},
  {Choi}, {Nishijima}, {Koshiba}, {Yokoyama}, {Goldsack}, {Martens}, {Murdoch},
  {Quilain}, {Suzuki}, {Vagins}, {Kuze}, {Okajima}, {Tanaka}, {Yoshida},
  {Ishitsuka}, {Matsumoto}, {Ohta}, {Martin}, {Nantais}, {Tanaka}, {Towstego},
  {Hartz}, {Konaka}, {de Perio}, {Chen}, {Jamieson}, {Walker}, {Minamino},
  {Okamoto}, {Pintaudi}, \& {Super-Kamiokande Collaboration}}]{Simpson2019}
{Simpson}, C., {Abe}, K., {Bronner}, C., {et~al.} 2019, \apj, 885, 133,
  \dodoi{10.3847/1538-4357/ab4883}

\bibitem[{SNOwGLoBES(: SuperNova Observatories with GLoBES)}]{snowglobes}
SNOwGLoBES. : SuperNova Observatories with GLoBES,
  \url{http://webhome.phy.duke.edu/~schol/snowglobes/}

\bibitem[{{Sugiura} {et~al.}(2022){Sugiura}, {Furusawa}, {Sumiyoshi}, \&
  {Yamada}}]{Sugiura2022}
{Sugiura}, K., {Furusawa}, S., {Sumiyoshi}, K., \& {Yamada}, S. 2022, Progress
  of Theoretical and Experimental Physics, 2022, 113E01,
  \dodoi{10.1093/ptep/ptac118}

\bibitem[{{Sumiyoshi} {et~al.}(2023){Sumiyoshi}, {Furusawa}, {Nagakura},
  {Harada}, {Togashi}, {Nakazato}, \& {Suzuki}}]{Sumiyoshi2023}
{Sumiyoshi}, K., {Furusawa}, S., {Nagakura}, H., {et~al.} 2023, Progress of
  Theoretical and Experimental Physics, 2023, 013E02,
  \dodoi{10.1093/ptep/ptac167}

\bibitem[{{Sumiyoshi} {et~al.}(2005){Sumiyoshi}, {Yamada}, {Suzuki}, {Shen},
  {Chiba}, \& {Toki}}]{Sumiyoshi2005}
{Sumiyoshi}, K., {Yamada}, S., {Suzuki}, H., {et~al.} 2005, \apj, 629, 922,
  \dodoi{10.1086/431788}

\bibitem[{{Suwa} {et~al.}(2019){Suwa}, {Sumiyoshi}, {Nakazato}, {Takahira},
  {Koshio}, {Mori}, \& {Wendell}}]{Suwa2019}
{Suwa}, Y., {Sumiyoshi}, K., {Nakazato}, K., {et~al.} 2019, \apj, 881, 139,
  \dodoi{10.3847/1538-4357/ab2e05}

\bibitem[{{Vartanyan} {et~al.}(2019){Vartanyan}, {Burrows}, \&
  {Radice}}]{Vartanyan2019}
{Vartanyan}, D., {Burrows}, A., \& {Radice}, D. 2019, \mnras, 489, 2227,
  \dodoi{10.1093/mnras/stz2307}

\bibitem[{{Vartanyan} {et~al.}(2022){Vartanyan}, {Coleman}, \&
  {Burrows}}]{Vartanyan2022}
{Vartanyan}, D., {Coleman}, M. S.~B., \& {Burrows}, A. 2022, \mnras, 510, 4689,
  \dodoi{10.1093/mnras/stab3702}

\bibitem[{{Yamasaki} \& {Yamada}(2005)}]{Yamasaki2005}
{Yamasaki}, T., \& {Yamada}, S. 2005, \apj, 623, 1000, \dodoi{10.1086/428496}

\bibitem[{{Yamasaki} \& {Yamada}(2006)}]{Yamasaki2006}
---. 2006, \apj, 650, 291, \dodoi{10.1086/507067}

\bibitem[{{Yamasaki} \& {Yamada}(2007)}]{Yamasaki2007}
{Yamasaki}, T., \& {Yamada}, S. 2007, in American Institute of Physics
  Conference Series, Vol. 937, Supernova 1987A: 20 Years After: Supernovae and
  Gamma-Ray Bursters, ed. S.~{Immler}, K.~{Weiler}, \& R.~{McCray}, 344--348,
  \dodoi{10.1063/1.3682927}

\bibitem[{{Zhang} {et~al.}(2008){Zhang}, {Woosley}, \& {Heger}}]{Zhang2008}
{Zhang}, W., {Woosley}, S.~E., \& {Heger}, A. 2008, \apj, 679, 639,
  \dodoi{10.1086/526404}

\end{thebibliography}
\bibliographystyle{aasjournal}



\end{document}